\documentclass[useAMS,fleqn,usenatbib]{mnras}

\usepackage{newtxtext,newtxmath}

\usepackage[T1]{fontenc}
\usepackage{ae,aecompl}

\usepackage{graphicx} 
\usepackage{amsmath}  
\usepackage{amssymb}  

\usepackage{chemformula} 
\usepackage{siunitx} 
\usepackage[outdir=./]{epstopdf}

\newcommand{\Av}{\ensuremath{\mathrm{A}_{\mathrm{v}}}}
\newcommand{\Avone}{\ensuremath{\mathrm{A}_{\mathrm{v},1}}}
\newcommand{\krome}{\textsc{krome}}
\newcommand{\ramses}{\textsc{ramses}}
\newcommand{\lampray}{\textsc{lampray}}
\newcommand{\cloudy}{\textsc{Cloudy}}

\newcommand{\enzo}{\textsc{Enzo}}
\newcommand{\Lampray}{\textsc{Lampray}}
\newcommand{\ctoray}{\textsc{c2ray}}
\newcommand{\healpix}{\textsc{healpix}}
\newcommand{\rayramses}{\textsc{rayramses}}
\newcommand{\mesa}{\textsc{mesa}}
\newcommand{\flash}{\textsc{flash}}
\newcommand{\treecol}{\textsc{treecol}}
\newcommand{\mpi}{\textsc{MPI}}
\newcommand{\openmp}{\textsc{OpenMP}}
\newcommand{\rsph}{\textsc{rsph}}
\newcommand{\ftte}{\textsc{ftte}}

\usepackage{color,soul}
\definecolor{dark-red}{rgb}{0.75, 0.00, 0.00}
\definecolor{hlcolor}{rgb}{1.00, 0.90, 0.85}\sethlcolor{hlcolor}
\definecolor{gray}{rgb}{0.7, 0.7, 0.7}
\definecolor{ctroels}{rgb}{0.65, 0.00, 0.65}

\newcommand{\KIDA}{\textsc{KIDA} database\footnote{\url{http://kida.obs.u-bordeaux1.fr}} \citep{Wakelam2012ApJS}}
\newcommand{\UMIST}{\textsc{UMIST} database\footnote{\url{http://udfa.ajmarkwick.net}} \citep{McElroy2013A&A}}
\newcommand{\ALADDIN}{Aladdin database\footnote{\url{https://www-amdis.iaea.org/ALADDIN}}}

\newcommand{\eqn}[1]{Eqn.(\ref{#1})}

\newcommand{\sect}[1]{Sect.~\ref{#1}}
\newcommand{\fig}[1]{Fig.~\ref{#1}}
\newcommand{\tab}[1]{Tab.~\ref{#1}}

\newcommand{\dd}{{\rm d}}
\newcommand{\chemicalNetworkTableCaption}{\caption{Chemical network in Test~4}}
\newcommand{\chemicalNetworkTableCaptionCont}{\contcaption{}}
\newcommand{\chemicalNetworkTableLabel}{tab:chemical_network}
\def\ontop#1{\vtop{\null\hbox{#1}}}

\newenvironment{chemical_network_table}[1]{
  \begin{table*}
      #1 }{
  \end{table*} }

\newenvironment{chemical_network_tabular}{
  \begin{tabular}{llr@{\:}lll}
    \hline
    No. & Reaction & \multicolumn{2}{l@{\:}}{Rate coefficient} & Notes & Ref.\\
    \hline }{
    \hline
  \end{tabular} }

\title[{\Lampray}: long characteristics adaptive radiation hydrodynamics]{{\Lampray}: Multi-group long characteristics ray tracing for adaptive mesh radiation hydrodynamics}

\author[Frostholm, Haugb{\o}lle, and Grassi]{
Troels Frostholm,$^{1}$\thanks{Contact e-mail: \href{mailto:frostholm@nbi.ku.dk}{frostholm@nbi.ku.dk}}
Troels Haugb{\o}lle,$^{1}$\thanks{Contact e-mail: \href{mailto:haugboel@nbi.ku.dk}{haugboel@nbi.ku.dk}} and
Tommaso Grassi,$^{1,2,3}$\thanks{Contact e-mail: \href{mailto:tgrassi@usm.lmu.de}{tgrassi@usm.lmu.de}}
\\
$^{1}$Centre for Star and Planet Formation, Niels Bohr Institute, and Natural History Museum of Denmark, University of Copenhagen,\\
{\O}ster Voldgade 5--7, DK--1350 Copenhagen, Denmark\\
$^{2}$Universit\"ats--Sternwarte M\"unchen, Scheinerstr.~1, DE--81679 M\"unchen, Germany\\
$^{3}$Excellence Cluster Origin and Structure of the Universe, Boltzmannstr.~2, DE--85748 Garching bei M\"unchen, Germany}

\date{in original form 2018 September XX}

\pubyear{2018}

\begin{document}

\label{firstpage}
\pagerange{\pageref{firstpage}--\pageref{lastpage}}
\maketitle

\begin{abstract}
We present {\lampray}: a multi-group long characteristics ray tracing method for adaptive mesh radiation hydrodynamics
in the {\ramses} code. It avoids diffusion, captures shadows, and treats colliding beams correctly, and therefore
complements existing moment-based ray tracing in {\ramses}.
{\lampray} includes different options for interpolation between ray and cell domain, and use either integral, Fourier, or an implicit
{\ctoray} method for hydrogen ionization to solve the radiative transfer.
The opacity can either be tabulated or computed through a coupling to the general non-equilibrium astro-chemistry framework {\krome}.
We use an H-He-C-O network with 36 species and 240 reactions to track the photo-chemistry in the interstellar medium across 6 and 10 orders of magnitude in
temperature and density. Self-shielding prescriptions for H$_2$ and CO are used together with a new model for the diffuse interstellar
UV-field. We also track the dust temperature, formation of H$_2$ on grains, and H$_2$O and CO ices in detail.
{\lampray} is tested against standard benchmarks for molecular cloud and star formation simulations, including the
formation of a Str\"{o}mgren sphere, the expansion of an ionization front, the photo-evaporation of a dense clump, and the
H-He-C-O chemistry in a static photo-dissociation front.
Efficient parallelisation is achieved with a separate
domain decomposition for rays where points along a ray reside in the same memory space, and data movement from
cell- to ray-domain is done with a direct hash-table lookup algorithm. 
Point sources are treated without splitting rays, and therefore the method currently only scales to a few point sources, while diffuse radiation
has excellent scaling.
\end{abstract}


\section{Introduction}
The interaction between radiation and matter is important for a majority of objects in the Universe, and often required for accurate,
quantitative numerical modelling in astrophysics. In the interstellar medium (ISM) the role of radiative stellar feedback versus feedback from
supernovae and galactic dynamics is debated \citep{Dale2013,KrumholzPPVI,Padoan2016,Peters2017,Butler2017,Howard2017}.
Numerical models that can address several of these issues at once are therefore needed to
explore the importance of the different effects.

Star forming regions harbour a complex filamentary distribution of gas and dust driven by supersonic turbulence. They can be observed through
the thermal emission from cold dust, and in atomic and molecular line transitions, but observations only provide a two-dimensional projected
snapshot in time. Due to the complex nature of the system synthetic observations of numerical models that can aid in interpreting the
observations and unravel what are the dominating physical processes at play, are highly desirable \citep{Haworth2018}.
Connecting numerical models to observations
requires accurate modelling of the chemistry and radiation field. Many tools exist to post process models creating dust and line radiation maps, but
either assumed or calculated chemical abundances have to be provided as input. In star forming regions, in particular close to pre-stellar cores,
the chemical equilibrium time scale can be longer than the dynamical time scale, and assuming chemical and radiative equilibrium is in many
cases not a good assumption. Consequently it is useful to treat the non-equilibrium chemical evolution directly in the numerical model.
At lower optical depths and in the vicinity of protostars the chemistry is to a large extend controlled by the radiation, and a proper model
therefore requires a solution of both radiative transfer and photo-chemistry.

On smaller scales the collapse of dense clouds, the formation of
protostellar systems and accretion on to the protostar is controlled by the environment in which it takes place \citep{Kuffmeier2017},
as well as by several physical processes including magnetic fields, radiative heating and cooling, and thermal and radiative ionization \citep{TanPPVI}. 
The formation of a possibly rotationally stable first core
is controlled by radiative cooling in changing opacity regimes, and its subsequent evolution is affected by radiative heating from the accretion shock
on the forming protostar \citep{Vaytet2017}.
The protoplanetary disc surface temperature and degree of ionization is to a large degree controlled by radiation from the
protostar, and the inner thermal structure by radiative and turbulent diffusive heat transport \citep{Dullemond2010}. 
Massive protostars are sources of copious high-energy
radiation, launch radiatively driven disc winds, and ionize their immediate surroundings creating HII regions \citep{Peters2010}.
Radiative transfer is therefore desirable for
detailed modelling of protostellar systems in different evolutionary stages and in particular at high masses.

In molecular clouds the energy cascades from large to small scales, generating a roughly self-similar structure down to the collapsing cores. This
process is multi-scale in nature, coupling the dynamics of molecular clouds at the tens of pc scale to proto-planetary systems through the
accretion history and magnetic field anchoring. The gravitational collapse channels gas through a disc to the star, in a delicate balance with the
environment. The connection of scales in the above processes can only be addressed with adaptive multi-scale mesh or meshless numerical
methods  \citep{Kuffmeier2017}.

{\ramses} \citep{Teyssier2002} is a multi-physics oct-based adaptive mesh refinement code for astrophysics, which is used by a large
community. An abundance of radiative transfer codes have been implemented in {\ramses}. \textsc{ATON} \citep{Aubert2008MNRAS} is a
time-dependent radiative transfer tool for cosmic reionization based on the M1 moment method, used in conjunction with {\ramses}. Both
flux-limited diffusion \citep{Gonzalez2015A&A,Commercon2014,Commercon2011A&A} and the M1 method \citep{Rosdahl2013,Rosdahl2015} have been
implemented on adaptive mesh for use in radiation-hydrodynamics. {\rayramses} \citep{Barreira2016JCAP} is a ray-tracing tool for post-processing of
cosmological simulations using curved rays to capture gravitational lensing, while \citet{Valdivia2014} provide an algorithm to estimate column densities
taking advantage of the fully threaded tree in {\ramses}.

In this paper we present a new algorithm for radiative transfer in {\ramses} that we call {\lampray}: Long characteristics AMr Parallel RAY tracing.
It stands out from the current radiative transfer algorithms in {\ramses} in two keys aspects: Firstly it is a multi-group long characteristics ray
tracing method, allowing for an accurate treatment of shadows and colliding beams. Secondly it is coupled to the non-equilibrium chemistry
code {\krome}, enabling relatively easy construction of complex photo-chemical models.

There already exists a number of algorithms for parallel ray tracing radiation hydrodynamics (RHD) on adaptive meshes. \citet{Wise2011MNRAS} introduced ray tracing RHD from point sources on the block adaptive mesh of the {\enzo} code, using adaptive ray tracing \citep{Abel2002MNRAS}, where the set of directions being traced from the source are split as the distance grows to maintain a high angular resolution. \citet{Rijkhorst2006A&A} did the same for the {\flash} code, also on a block AMR grid, using so-called hybrid characteristics, where the local radiation field is found on blocks using long characteristics, and these are connected with short characteristics between blocks. \citet{Buntemeyer2016NewA} extended this method to the diffuse radiation field component. These methods are however unsuited for {\ramses}' octree grid due to the high computational cost of looking up neighbour cells. Also worth mentioning is the {\treecol} method by \citet{Clark2012MNRAS}, which estimates column depths by tracing along {\healpix} directions in an octree. 
{\lampray} has been carefully optimised for running at scale with a hash-table based ray tracing algorithm and uses a dual parallel decomposition where
the solution of the radiative transfer is performed in a separate ray-centric domain with near-perfect parallel load-balancing. The implementation
is modular and contains several different modules for solving the radiative transfer along the rays, for computing the opacities, and for applying the result 
to gas heating/cooling, chemistry etc.

In section \ref{sec:Numerical methods} we describe the numerical methods used for solving radiation hydrodynamics with an emphasis on
the equation of radiative transfer, photo-chemistry, and gas heating/cooling. Section \ref{sec:implementation} describes the implementation
of ray tracing, how the change of domain decomposition and scaling is achieved, and how boundary conditions are handled. Section \ref{sec:tests} 
checks the validity of the code in a number of benchmark simulations, performance, and scaling is reported in section \ref{sec:performance},
while section \ref{sec:summary} summarises our method and discusses future applications.

\section{Numerical methods}\label{sec:Numerical methods}
Radiation hydrodynamics model the interaction and evolution of radiation and matter with a fluid description. Radiation can transport energy
and momentum through emission, absorption, and scattering. In addition, when high-energy photons are present radiation has important consequences
for the chemical composition of the gas, and the photo-chemistry can drive the gas out of local thermal equilibrium. A comprehensive description of
radiation hydrodynamics therefore contains three main components: \emph{hydrodynamics}, \emph{radiative transfer}, and \emph{chemical evolution}.

\emph{Hydrodynamics} describes the bulk transport and mixing of matter and is captured in the equations for conservation of mass and species,
balance of energy, and the Navier-Stokes equation for the momentum. Written in conservative form they are
\begin{align}
\frac{\partial\rho}{\partial t} + \boldsymbol{\nabla}\cdot [ \rho\mathbf{u} ] & = 0 \\
\frac{\partial\rho X_s}{\partial t} + \boldsymbol{\nabla}\cdot [ \rho X_s \mathbf{u} ] & = m_s (F_s - D_s) \\
\frac{\partial\rho\mathbf{u}}{\partial t} + \boldsymbol{\nabla}\cdot [ \rho\mathbf{u}\otimes\mathbf{u} + P\mathbb{I} ] & = 0 \\
\frac{\partial \rho e_\mathrm{tot}}{\partial t} + \boldsymbol{\nabla}\cdot [ (\rho e_\mathrm{tot} + P) \mathbf{u} ] & = \Gamma_\mathrm{gas} - \Lambda_\mathrm{gas}\,,
\end{align}
where $\rho$ is the density, $\mathbf{u}$ the velocity, $X_s$ the mass abundance, $m_s$ the mass, and $F_s$, $D_s$ the
rates of formation and destruction of species $s$. The corresponding number density is $n_s = \rho X_s / m_s$.
$P$ is the gas pressure, $\rho e_\mathrm{tot}$ is the total energy density of the fluid,
which is the sum of the internal energy density $e_\mathrm{int}$ and the kinetic energy density
\begin{equation}
\rho e_\mathrm{tot} = \rho e_\mathrm{int} + \frac{1}{2}\rho u^2\,,
\end{equation}
and $\Gamma_\mathrm{gas}$, $\Lambda_\mathrm{gas}$ are the total volumetric heating and cooling rates describing the exchange of
energy from other processes, such as release and storage of chemical energy, and absorption or emission of radiation.
We assume that the equation of state is that of an ideal gas relating the pressure and internal
energy as $P = (\gamma - 1)e_\mathrm{int}$, but allow the adiabatic index to be a function of the state and composition of the gas
$\gamma(\rho, T, X_s)$. The gas temperature is related to the pressure as $P = n_\mathrm{gas} k_B T$, and the total number density can be found
from the density and abundances as $\rho = \mu\, n_\mathrm{gas}$. The molecular weight of the gas is $\mu = 1 / (\sum X_s / m_s)$.

{\lampray} does not compute the radiative pressure and momentum, and we have for clarity
left out other contributions from e.g.~magnetic fields, self-gravity, and external forces, which are included in {\ramses}.

The dynamical time-scale of the fluid is dictated by the bulk motion and the speed of sound. The chemistry, contained in the terms
$F_s$, $D_s$, often evolves on a different and faster time-scale, and the same is true for the radiation, which affects the
chemistry. The chemistry and the radiation determine the heating and cooling. In {\ramses} we assume a decoupling between
the three components and use operator splitting, to avoid evolving the hydrodynamics on the much shorter time-scale of the other processes,
which would be prohibitively expensive in three dimensional models. First, the \emph{hydrodynamics} is evolved with a single explicit hydrodynamical timestep, disregarding the source terms, corresponding to setting $F_s = D_s = \Lambda_\mathrm{gas} = \Gamma_\mathrm{gas} = 0$ in
the equations above. Then, the \emph{radiative transfer} is calculated with {\lampray}. The chemical species, $\rho X_s$, and the average
radiation field together
with the temperature in each computational cell is then used as an input for the \emph{chemical evolution} and the heating and cooling.
{\krome} is used to evolve the species, $\rho X_s$, and compute the change in the temperature, $T$.
If a model is run without chemistry, $\Gamma_\mathrm{gas} - \Lambda_\mathrm{gas}$ is obtained directly from the radiative transfer.

\emph{Radiative transfer} is the flow of photons through a medium governed by a transport equation, which includes the advection of photons
in a given direction $\boldsymbol{n}$ and the absorption and emission of new photons, described by the extinction coefficient $\alpha_{\nu}$ and
emissivity $j_{*,\nu}$ at frequency $\nu$. We are interested in radiation in the ISM and around new-born stars, and the 
light-crossing time is much lower than the dynamical time, which in a non-cosmological setting corresponds to having non-relativistic velocities.
{\lampray} therefore solves the time-independent equation of radiative transfer
\begin{equation}\label{eq:radiative-transfer}
 \boldsymbol{n}\left(\boldsymbol{x}\right)\cdot\boldsymbol{\nabla}I_{\nu}\left(\boldsymbol{x},\Omega\right)=j_{*,\nu}\left(\boldsymbol{x}\right)-\alpha_{\nu}\left(\boldsymbol{x}\right)I_{\nu}\left(\boldsymbol{x},\Omega\right)\,,
\end{equation}
where $I_{\nu}\left(\boldsymbol{x},\Omega\right)$ is the intensity in a direction $\Omega$ at a point $\boldsymbol{x}$, and we have 
neglected scattering. The mean intensity, which is the main input to compute the effect of radiation on matter, is 
\begin{equation}
 J_{\nu}\left(\boldsymbol{x}\right)=\frac{1}{4\pi}\int_{0}^{4\pi}I_{\nu}\left(\boldsymbol{x},\Omega\right)\mathrm{d}\Omega\,,
\end{equation}
while the intensity of the emitted radiation is given by the source function
\begin{equation}
 S_{\nu}\left(\boldsymbol{x}\right)=\frac{j_{*,\nu}\left(\boldsymbol{x}\right)}{\alpha_{\nu}\left(\boldsymbol{x}\right)}\,.
\end{equation}
Using the non-dimensional optical depth $\mathrm{d}\tau_\nu = \alpha_{\nu} \mathrm{d}x$ that measures the fractional decrease in intensity
we can rewrite \eqn{eq:radiative-transfer} along a given direction $\mathbf{n}$ as
\begin{equation}\label{eq:radiative-transfer2}
 \frac{\mathrm{d}I_{\nu}\left(\tau_\nu\right)}{\mathrm{d}\tau_\nu}=S_{\nu}\left(\tau_\nu\right) - I_{\nu}\left(\tau_\nu\right)\,.
\end{equation}
This looks deceptively simple, but in principle has to be solved along every ray connecting points inside the model for each frequency,
making it a six dimensional problem. Further contributing to the problem, the absorption coefficients depend on the microphysics, including
the abundance and state of the gas, and the amount and nature of the dust. A popular approach is to tabulate the absorption as a function
of density and temperature using a table specifically crafted for the problem at hand. An alternative is to compute it directly from
the state of the system $(\rho X_s, T)$. To do this we need to know the chemical composition.

The \emph{chemical evolution} is found by solving the rate equations individually in each cell of the simulation domain, assuming that no mixing
or advection takes place. This is most conveniently formulated in terms of the number densities $n_s$
\begin{equation}\label{eq:chemical-evolution}
  \frac{\mathrm{d}n_s}{\mathrm{d}t} = F_s - D_s\,,
\end{equation}
where the formation and destruction rates of a species are
\begin{align}\label{eq:chemical-evolution2}
  F_s & = \sum_{j \in \mathcal{F}_s} \left( k_j \prod_{r \in \mathcal{R}_j} n_{r} \right) \\
  D_s & = \sum_{j \in \mathcal{D}_s} \left( k_j \prod_{r \in \mathcal{R}_j} n_{r} \right)\,,
\end{align}
$\mathcal{R}_j$ is the set of reactants for reaction $j$ of the form $r_1 + r_2 + ... \rightarrow p_1 + p_2 + ... $, and the sums are over reactions
where species $s$ is formed ($\mathcal{F}_s$) or destroyed ($\mathcal{D}_s$). The rate coefficients $k_j$ may depend on temperature 
and/or radiation field among other factors. In particular, for a photo-ionization or photo-dissociation reaction 
\begin{equation}\label{eq:photo-rate-coefficients}
  k_{j} = 4\pi \int ^\infty_{\nu_{\mathrm{th}}} \frac{\sigma_{\mathrm{j},\nu} J_\nu}{h\nu} \mathrm{d}\nu,
\end{equation}
where $\sigma_{j,\nu}$ is the frequency dependent photo cross section for reaction j and $\nu_{\mathrm{th}}$ is the energy threshold of the reaction.
The removal of photons by these processes along with the extinction coefficient of dust $\alpha_\nu^{\textrm{d}}=\kappa_\nu^{\textrm{d}} f_d \rho $ gives a total extinction coefficient
\begin{equation}\label{eq:opacity}
  \alpha_\nu = \alpha_\nu^{\textrm{d}} + \sum_{j\in \mathcal{P}} \sigma_{j,\nu} \prod_{r \in \mathcal{R}_j} n_{r}\, ,
\end{equation}
where $\mathcal{P}$ is the set of all photo-chemical reactions, $f_d$ is the dust to gas mass ratio and $\kappa_\nu^{\textrm{d}}$ the dust opacity. 
In addition to evolving the chemistry, {\krome} also evolves the temperature inside each cell accounting
for heating and cooling because of the radiation field or because of other processes such as e.g.~exo/endothermic reactions.
The temperature evolves according to
\begin{equation}
k_B \frac{\mathrm{d}T}{\mathrm{d}t} = (\gamma - 1)\frac{\Gamma_\mathrm{gas} - \Lambda_\mathrm{gas}}{n_\mathrm{gas}}\,.
\end{equation}
In this paper, our main concern is {\lampray}, and we refer to \citet{Grassi2014MNRAS,Grassi2017MNRAS} for an overview of the different
cooling and heating processes that are included in {\krome}. 

\subsection{Radiative transfer}
The radiative transfer equation is solved explicitly using long characteristics ray tracing. We discretise the intensity in frequency into a set of bins
so $I_{\textrm{bin}} = \int ^{\nu_{u\left(\textrm{bin}\right)}} _{\nu_{l\left(\textrm{bin}\right)}} I_\nu \textrm{d}\nu$, where $\nu_{u\left(\textrm{bin}\right)}$
and $\nu_{l\left(\textrm{bin}\right)}$ are the upper and lower frequency limits of the bin.
In direction and space, we discretise along individual, global, rays with constant $\boldsymbol{n}$ along which \eqn{eq:radiative-transfer2} takes the form
\begin{equation}\label{eq:RT-eq-1D}
 \frac{dI_{\textrm{bin}}}{d\tau_\textrm{bin}}=S_{\textrm{bin}}-I_{\textrm{bin}}\,,
\end{equation}
with source function $S_{\textrm{bin}}=j_{*,\textrm{bin}} / \alpha_{\textrm{bin}}$ and optical depth element $d\tau_{\textrm{bin}}=\alpha_{\textrm{bin}} dl$,
where $l$ is the coordinate along the ray. In {\ramses} the computational domain is covered by an adaptive mesh described with a fully threaded octree,
and the rays are considered to go through leaf cells on all levels of the mesh, and each extend from one end of the simulation box to the other. Each ray
is discretised into points that lie centred on the coordinate axis of fastest propagation within each leaf cell. The radiation field is separated into two
components: a diffuse field and
a point source field. In both cases, the discretisation in direction uses the {\healpix} scheme, which provides $N_{\textrm{dir}}=12 N_{\textrm{side}}^2$
equal area directions at resolution $N_{\textrm{side}}$. In the diffuse field, for a given direction a set of rays are chosen such that at a given level of
refinement the spacing between rays is equal, and such that every leaf cell is covered by at least one ray. In the point source field, every point source is
simply intersected by one ray in each direction (see \sect{sec:ray-tracing}). 

Several algorithms are provided for solving \eqn{eq:RT-eq-1D} including Feautriers method \citep{Feautrier1964SAOSR,1971MNRAS.152....1H} and the integral method of orders 0, 1 and 2. Since rays are not discretized at cell centres, interpolation is required of $\alpha_{\textrm{bin}}$ and $j_{*,\textrm{bin}}$ from cell centres to ray points, and of $I_{\textrm{bin}}$ from the ray points to the cell centres (see \sect{sec:ray-geometry-and-interpolation}). The intensity is finally averaged over $N_{dir}$ directions

\begin{equation}
 J_{\textrm{bin}}=\frac{1}{N_{\textrm{dir}}}\sum_{i=1}^{N_{\textrm{dir}}}I_{\textrm{bin},i}\,,
\end{equation}
where we have exploited that by construction {\healpix} is an equal-area covering of the sphere.

The three components, radiative transfer, chemistry, and hydrodynamics are treated as independent within a time step, and are combined using
operator splitting as follows. In a single timestep, first {\ramses} evolves the hydrodynamics, then, the radiative transfer equation is solved
with the updated densities and temperatures, and the chemical compositions inside the cells of the previous time step.
Finally, the chemical network is evolved in each cell ignoring any $PdV$ work using the new radiation field in each cell, giving a net heating / cooling
term by which the total energy is changed. A new mean molecular weight and adiabatic index is also computed after the chemistry update.
The timestep can be adaptively sub-cycled, proportional to the cell size. Since radiation propagation is treated as instantaneous, it connects
regions with short time steps separated by regions with long time steps. Therefore, in the current version of {\lampray}, the radiative transfer
is solved globally at shortest timestep, updated in the leaf cells, and then propagated by averaging to the coarser levels.

\subsection{Coupling to chemistry}
We have implemented two methods that solve \eqn{eq:chemical-evolution}, \eqn{eq:photo-rate-coefficients} and \eqn{eq:opacity}. The first method
(\sect{photo-chemistry-with-krome}) solves it for any set of photo-reactions and in connection with an arbitrary chemical network using the {\krome}
package. It however requires a fairly restrictive time step if a correct propagation speed of fast ionization fronts is required, though it should be
emphasized that it is photon-conserving and converges to the right radius once front has propagated. The second method
(\sect{hydrogen-ionization-with-ctoray}) is well suited for solving the propagation of fast moving ionization fronts, but only treats ionization of H,
and was used for the StarBench benchmark \citep{Bisbas2015MNRAS}.

\subsubsection{Photo-chemistry with KROME}\label{photo-chemistry-with-krome}
{\krome} \citep{Grassi2014MNRAS} is a framework that generates code from a description of a chemical network to efficiently solve its time-dependent evolution. It is able to compute photo-chemical rate coefficients based on frequency-binned mean intensities and on frequency-dependent photo cross-sections where available, and include the latter to evaluate the total frequency-binned opacity. Additionally, dust opacities may also be applied through the use of well-established lookup tables\footnote{\url{https://www.astro.princeton.edu/~draine/dust/dust.diel.html}}.

\begin{figure}
\begin{center}
\includegraphics[width=0.3\textwidth]{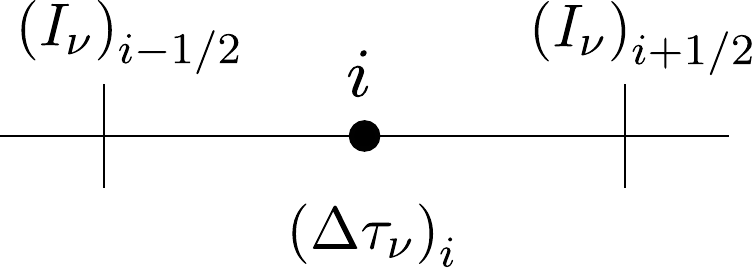}
\caption{Discretisation along a ray. \label{fig:discretisation-along-ray}}
\end{center}
\end{figure}

\citet{Mellema2006NewA} describes a very stable and accurate implicit method for solving the problem of ionization front propagation (see \sect{hydrogen-ionization-with-ctoray}). Unfortunately we cannot easily use it with {\krome}, since it relies on formulating analytical expressions for the ionization fraction assuming a constant ionization rate coefficient during a single time step, and this is not possible for a general chemical network that may also include arbitrary non-photochemical reactions. We can, however, use the part of the method \citep[originally from][]{Abel1999ApJ} where the photo-chemical rate is set directly by the energy lost to extinction. It is equivalent to using the average intensity over a ray element
optical depth $\Delta \tau_\nu$, as shown in the following. Consider the discretisation of intensity and optical depth along a ray given in \fig{fig:discretisation-along-ray}. 

The ionization rate per frequency $\Gamma_{\nu}$ at frequency $\nu$ can be written
\begin{equation}
 \Gamma_{\nu}=\sigma_{\nu}\frac{\left\langle I_{\nu}\right\rangle _{i}}{h\nu}\,,
\end{equation}
with $\sigma_{\nu}$ the cross section of the photo-chemical reaction in question, and $\left\langle I_{\nu}\right\rangle _{i}$ the average intensity over ray element $i$. Within ray element $i$, the intensity, subject to extinction due to ionization events, is 
\begin{equation}
I_{\nu}\left(\tau_{\nu}-\left(\tau_{\nu}\right)_{i-1/2}\right) = \left(I_{\nu}\right)_{i-1/2} e^{-\left(\tau_{\nu}-\left(\tau_{\nu}\right)_{i-1/2}\right)},
\end{equation}
so the average intensity over the optical depth element is
\begin{equation}
 \left\langle I_{\nu}\right\rangle _{i}=\frac{\left(I_{\nu}\right)_{i-1/2}}{\Delta\tau_{\nu}}\int_{0}^{\Delta\tau_{\nu}}e^{-\tau_{\nu}}\dd\tau_{\nu}\,.
\end{equation}  
Given reactant number density $n$ and ray element size $\Delta l$, the ray element optical depth is 
$\Delta\tau_{\nu}=\sigma_{\nu}n\Delta l$, so we have
\begin{equation}
 \Gamma_{\nu}=\frac{\left(I_{\nu}\right)_{i-1/2}}{h\nu}\frac{1-e^{-\Delta\tau_{\nu}}}{n\Delta l}\,.
\end{equation}
The final expression corresponds to the one suggested by \citet{Abel1999ApJ}, but for Cartesian geometry instead of spherical-polar.
The expression also holds for more than one photo-reaction. In this case, the optical depth element is
\begin{equation}
 \Delta\tau_{\nu}=\sum_{\mathrm{r}\in\mathcal{P}}\sigma_{\nu,\mathrm{r}}n_{s(\mathrm{r})}\Delta l\,,
\end{equation}
where the sum is over all photo-reactions $\mathcal{P}: s(\mathrm{r}) + \gamma \rightarrow p(\mathrm{r})$. $s(\mathrm{r})$ is the reactant and
$p(r)$ is the product in reaction r.
The per-volume photon absorption rate is then
\begin{eqnarray}
\dot{n}_{\textrm{photon},\nu} & = & \sum_{\mathrm{r}\in\mathcal{P}}\Gamma_{\mathrm{r},\nu}n_{s(\mathrm{r})}\\
 & = & \frac{\sum_{\mathrm{r}\in\mathcal{P}}\sigma_{\nu,\mathrm{r}}n_{s(\mathrm{r})}\left\langle I_{\nu}\right\rangle _{i}}{h\nu}\\
 & = & \frac{\left(I_{\nu}\right)_{i-1/2}}{h\nu}\frac{1-e^{-\Delta\tau_{\nu}}}{\Delta l},
\end{eqnarray}
where the cross sections are assumed to be zero below the threshold frequency. Using the average intensity thus allows us to interpolate
frequency-binned opacities to the rays, and compute optical-depth averaged intensities there. This is crucial because the cost of
interpolation and communication is decided by the choice of frequency binning, instead of being dictated by the often large number of species. 
Photons can also be lost to excitations of molecules that do not result in a reactive process, and to take into account such non-reactive cross-sections, these interactions can be included as dummy reactions in the chemical network. In this paper we ignore such extra processes, but we do include \ch{H2} and \ch{CO} self-shielding (see \sect{sec:selfshielding}), since their contribution is important in photo-dissociation regions (PDRs).

Assuming a constant source function over a ray element, the average intensity over $\Delta\tau$ is (dropping the explicit dependence on frequency $\nu$)
\begin{equation}\label{eq:tau-averaged-I}
 \left\langle I\right\rangle _{i}=\frac{1}{2}\left(I_{i-1/2}^{+}+I_{i+1/2}^{-}-2S_{i}\right)\frac{1-e^{-\Delta\tau}}{\Delta\tau}+S_{i}\,.
\end{equation}
We solve for the staggered intensity using a 0th order integral method, since this is stable even for unresolved ionization- and dissociation fronts. This works in the cases we consider, where the emission comes entirely from optical thin regions. To use the method for e.g.~an emitting atmosphere, \eqn{eq:tau-averaged-I} could still be used with a solver that is more accurate in thick regions, as long as the optical surface is resolved, since \eqn{eq:tau-averaged-I} goes to the correct limit in the optically thick case. 

\begin{figure}
\begin{center}
\includegraphics[width=0.48\textwidth]{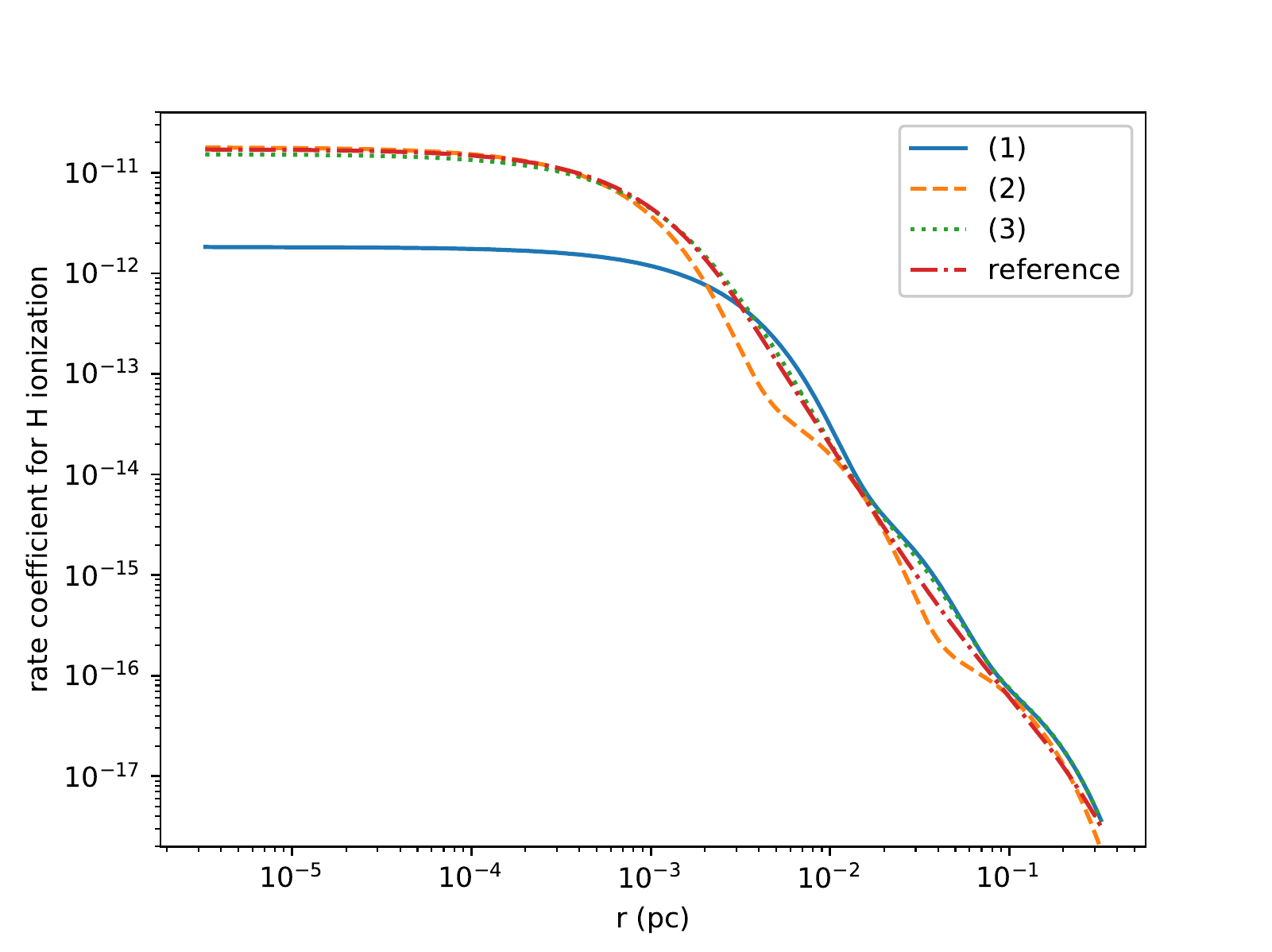}
\caption{Test~4: Three approaches to frequency binning and average photo cross section. (1) Bin limits 13.6, 24.6, 54.4, 100 eV and unweighted average. (2) Same bin limits, but cross section according to \eqn{eq:intensity-weighted-cross-section}. (3) Direct average, with two extra bins with limits 16.50 and 31.43 eV. Reference has direct average and 200 logarithmic bins from 5 to 100 eV. The test is in a pure n$_{\mathrm{H}}=100$ medium using the ISRF from Black.}\label{fig:PDR-frequency-bins}
\end{center}
\end{figure}

When discetising in frequency, we need to define an appropriate frequency-mean opacity. In our example calculation of the structure of a photo-ionized region irradiated with the spectrum from \citet{Black1987ASSL} (\sect{sec:PDR-tests}), we find that the most accurate method is to use the unweighted average, combined with careful placement of bin limits. To describe the ionization transitions of \ch{H} and \ch{He} as accurately as possible with few frequency bins, we compared the following
three approaches to a reference solution using 100 logarithmically spaced bins.
(1) Three bins above the ionization threshold of \ch{H} (13.6 eV) that includes separations corresponding to the ionization threshold of \ch{H}, \ch{He}, and \ch{He+}, giving bin limits in eV: (13.6, 24.6, 54.4, 100). Photo cross sections are evaluated at the midpoint value in each bin. 
(2) The same frequency bins, but using the convolution of the cross section with the ISRF, i.e. 
\begin{equation}
\left<\sigma\right>_{\mathrm{bin}} \equiv \frac{\left< \sigma J_\mathrm{Black} \right>_{\mathrm{bin}}}{\left< J_\mathrm{Black} \right>_{\mathrm{bin}}}. \label{eq:intensity-weighted-cross-section}
\end{equation}
(3) The cross section at the midpoint, but with an extra bin in the \ch{H} and \ch{He} regions to improve these two important transitions. The location of the extra bin limits were optimised to minimize the error in the optically thin rate coefficients, resulting in two additional bin limits of 16.50 and 31.43 eV. The approaches were compared to a uniform atomic hydrogen only medium with n$_{\ch{H}}=100\, \mathrm{cm}^{-3}$. As seen in \fig{fig:PDR-frequency-bins}, approach (1) underestimates the rate coefficient of H ionization by an order of magnitude, while (2) by definition is correct in the optically thin region, but overestimates the opacity at larger optical depths, which should in reality fall as energy would preferentially be removed where the cross section is large. The extra bins in (3) however reduce the optically thin error to roughly 20\%, and is even better at large optical depth, so this is the approach that is used in the following. 

For some photo-reactions, a cross section is not available in the literature, but instead a parameterisation of the rate coefficient in terms of visual
extinction ({\Av}) and external field strength ($G_0$) in units of the Draine Interstellar Radiation Field (ISRF) \citep{Draine1978ApJS} is available.
In that case we parameterise the actual radiation field in terms of $G_0$ and {\Av} as
\begin{equation}
J_{\mathrm{bin}} = G_0\, J_{\mathrm{bin,Draine}}\exp{\left(-\tau_{\mathrm{bin}}\right)}\, ,\label{eq:I-Av}
\end{equation}
where $J_{\mathrm{bin,Draine}}$ is the ISRF and we assume $G_0$ to be constant. The optical depth can be expressed in terms of the
dust column density, as well as the visual extinction as
\begin{equation}
\tau_{\mathrm{bin}} = \kappa_{\mathrm{bin,d}}f_{\mathrm{d}}\frac{N_{\mathrm{H}}\,m_\mathrm{p}}{ x_{\mathrm{H}}} = \gamma_\mathrm{bin}A_\mathrm{v}\, .
\end{equation}
Here, $f_{\mathrm{d}}$ is the dust mass fraction, $N_{\mathrm{H}}$ the column density of hydrogen, $m_{\mathrm{p}}$ the proton mass, and $x_{\mathrm{H}}$ the hydrogen mass fraction. Observationally, the visual extinction is approximately $A_{\mathrm{v}}=\frac{N_{\mathrm{H}}}{1.8 \times 10^{21}}\, \mathrm{mag}$. Using this relation we have
\begin{equation}
\gamma_\mathrm{bin} = 1.8 \times 10^{21}\kappa_{\mathrm{bin,d}}f_{\mathrm{d}} \frac{m_\mathrm{p}}{ x_{\mathrm{H}}}\, .
\end{equation}
Taking the logarithm of \eqn{eq:I-Av} gives
\begin{equation}
\ln{\left( J_{\mathrm{bin}} \right) } - \ln{\left(J_{\mathrm{bin,Draine}}\right)} = \ln{\left(G_0\right)} - \gamma_\mathrm{bin}A_\mathrm{v}
\end{equation}
To determine $\ln{\left(G_0\right)}$ and $A_\mathrm{v}$ needed for the reaction rates we make a linear fit by least squares to the above
equation over the relevant frequency bins with a photon energy between 5 and 13.6 eV.

\subsubsection{Hydrogen ionization with {\ctoray}}\label{hydrogen-ionization-with-ctoray}
In an explicit method, the evolution of radiatively driven ionization fronts can require a very small time step and cell size to resolve
correctly the ionization front, and obtain the right propagation speed. {\ctoray} \citep{Mellema2006NewA} is an explicitly photon-conserving
and causally implicit photo-ionization method that takes the ionization rate to be explicitly set by the flux attenuation, and uses of an analytical
expression for the time evolution to compute the time-step averaged ionization fraction. This makes it a robust method that converges with a more
or less arbitrary time step and spatial resolution. It is implemented here for the ionization of pure hydrogen, by interpolating, not opacity and emissivity, but the abundance of \ch{H} and \ch{H+} to the rays, solving along the rays as described in \citet{Mellema2006NewA}, and interpolating back the resulting ionization and recombination rate to the cell centres, where they are applied. The entire procedure is repeated until the neutral fraction converges.
While robust and elegant, this approach is very specific to the case of hydrogen ionization.

\subsection{Interstellar radiation field as diffuse emission}\label{sec:external-isrf-as-diffuse-emission}
In earlier 3D dynamical models of PDRs, the external radiation field has been simulated as an ISRF, attenuated through
a column density measured from the edge of the simulation box \citep[e.g.~][]{Nelson1997,Glover2007a}. While this works well when
modelling an isolated cloud that stays more or less centred in the simulated box, when simulating turbulent clouds in a periodic box, this
choice introduces an artificial asymmetry by predominantly irradiating regions closer to the boundaries. To respect the periodicity of the box
we instead choose to simulate the external field as an artificial emission in the hot, dilute inter-cloud medium. This is somewhat similar to
the proposal by \citet{Walch2015}, where each point is irradiated from a constant distance, but does not introduce such a distance cut-off.

For the emission, we use the spectral energy distribution due to \citet{Black1987ASSL}. The criteria for which cells will contain a source is
adhoc in nature. We have chosen to use a chemical cutoff; we place an emission
$j_{*,\mathrm{c},\nu}=j_{*,0,\nu} V \frac{w_{\mathrm{c}}}{w_{\mathrm{tot}}}$ in sufficiently ionized cells
according to a
weighting function
\begin{equation}
 w_{\mathrm{c}}=\begin{cases}
\rho^{2} & n_{\ch{H}}\le n_{\mathrm{th}}\\
0 & n_{\ch{H}}>n_{\mathrm{th}}\,,
\end{cases}
\end{equation}
where $n_{\mathrm{th}}$ is a threshold neutral number density, $V$ is entire the simulation volume, and
$w_{\mathrm{tot}}=\sum_{\mathrm{c}} \Delta V_c w_{\mathrm{c}}$ is a normalising factor, with $\Delta V_\mathrm{c}$
the cell volume. By weighting with $\rho^2$ we balance the recombination rate equally well in low and high density regions in a simulation,
and do not artificially drive the expansion or collapse of HII regions. The normalisation corresponds to assuming a constant amount of photons
to be created per unit time, and therefore help in avoiding feedback loops. In principle this could be refined to depend on e.g.~the instantaneous
star formation rate, but we find the constant to be a good comprise given that a diffuse radiation field really represent the radiation \emph{external}
to the region coming from the larger galactic environment, and not a self-consistently created radiation field generated by the stellar feedback
inside the box.

\subsection{Self-shielding of \ch{H2} and \ch{CO}}\label{sec:selfshielding}
The self-shielding of $\textrm{H}_2$ and CO needs to be treated carefully to get qualitatively correct results for the two most abundant
molecules in the ISM. Since this requires that one follows in detail their excited states and the associated frequencies \citep{Visser2009A&A},
this is not currently achievable in 3D radiation hydrodynamics calculations. We follow a common approach
\citep{Glover2007a, Safranek-Shrader2017MNRAS} where a prescription for self-shielding as a function of column density and
temperature is used to find the dissociation rate.  The prescription used here is from \citet{Richings2014II}, and is derived from
detailed calculations with the 1D photochemistry code \cloudy{} \citep{Ferland1998,Ferland2013}. In it, the dissociation rates in optically
thick regions are found as \cite[Eqn. 3.18 and 3.19]{Richings2014II}
\begin{equation}
\Gamma_{\ch{H2},\mathrm{thick}}=\Gamma_{\ch{H2},\mathrm{thin}} f_{\mathrm{s}}^{\ch{H2}}
\end{equation}
and 
\begin{equation}
\Gamma_{\ch{CO},\mathrm{thick}}=\Gamma_{\ch{H2},\mathrm{thin}} f_{\mathrm{s}}^{\ch{CO}}
\end{equation}
where $\Gamma_{\mathrm{x},\mathrm{thin}}$ is the dissociation rate of species x given an unattenuated ISRF, and $f_{\mathrm{s}}^{\mathrm{x}}$ is a total shielding factor for that species, given by
\begin{equation}
f_s^{\ch{H2}}=\mathcal{S}_\mathrm{d}^{\ch{H2}}\left(N_\mathrm{dust}\right)
\mathcal{S}_\mathrm{self}^{\ch{H2}}\left(N_{\ch{H2}}\right)
\end{equation}
and 
\begin{equation}
f_s^{\ch{CO}}=\mathcal{S}_\mathrm{d}^{\ch{CO}}\left(N_\mathrm{dust}\right)
\mathcal{S}_{\mathrm{self},\ch{CO}}^{\ch{CO}}\left(N_{\ch{CO}},N_{\ch{H2}}\right),
\end{equation}
where $N_x$ is column density of dust, $\ch{H2}$, and $\ch{CO}$ respectively. The dust column density is proportional to the total Hydrogen
column density and the dust to gas mass fraction. For the individual shielding factors for dust $\mathcal{S}_\mathrm{d}^{\mathrm{x}}$ and self-shielding $\mathcal{S}_\mathrm{self}^{\mathrm{x}}$ we refer to \citet{Richings2014II}.
As noted in \citet{Glover2007a}, turbulent Doppler-shift of lines is not taken into account and this approximation will tend to overestimate
self-shielding. Another reservation is that the prescription is based on the equilibrium solutions to a static dissociation front, while we intent to use
it in the non-equilibrium, dynamical environment of a turbulent ISM. 

Our approach is similar to how \citet{Safranek-Shrader2017MNRAS} post-process simulations by ray-tracing the
galactic disc using a uniform emission in the galactic plane that represents the stellar radiation field. They replace the usual solution to the 1D radiative
transfer equation by the following double sum (their Eqn.~23, slightly relabelled here)
\begin{equation}\label{eq:double-sum}
 I_{\mathrm{bin}}\left(n\right)=\sum_{i=0}^{n-1}\Delta l_{i}j_{*,\mathrm{bin},i}
 \mathcal{S}_{\mathrm{bin}}\left(\sum_{j=i}^{n}\mathbf{n}_{spec,j}\Delta l_{j}\right)\,,
\end{equation}
 where $I_{\mathrm{bin}}\left(n\right)$
  is the intensity in the frequency range ``bin'' in cell $n$, $\Delta l_{i}$
  is the path length of the ray through cell $i$
 , $j_{*,\mathrm{bin},i}$
  is the emissivity in cell $i$
  in the frequency bin, $\mathcal{S}_{\mathrm{bin}}$
  is the shielding function and $\mathbf{n}_{spec,i}$
  is a vector of the number densities of the relevant species in cell $i$
 .
The double sum must be computed in every cell. If the sum over column depth is performed as a series of running sums, one for each source, this method has complexity $\mathcal{O}(N^2)$ in the number of points along the ray. 
Since in our case the sources are by definition placed in optically thin regions, we can to a good approximation gather a set of neighbouring sources into a single source. 
Computing only running sums from cells with a source, the complexity reduces to $\mathcal{O}(N N_r)$ where $N_r$ is the number of emitting regions along the ray, which is usually low in realistic situations. 
In this method, it is n$_{\ch{H2}}$, n$_{\ch{CO}}$, n$_{\ch{H}_{\mathrm{tot}}}$, $T_{\mathrm{gas}}$ and $j_{*,c} = \int_0^\infty j_{*,c,\nu} \mathrm{d}\nu$ that are interpolated from cell centres to rays. 
The resulting frequency-integrated solid-angle mean intensity $J^{\mathrm{dis}}$ is interpolated back to cell centres, and the dissociation rate per molecule is found as 
\begin{equation}
  \Gamma_{\mathrm{dis,thick}} = \frac{J^{\mathrm{dis}}}{\int_0^\infty J^{\mathrm{Black}}_\nu \mathrm{d}\nu}\Gamma_{\mathrm{dis,thin}},
\end{equation}
where $\Gamma_{\mathrm{dis,thin}}$ is the optically thin dissociation rate given the Black ISRF, with values $\Gamma_{\ch{H2},\mathrm{thin}}=7.5\times10^{-11}\, \mathrm{s}^{-1}$ \citep{Richings2014I}, and $\Gamma_{\ch{CO},\mathrm{thin}}=2.6\times10^{-10}\, \mathrm{s}^{-1}$ \citep{Visser2009A&A}. 

\subsection{Thermal balance of dust grains}\label{sec:thermal-balance-dust-grains}
Grain surface chemistry depends on the temperature of the dust grains, and is used in our chemical network for calculating
the $\ch{H2}$ formation rate, and the freeze-out and evaporation of $\ch{H2O}$ and $\ch{CO}$ ices. The kinetic cooling or
heating of the gas also depends on the temperature differential between dust and gas. In \citet{Grassi2017MNRAS} we
demonstrated how these functions can be tabulated given a radiation environment. The core assumption is that the time-scale for
dust grains to be in thermal equilibrium is shorter than any relevant dynamical timescales, which is usually the case. Then
Kirchhoff's law holds. The radiation absorbed ($\Gamma_\mathrm{abs}$) by the dust grain is equal to the emitted radiation
($\Gamma_\mathrm{em}$) plus the kinetic cooling due to interaction with gas molecules ($\Lambda^\mathrm{col}_\mathrm{gas}$)
\begin{equation}
\Gamma_\mathrm{abs} = \Gamma_\mathrm{em} + \Lambda^\mathrm{col}_\mathrm{gas}\,.
\end{equation}
Integrating over a distribution of dust grains embedded in a gas in the \emph{optically thin regime} this becomes \citep{TielensBook,Grassi2017MNRAS}
\begin{align}\nonumber
 \Gamma_\mathrm{abs} & = \int_{0}^{\infty} \dd E\, \int \dd a\, \pi a^2 \varphi(a) \frac{Q_\mathrm{abs}(a,E) J(E)}{h} \\
 \Gamma_\mathrm{em} & = \int_{0}^{\infty} \dd E\, \int \dd a\, \pi a^2 \varphi(a) \frac{Q_\mathrm{abs}(a,E) B\left[E,T_d(a)\right]}{h}\\
 \Lambda^\mathrm{col}_\mathrm{gas} & = 2 f v_\mathrm{gas} n_\mathrm{gas}\int \dd a\, \pi a^2\varphi(a) k_B\left[T-T_d(a)\right] \,, \nonumber
\end{align}
where $\pi a^2$ is the geometrical cross section of a dust grain with radius $a$, and $\varphi(a)$ is the distribution function,
e.g.~the number density of grains per grain size. $Q_{abs}(a,E)$ is the absorption coefficient of the grains,
$B\left[E,T_d(a)\right]$ is the black-body spectral radiance with dust grain temperature $T_d(a)$, $J(E)$ is the mean intensity at energy $E$,
and $v_\mathrm{gas}$ is the thermal gas speed.
The factor $f$ accounts for gas-grain collisions with atoms and molecules other than atomic hydrogen, and it depends not only
on the gas composition, but also on the grain charge distribution (see \citealt{Draine2009} sect.~24.1.2). We will assume it to
be $f = 0.5$, which corresponds to a cold partially molecular gas. In this paper, we model the radiation as being constant inside each
bin: $J(E) = J_i$, if $E_\mathrm{min, i} < E < E_\mathrm{max, i}$. Then the radiation absorbed by the dust grains can be calculated as
\begin{equation}
 \Gamma_\mathrm{abs} = \sum_{i \in \mathrm{bins}} J_i \tilde\Gamma_\mathrm{abs,i}\,,
\end{equation}
where
\begin{equation}
 \tilde\Gamma_\mathrm{abs,i} = \int_{E_\mathrm{min, i}}^{E_\mathrm{max, i}} \dd E\, \int \dd a\, \pi a^2 \varphi(a) \frac{Q_\mathrm{abs}(a,E)}{h}\,.
\end{equation}
To implement this in actual code we precalculate $\tilde\Gamma_\mathrm{abs,i}$, and then construct log-spaced 3D tables for the $\ch{H2}$
formation rate, surface averaged dust temperature $T_\mathrm{d}$, and gas cooling $\Lambda^\mathrm{col}_\mathrm{gas}$ as a function of
$\Gamma_\mathrm{abs}$, $T$, and $n_\mathrm{gas}$, following the methodology of \citet{Grassi2017MNRAS}.
The formalism allow the $\ch{H2}$ formation rate to be tabulated for the specific dust
 distribution with individual dust temperatures for each species and size bins (we used 20 logarithmically spaced size bins for the distribution with
 a mixture of silicate and carbonaceous grains). But
to calculate the absorption and desorption of $\ch{H2O}$ and $\ch{CO}$ ices we need a single averaged dust temperature in the network.
For that we chose to use a number density weighted temperature
\begin{equation}
T_\textrm{d} = \frac{\int \dd a\, \varphi(a) T_\textrm{d}(a)}{\int \dd a\, \varphi(a)}\,.
\end{equation}
Notice that to be fully consistent one would need to calculate the ice to gas rates based on the temperatures of the individual bins, which
could be done by introducing additional lookup tables. Alternative
definitions, such as a surface averaged or emission averaged temperatures could be used, though this does only change the
dust temperature with up to a few degrees, and therefore only lead to very minor changes in the results. We also note that in very dense and
dusty regions the assumption that the thermal radiation is optical thin may not hold. this can be dealt with either by a suitable parameterisation
of the self-absorption and escape fraction inside a single cell \citep{Grassi2017MNRAS} or explicitly through the introduction of infrared frequency bins
and an account of scattered light \citep{Rosdahl2015}.

Ideally, each set of tables should be made for the specific environment, and we caution that the 3D tables currently available through
{\krome} are made for the tests in this paper. In all our tests we have used a MRN power law grain size distribution with $\varphi(a)\propto a^{-3.5}$ \citep{Mathis1977}, and a size range from $5\times10^{-7}$ cm to $2.5\times10^{-5}$ cm. We use a mixture of carbonaceous and silicate grains
in the ratio 9:1 (same as the ratio of the key elements $\ch{C}$ and $\ch{Si}$ at Solar metallicity) and assume a solar dust to gas mass ratio of 0.00934.

\subsection{Adsorption and desorption of CO and H$_2$O ices}\label{sect:freeze_out}
\newcommand{\ith}{i$th$}
For the adsorption reaction rate coefficients per gas molecule for species $s$, $k_\mathrm{ads,s}$, we follow \citet{Hollenbach1979, Hocuk2015}.
\begin{equation}\label{eqn:freeze_out_distribution}
	k_\mathrm{ads,s} = \sigma_s v_\mathrm{gas,s} n_\mathrm{dust}\,,
\end{equation}
where $\sigma_s$ is the cross section, $v_\mathrm{gas,s}$ is the thermal velocity of species $s$, and $n_\mathrm{dust}$ is the number
density of dust grains. The cross section is given as the product of a sticking rate $S(T,T_d)$ and the surface area
of the dust. Assume an MNR dust distribution with a power law exponent $p=-3.5$, minimum and maximum grain sizes
$a_{\rm min}$ and $a_{\rm max}$, then we have
\begin{equation}\label{eqn:freeze_out_distribution2}
\sigma_s n_\mathrm{dust} = S(T,T_d) \, \frac{f_d\rho_g}{4/3\,\rho_\mathrm{m}} \frac{p+4}{p+3} \times \frac{a_{\rm max}^{p+3}-a_{\rm min}^{p+3}}{a_{\rm max}^{p+4}-a_{\rm min}^{p+4}}\,,
\end{equation}
where $\rho_\mathrm{m}=3$~g~cm$^{-3}$ is the material density of the dust, and we parametrize the sticking coefficient as
\begin{equation}
	S(T, T_d) =  \left[1 + 0.04 \sqrt{T+T_d} + 0.002\,T + 8\times10^{-6} T^2\right]^{-1}\,.
\end{equation}

For the desorption we include three contributions: thermal, non-thermal, and cosmic ray desorption. The total rate is then
\begin{equation}
 	k_\mathrm{des,s} = k_\mathrm{th-des,s} + k_\mathrm{ph-des,s} + k_\mathrm{cr-des,s}\,.
\end{equation}
We assume that the thermal desorption of species $s$ follows the Wigner-Polyany mechanism
\begin{equation}\label{eqn:thermal_evaporation}
	k_\mathrm{th-des,s}(T_d) = \nu_0 \exp\left({-\frac{E_s}{k_b T_d}}\right)\,,
\end{equation}
where $\nu_0=10^{12}$~s$^{-1}$ and $E_s$ is the binding energy of the ice.
We include non-thermal desorption due to radiation as \citep{Hollenbach2009,Hocuk2015}
\begin{equation}
	k_\mathrm{ph-des,s} = G_0\,F_0\,a_p^2\,Y_s\exp\left(-1.8 \Av\right)\,,
\end{equation}
where $F_0=10^8$~cm$^{-2}$~s$^{-1}$ is the number of photodesorbing photons at $G_0=1$ and $\Av = 0$,
$a_p=3\times10^{-8}$~cm, $Y_{\rm CO}=10^{-3}$, and $Y_{\rm H_2O}=2\times 10^{-3}$.
The cosmic-ray photodesorption  rate \citep{Hasegawa1993,Reboussin2014} is
\begin{equation}
	k_\mathrm{cr-des,s} = f(70\,{\rm K})\,k_\mathrm{th-des,s}(70\,{\rm K})\,,
\end{equation}
where the fraction of time spent by the grain at 70~K is $f(70\,{\rm K})=3.16\times10^{-19}\zeta/\zeta_0$,
$\zeta_0=1.3\times10^{-17}$~s$^{-1}$, and $\zeta$ is the cosmic ray ionization rate. Although for the adsorption we
used a size distribution in accordance with our general model for the dust grains, in this expression assumes
dust grains to be of size $a=0.1$~$\mu$m \citep{Leger1985} introducing a small inconsistency.

Including ice evaporation in the chemical network increases the stiffness of the differential equations, in particular when the dust temperature is
time-dependent. This is because of the exponential dependence of the ice evaporation on the dust temperature. If the equations are reformulated
in terms of the total number density of a species, $n_{\rm tot} = n_{\rm gas} + n_{\rm ice}$, and the gas phase number density, $n_{\rm gas}$,
the stiffness is significantly reduced (W-F.~Thi, private communication). For the ices we solve
\begin{eqnarray}
 \dot n_{\rm tot} &=& \dot n_{\rm gas} + \dot n_{\rm ice} \equiv \dot n_{\rm chem}\\
 \dot n_{\rm gas} &=& \dot n_{\rm chem} - \left(k_{\rm des} + k_{\rm ads}\right) n_{\rm gas} + k_{\rm des} n_{\rm tot}\,,
\end{eqnarray}
where $\dot n_{\rm chem}$ is the change in the species due to gas-phase chemical reactions,  and $k_{\rm des}$ and
$k_{\rm ads}$ are the evaporation and freeze-out rate coefficients. Apart from thermal evaporation, in the case of CO
we also include non-thermal evaporation by UV photons that can dominate the evaporation rate in low-density regions
with infrequent collisions, where dust cools efficiently and keeps cool even when the gas is hot.

\section{Implementation}\label{sec:implementation}
The hybrid-characteristics method described for point-sources in \citet{Rijkhorst2006A&A} and diffuse radiation in \citet{Buntemeyer2016NewA} works well on patch-based hierarchical adaptive meshes. {\ramses}, however, has a cell-based octree, based on the fully threaded tree by \citet{Khokhlov1998}, and the ``patches'' that would be connected by short-characteristics rays, would consists of only $2\times 2\times 2$ cells. In this structure, neighbour cells are found by first accessing the coarser parent cell of an oct, then finding the neighbour, and then looking for refined cells in the neighbour to the parent oct. This makes repeated access to neighbour cells required by the method fairly expensive, and implies that a short-characteristics-based method would not perform well on a fully threaded tree.

\begin{figure}
\begin{center}

\includegraphics[width=0.36\textwidth]{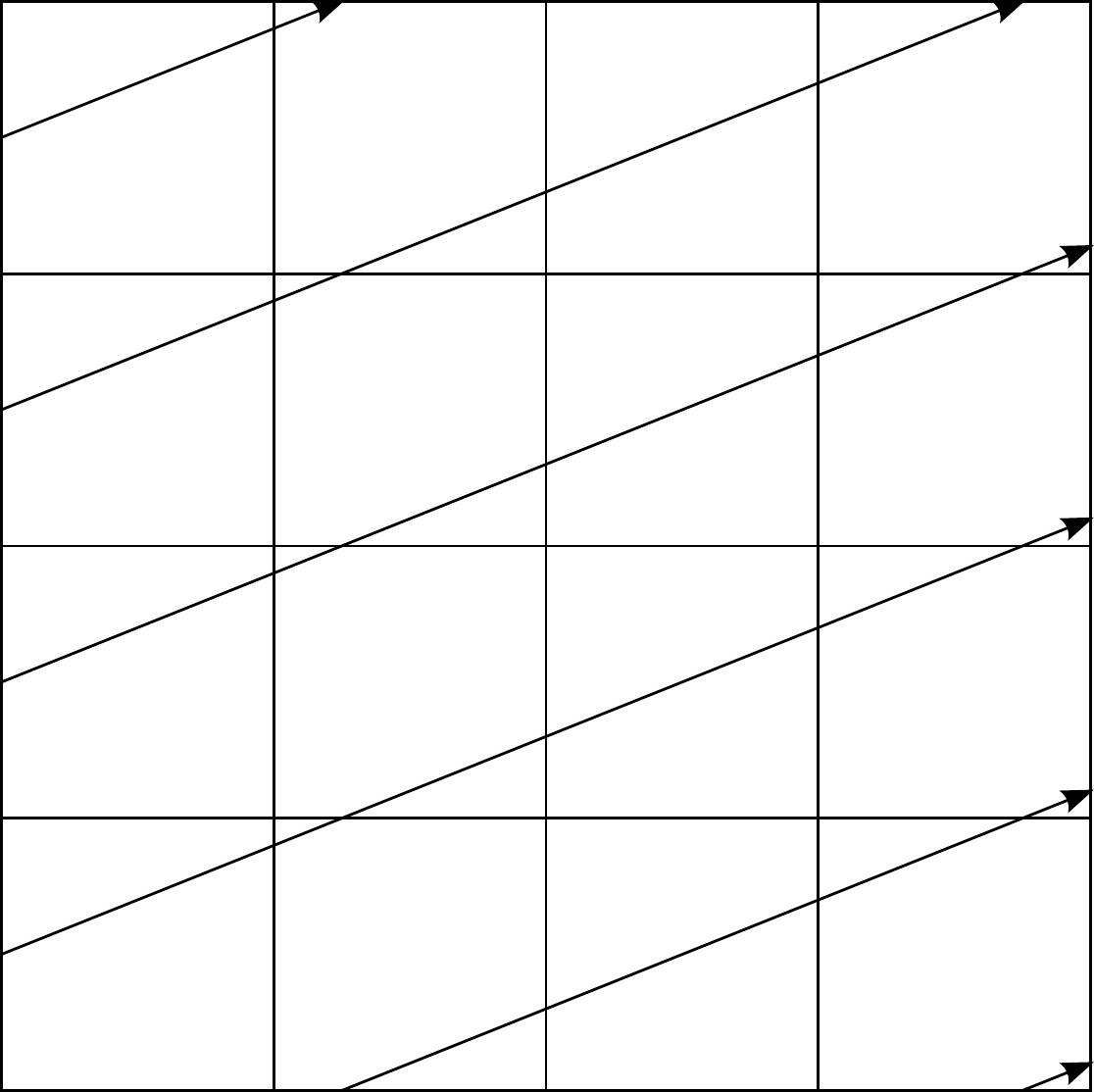}

\caption{Long characteristics rays as lined-up short characteristics. Points along the rays are connected to cell centres by interpolation. \label{fig:economic-long-characteristics}}

\end{center}
\end{figure}

\begin{figure}
\begin{center}

\includegraphics[width=0.5\textwidth]{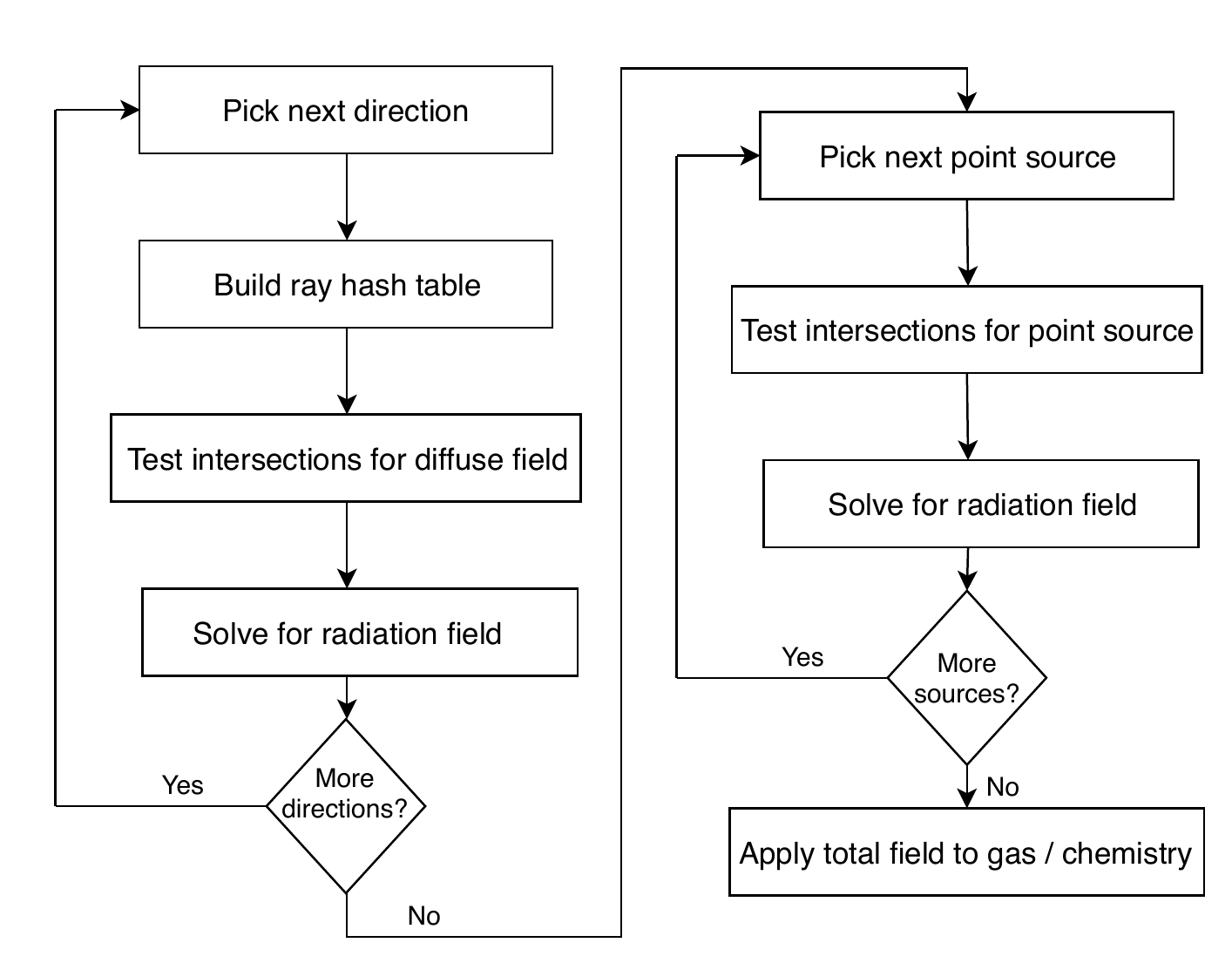}

\caption{Outline of {\lampray} in flow chart form. See the text for an explanation of the steps. }\label{fig:lampray-flow-chart}

\end{center}
\end{figure}

\begin{figure}
\begin{center}

\includegraphics[width=0.48\textwidth]{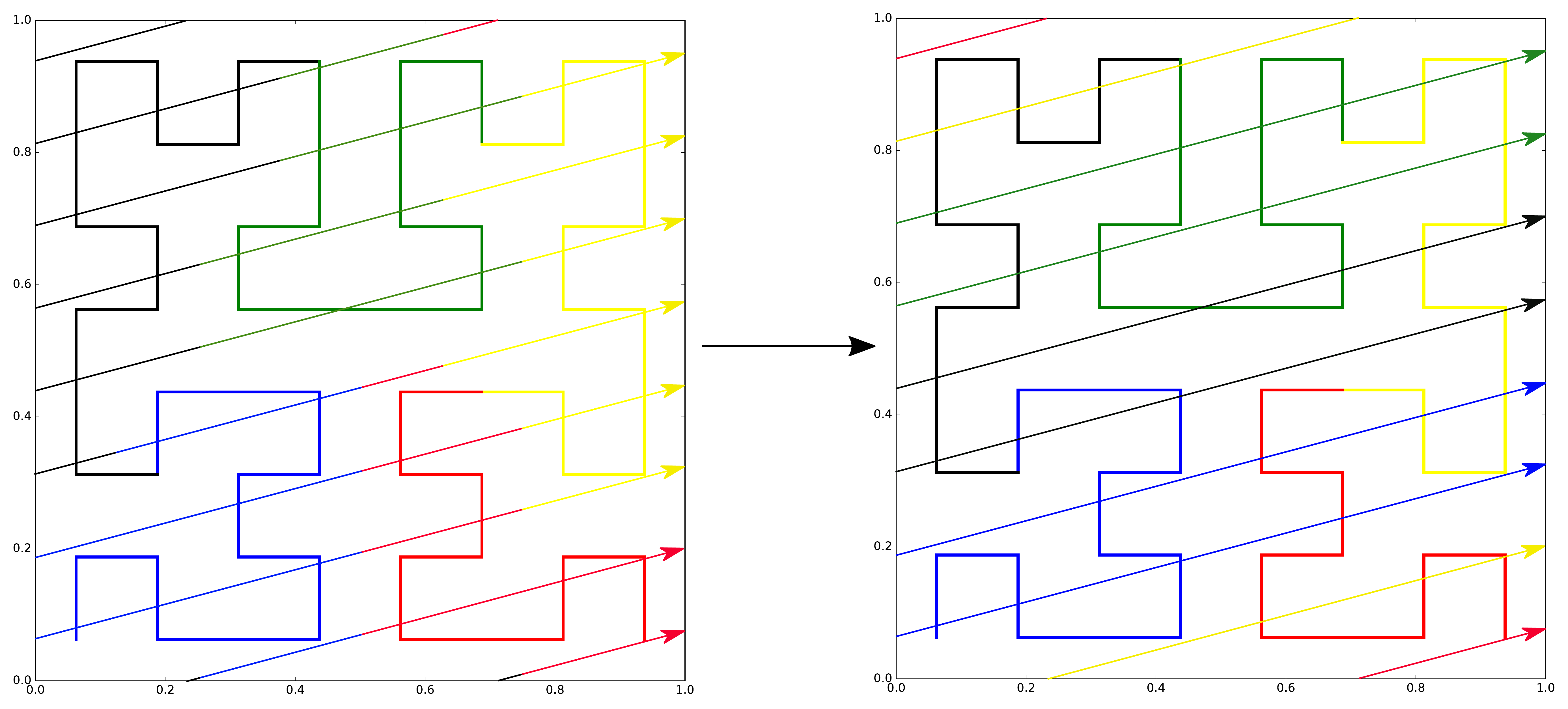}

\caption{The domain decomposition is changed. Colours denote {\mpi} rank ownership. }\label{fig:ray-and-cell-domain}

\end{center}
\end{figure}

Our solution is instead to use long characteristics rays that cover the entire domain. These are organized in a sense as short characteristics rays that line up and connect the computational domain from one end to the other, so repeated calculation of optical depth through similar paths is avoided (see \fig{fig:economic-long-characteristics}). Quantities required by radiative transfer are interpolated to the points along the long rays. This happens in the domain decomposition used for the hydrodynamics, which in {\ramses} means that an {\mpi} rank owns cells along one or more slices of a Hilbert space-filling curve in physical space. A single long ray therefore typically crosses a number of {\mpi} ranks. 
To compute efficiently the solution to the radiative transfer along rays we change the domain decomposition for the ray points, such that, for the purpose of radiative transfer, all points along a given ray belongs to the same {\mpi} rank, and all ranks own a similar number of ray points. In the new \emph{ray domain} decomposition, the solution to the one-dimensional radiative transfer problem can be obtained efficiently, with near-perfect vectorisation, cache usage and load balancing. Once the solution is known at the ray points, it is communicated back to the cell domain in a similar but reverse exchange of information between the ray and cell decomposition, and interpolated to the cell centres (see \fig{fig:ray-and-cell-domain}).

Although the communication step is extensive, the cost is not as high as one might expect, and turn out to be worth the gains that comes from the decoupling it provides. 

This underlying idea was first presented in the masters thesis of \citet{Frostholm2014}. In this work most parts of the implementation have been completely replaced, including methods for ray tracing, interpolation, and handling of periodic boundaries. The photo-chemistry solver is completely new. 

The rest of this section will explain how rays are laid out and traced to properly cover the AMR grid, how the interpolation is done between cells and rays, and finally how the {\mpi} parallel nature of {\ramses} is handled when tracing rays and by changing domain decomposition. The overall method is sketched in \fig{fig:lampray-flow-chart}.

\subsection{Ray Tracing}\label{sec:ray-tracing}
\begin{figure}
\begin{center}

\includegraphics[width=0.48\textwidth]{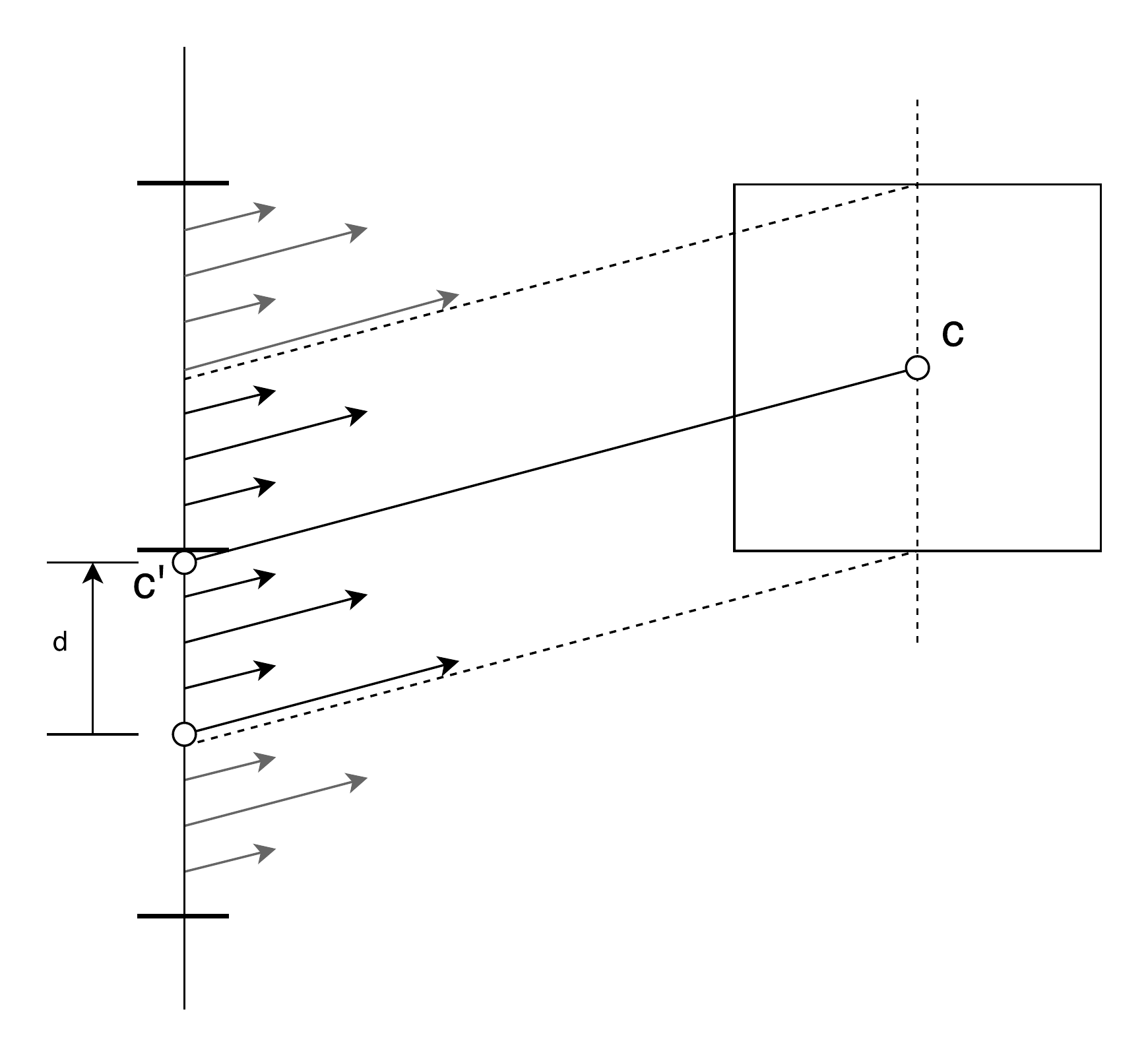}

\caption{Given a ray direction, C' is the projection of cell centre C along this direction to the ray plane (solid vertical line), and is located a distance d from the nearest ray origin at that level of refinement on the ray plane (bottom open circle). Rays at lower level of refinement that intersect the cell originate at most half a cell width away from C'. }\label{fig:picking-the-child-rays-recursively}

\end{center}
\end{figure}

The radiation field is split into two components: A diffuse field for extended sources of emission, and a point source field. The diffuse field is discretised in solid angle into a set of equal-solid-angle directions using the {\healpix} scheme \citep{Gorski2005}, and for each direction into a set of parallel rays that cover the computational domain end-to-end. For the diffuse field, the radiative transfer problem is solved one direction at a time to limit memory usage. 

In order to cover leaf cells at all levels of refinement, rays are chosen as follows.
Consider a single ray direction $\boldsymbol{r}$. Let $a_z$ be the coordinate axis where $\boldsymbol{r}$ has it's largest component. We will refer to the plane defined by $a_z=0$ as the ray plane. At a given refinement level $\ell$ we define a Cartesian grid on the ray plane, with cell centres $c^{\textrm{rp}}_{i,j}=0.5^\ell L (i+0.5)$ where $L$ is the box length. At each level, for all $(i,j)$, we choose rays with origin $c^{\textrm{rp}}_{i,j}$. We should however only include rays that intersects leaf cells at level $\ell$. Rays at level $\ell$ that are surrounded in the ray plane by rays originating from all four ray plane child cells at level $\ell+1$ can however be excluded. And finally we should honor intersections of rays at level $\ell$ with leaf cells at lower levels. 

The ray plane grids occupied by rays for each level are sparse structures. We choose to represent these as a single hash table, with the triple $(\ell,i,j)$ as hash key. The table is constructed as follows. For each leaf cell, we project the cell centre along the ray direction on to the ray plane and round the projected coordinate to get integer indices $i$ and $j$. The ray plane cell is marked as containing a ray by placing the arbitrary value 1 under the key $(\ell,i,j)$. Exclusion of surrounded rays is deferred to when ray points are constructed. Once the table has been built, the keys are sorted and the table values are replaced by unique ray ids that enumerate the keys. 

A point along the ray is defined as the structure that contains the ray id, the id of the intersected cell, the distance along the ray from the ray plane to the back face of the intersected cell, and the values of radiative transfer variables at the ray point. Points along rays are constructed as follows. 1) For each leaf cell we find $i$ and $j$ by projecting on the ray plane as before. By construction, the corresponding ray exists. 2) The ray plane cell has four children with keys $(\ell+1, 2i, 2j)$, $(\ell+1, 2i+1, 2j)$, $(\ell+1, 2i, 2j+1)$ and $(\ell+1, 2i+1, 2j+1)$. These are looked up to see if they are occupied by rays. If they all are, the ray at level $\ell$ is excluded, so nothing is done. Otherwise a ray point is created. 3) In any case, rays on level $\ell+1$ may also intersect our leaf cell. These may not be the same as the four child rays. If the projected cell centre is more than a quarter cell size above (below) the ray plane cell centre in x, the intersecting rays at level $\ell+1$ will have indices $2i+1$ and $2i+2$ ($2i-1$ and $2i$), and similarly in y. For those rays that exist, ray points are constructed recursively from point~2. 

There is a subtlety in picking out the intersecting rays at the following level at every recursive step. Going from level $\ell$ to $\ell+1$ we can proceed as just described. Lets label by $d$ the distance from the ray plane cell centre to the projected cell centre in units of the cell size at level $\ell$, and the vector of indices at level $\ell+1$ by $\boldsymbol{i}_{\ell+1}=(2i, 2i+1)+q_{\ell+1}$, where 
\begin{equation}q_{l+1}=
\begin{cases}
1 & d > \frac{1}{4}\\
0 & \frac{1}{4} \geq d > 0\\
0 & 0 \geq d > -\frac{1}{4}\\
-1 & -\frac{1}{4}  \geq d
\end{cases}.
\end{equation} Going from level $\ell+1$ to $\ell+2$ we have $\boldsymbol{i}_{\ell+2}=(4i, 4i+1, 4i+2, 4i+3)+q_{\ell+1}+q_{\ell+2}$. Another way to say this is that at level $\ell+2$ we have for each parent index $i^p \in \boldsymbol{i}_{\ell+1}$ two child indices $\boldsymbol{i}^c_{\ell+2}=(2i^p, 2i^p+1)+q_{\ell+2}$. Here $q_{\ell+2}$ is determined by the following eight cases (see \fig{fig:picking-the-child-rays-recursively}). 
\begin{equation}q_{\ell+2}=
\begin{cases}
0 & d > \frac{3}{8}\\
-1 & \frac{3}{8} \geq d > \frac{1}{4}\\
1 & \frac{1}{4}  \geq d > \frac{1}{8}\\
0 & \frac{1}{8}  \geq d > 0\\
0 & 0  \geq d > -\frac{1}{8}\\
-1 & -\frac{1}{8}  \geq d > -\frac{1}{4}\\
1 & -\frac{1}{4}  \geq d > -\frac{3}{8}\\
0 & -\frac{3}{8}  \geq d
\end{cases}.
\end{equation}
At this point a pattern emerges. If we were to continue to level $\ell+3$, we would get the offsets $0, -1, 1, 0$ repeated 4 times, at level $\ell+4$ repeated 8 times etc. So if we define $\boldsymbol{Q}=(0,-1,1,0)$ and at child level $\ell + L$ we define $d'_{L}=\lceil(\frac{1}{2})^{L} d - 1\rceil \mod{4}$, we can write $q_{L}=Q_{(d'_L)}$.

For point sources, radiative transfer is solved one source at a time. The point source field is covered by the rays given by vectors at a given {\healpix} resolution. Using the {\healpix} library, tracing point source rays is simple. 
Given a cell located at $\boldsymbol{c}$ and a point source at $\boldsymbol{p}$, 
we translate $\boldsymbol{c} \rightarrow \boldsymbol{c}' = \boldsymbol{c} - \boldsymbol{p}$, 
and use the \verb!query_polygon! {\healpix} routine 
to find the {\healpix} vectors that intersect the square defined by the intersection between the plane perpendicular to the largest component of $\boldsymbol{c}'$ and the cell outline around $\boldsymbol{c}'$. The ray segment centres are chosen as the intersection between this plane and the {\healpix} vectors. We use the {\healpix} nested pixel index as unique ray id for matching up ray points and load balancing rays.

\subsection{Interpolation}\label{sec:ray-geometry-and-interpolation}
When interpolating from cells to rays, we are interpolating in three dimensions from a uniform grid so we can get a second-order accurate, conservative interpolation with triangular-shaped cloud (TSC) interpolation. Where support points are missing, they are reconstructed from the eight surrounding parent cells using cell-in-cloud (CIC) interpolation. 

When the radiative transfer problem has been solved along the rays, and we need to interpolate back from rays to cells, we interpolate from a less structured set of points. As interpolation errors can be hard to avoid at refinement boundaries, we made a few preliminary tests in two dimensions. The setup is as follows. We have a 2D simulation box $16\times 16$ cells large with the central $8\times 8$ cells each split into four. The physical size is $1\times1$ in arbitrary units. The extinction coefficient is uniform at $\alpha=1$ and a radiation field with intensity $I_0=1$ is applied at one face, with a single ray direction with angle $\theta$ with the x-axis. We use two ray coverage strategies: R1) The fine region is covered by rays as described above, so they extend to the coarse region and cover the coarse cells twice as densely. R2) The same as R1, but with dense coverage extended by one cell in each direction. Coarse rays are considered to cover a width of $\Delta a = \sin(\Theta)\Delta x$ with $\Delta x$ the coarse cell size, and fine rays cover half that width. The quantity that is interpolated is $Q = I \Delta a \Delta l$ where $\Delta l$ is the length of the ray segment through the cell. After interpolation, an intensity is recovered by dividing by the cell area $I = Q / (\Delta x)^2$. 

Interpolation on refinement boundaries is treated as follows. A ray point in a fine boundary cell gets interpolated to either fine cell centres, or -- where those points are in a coarse cell -- to a virtual fine cell in the coarse cell. From here it is further interpolated by one of the mentioned conservative methods to the surrounding coarse cell centres. The highest order method for deposition from virtual fine cells that has support is CIC. A ray point in a coarse boundary cell is interpolated to a fine cell by first interpolating to the coarse parent cell, and then adding the contribution to the fine cell. 

For each interpolation method, we try out two cases: Firstly interpolation in two dimensions on our 2D grid (corresponding to 3D if implemented in {\ramses}). Secondly interpolation in one dimension, namely along the line perpendicular to the main propagation axis (corresponding to 2D in the plane perpendicular to the main propagation axis in {\ramses}).

\begin{figure}
\begin{center}
\includegraphics[width=0.48\textwidth]{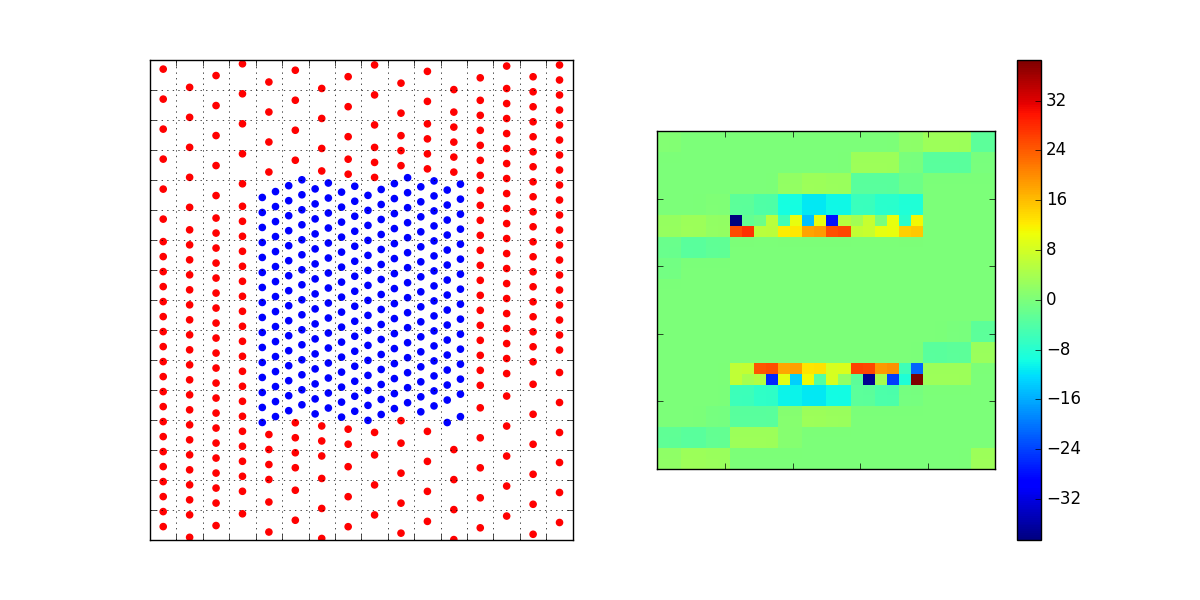}
\includegraphics[width=0.48\textwidth]{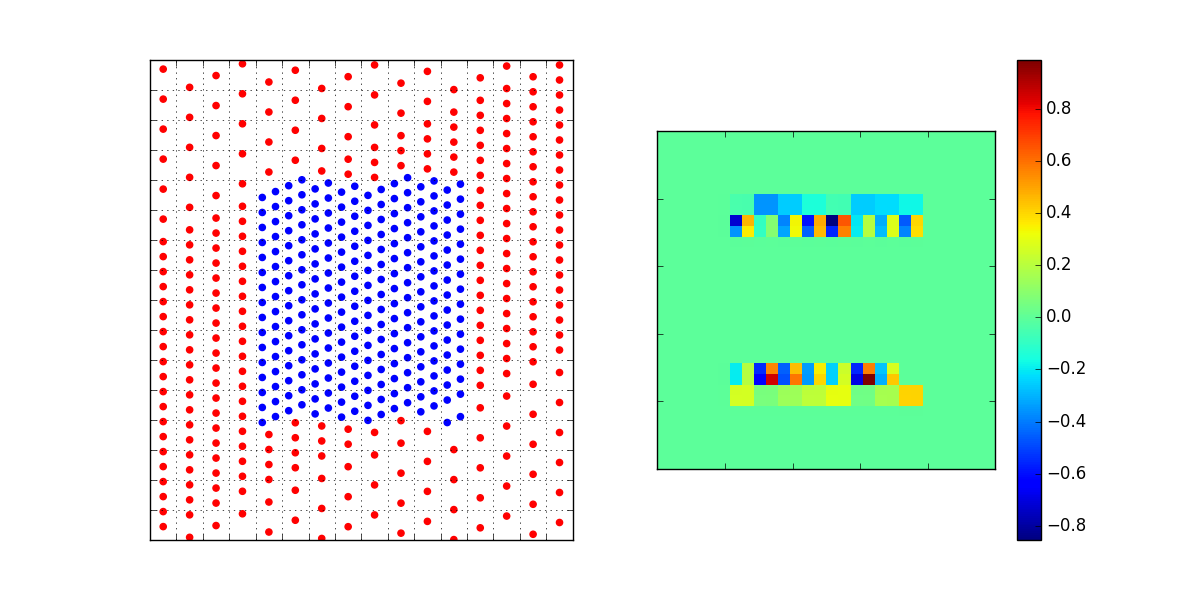}
\caption{Ray discretisation (left) and relative interpolation errors in percent (right) for the tests of interpolation method labelled I3 (top) and I8 (bottom).}\label{fig:interpolation}
\end{center}
\end{figure}

In the test, all combinations of the described methods conserve energy to machine precision as expected, but none give a maximal interpolation error below $30\%$. We therefore also try to relax the conservation requirement and use an irregular mesh interpolation method. This consists of interpolating as described above, but dividing the result by the total weight. Because of the normalisation, the intensity is interpolated directly without weighting by volume. With the irregular method using TSC, with NGP used for fine-coarse virtual grid propagation, the R1 coverage strategy, and the interpolation done in 1D, the maximal error reduces to just below $1\%$ with a relative deficit of total energy of less than $5\times10^{-7}$. Both maximal error and energy conservation error depend on the field gradient, and grow roughly linearly with $\alpha$. Selected results are summarised in \tab{table:interpolation-test-summary}. Ray discretisation and error for tests I3 and I8 are shown in \fig{fig:interpolation}.

Based on these tests, we choose to use the unstructured variant of TSC interpolation in the plane perpendicular to the main ray direction and the R1 coverage strategy. 

\begin{table*}
\begin{tabular}{|c|c|c|c|c|c|}
\hline 
Test & Interpolation & Reconstruction on boundaries & Ray coverage & Maximum relative error & Relative conservation error\tabularnewline
\hline 
\hline 
I1 & TSC (2D) & NGP & R1 & 34.99 \% & $1.8\times10^{-14} \%$\tabularnewline
\hline 
I2 & TSC (2D) & CIC & R1 & 40.35 \% & $5.4\times10^{-14} \%$\tabularnewline
\hline 
I3 & TSC (1D) & NGP & R1 & 38.57 \% & $1.8\times10^{-14} \%$\tabularnewline
\hline 
I4 & TSC (1D) & NGP & R2 & 35.50\% & $5.4\times10^{-14} \%$\tabularnewline
\hline 
I5 & CIC (1D) & NGP & R1 & 44.61\% & $7.2\times10^{-14} \%$\tabularnewline
\hline 
I6 & QSP (2D) & NGP & R1 & 63.74\% & $1.8\times10^{-14} \%$\tabularnewline
\hline 
I7 & TSC unstructured (2D) & NGP & R1 & 1.76\% & $1.2\times10^{-2} \%$\tabularnewline
\hline 
I8 & TSC unstructured (1D) & NGP & R1 & 0.99\% & $4.2\times10^{-5} \%$\tabularnewline
\hline 
\end{tabular}

\caption{Summary of tests of interpolation method and ray coverage strategy. NGP, CIC, TSC, and QSP are B-spline interpolation
of order zero, to four, respectively.}\label{table:interpolation-test-summary}
\end{table*}

\subsection{Parallelisation}\label{sec:ray-centric-domain-decomposition}

\begin{figure}
\begin{center}

\includegraphics[width=0.3\textwidth]{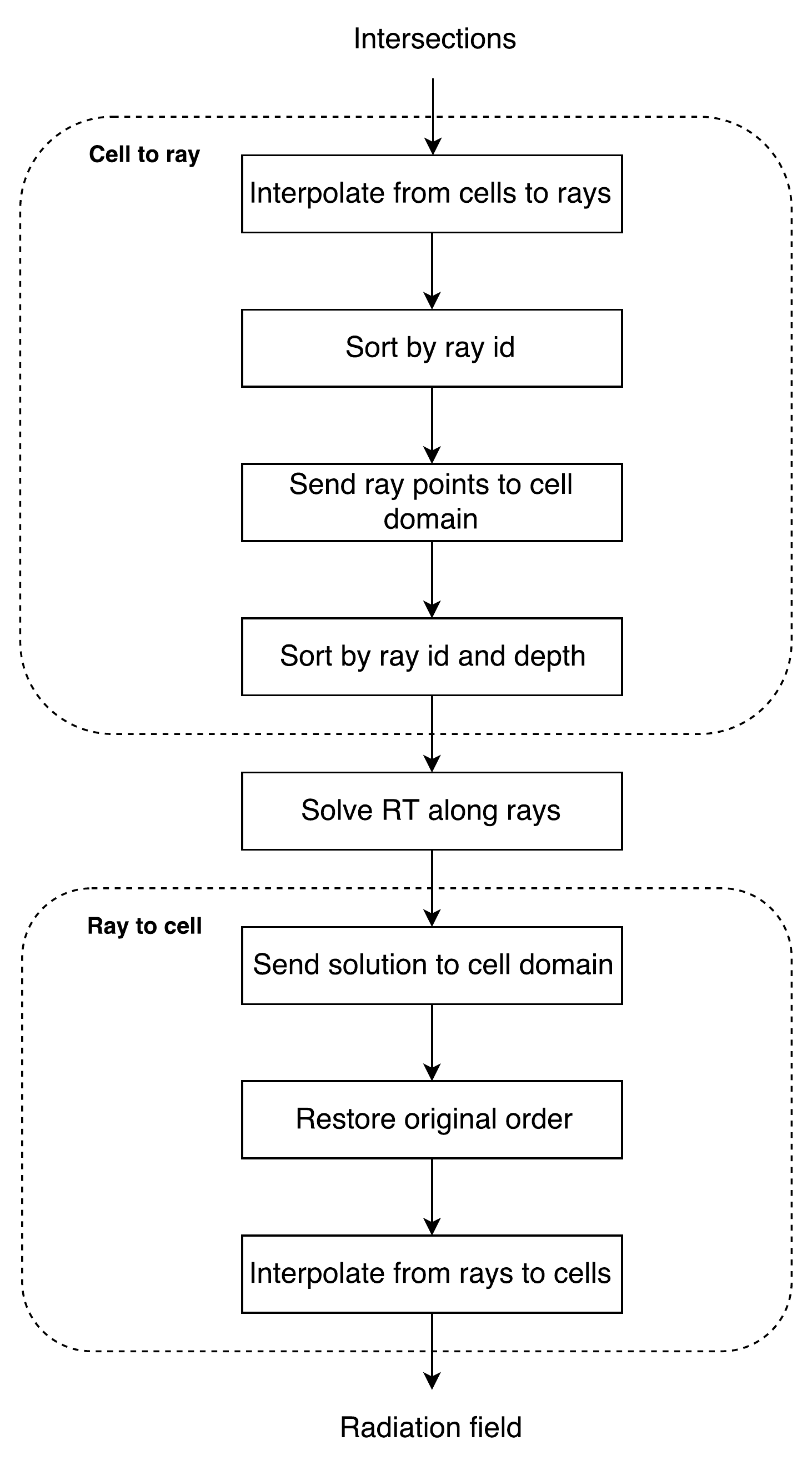}

\caption{Flow chart indicating the steps involved in changing the domain decomposition from cell-based to ray-based and back. }\label{fig:solve-rt-flow-chart}

\end{center}
\end{figure}

The {\ramses} grid is distributed over different {{\mpi}} ranks, and when constructing the hash table of rays in \sect{sec:ray-tracing}, we need to merge them before constructing ray points. This is done by extracting an array of the keys and values on each rank, gathering them on a single master rank, inserting them into the master rank's table, and then scattering keys and values to all ranks. 

When ray points have been constructed, and radiation variables interpolated to them, they represent ray intersections with the leaf cells present at a given rank, and they have the same cell-centric domain decomposition as {\ramses}' grid. We now change to a ray-centric domain decomposition. The load is balanced between ranks by dividing rays into equally sized chunks according to ray id, and sending ray points to the ranks that correspond to the ray they belong to. 

Before sending data from cell- to ray domain, the ray points, initially gathered in the order they were created, are sorted in memory according to ray id, so continuous blocks can be sent. 
The switch consists of a large communication step, where a number of ray points (typically on the order of the number of leaf cells on one rank) are exchanged, and all domains potentially could need to communicate with each other. 
Once received in the ray domain, the ray points must be sorted into the order in which they will be traversed when solving the radiative transfer equation along rays according to ray id and distance along the ray. Sorting is accomplished with the Quicksort algorithm. The resulting order of the two sorting operations is saved and reused to get the solution returned to the cell domain back into the original order. 

For some applications, the communication step can become too large to be handled correctly by a variant of an \verb+MPI_ALLTOALL+ operation, and is therefore split into a series of smaller communications. The domain decomposition switch is summarised in \fig{fig:solve-rt-flow-chart}.

\subsection{Scaling considerations}
As described, each rank must have access to a global hash table of all rays. Inevitably, for a large enough number of ranks, merging of tables will dominate computational time, and holding the table will dominate memory usage. Here we describe a way to solve this problem. The method is however not currently implemented into the code.  

For the ray plane cell at level $L$ with position $(i_x,i_y)$ counted in number of cells, we define the unique index
\begin{equation}
 i = \left[\sum^{L-1}_{l=L_{min}}4^l\right]\, + i_x + 2^L i_y + 1\,,
\end{equation}
where $L_{min}$ is the level of {\ramses}' uniform root grid. Each rank constructs a ray hash table as previously described, but using $i$ as table value instead of an enumeration of the rays. Consider an arbitrary rank $R$. {\ramses} holds a set of Hilbert key intervals that describe which cells belongs to each rank. We proceed to compute a set of bounding boxes for the cells on rank $R$ from the Hilbert key intervals. Sets of bounding boxes are likewise computed on $R$ for the remaining ranks. The projection of such a bounding box on the ray plane corresponds to a set of hash keys. For each rank $r$, we can take the union of its projected bounding boxes with those of $R$. These areas are mapped to a set of hash keys. Those keys and their values are extracted and sent to rank $r$. After applying this operation on all ranks, every rank has a table of all rays that intersects it's cells. Ray points are generated, and interpolated to, as previously described. We can however not achieve uniform load balancing by taking equal size domains in the ray index, since it is now sparse. We can instead get a statistically approximately uniform load balancing by shuffling the rays over the ranks. This can be done in many ways. An example is to assign ray $i$ to rank $r$ according to 
$r = i \mod{n_{\textrm{ranks}}}$, where $n_{\textrm{ranks}}$ is the total number of ranks. 

With these modifications the rest of the method can proceed as previously described.
Note that in order to support more than 21 levels of refinement, an 8 byte integer is not enough to represent the ray index, so one must construct an integer type of 16 bytes. Likewise the hash keys must be triples of 8-bit integers. 

\subsection{Boundary conditions}
We have implemented the following boundary conditions for the radiation. A prescribed incoming intensity ($I(0)=I_{0}$, $I(L)=I_{L}$), which is useful in test problems, radiative equilibrium ($I(0)=S(0)$, $I(L)=S(L)$), and periodic boundaries for the radiation. Periodicity is useful in situations with periodic boundaries for the hydrodynamics, when the optical depth is also a fraction of the domain size. 
Intersection between a cell, given by its centre $\boldsymbol{c}$ and size $\Delta x$, and a ray, given by a direction vector $\boldsymbol{r}$ and a displacement vector $\boldsymbol{p}$, is tested by transporting $\boldsymbol{c}$ along $\boldsymbol{r}$ to the ray plane and looking up the ray at the location in a hash table, as described in \sect{sec:ray-tracing}. Let $a_z$ denote the axis along which $\boldsymbol{r}$ has the largest component, and $\boldsymbol{c'}$ the transported cell centre. To get periodicity around the four faces perpendicular to $a_z$, we can simply take $\boldsymbol{c'} \rightarrow \left(\boldsymbol{c'} \mod{L}\right)$ before consulting the hash table. 
To also get periodicity along axis $a_z$, we repeat the intersection test twice more, with $\boldsymbol{c}$ translated one box length along axis $a_z$ in the positive and negative direction, i.e.~$c''_a=c_a \pm L_a$. The intersection is further discarded if $\boldsymbol{c''}$ is more than a fraction $f$ of $L$ outside the box, where $f$ is an adjustable parameter. 
The solution on the part of the ray that is outside the box along axis $a_z$ is discarded, as the domain is already covered entirely by rays periodic over four faces. The discarded part only acts to provide a similar environment, which will be the case if the optical depth over the distance $f \cdot L$ is large. A boundary condition is however still required for the ray, and can be one of those mentioned in the beginning of the section. If a significant fraction of the volume is optically thin, this will not accomplish a perfectly periodic box, but still be a very good approximation. If this is critical for a model convergence can be checked by running several simulations with different values of $f$.

\section{Tests}\label{sec:tests}
{\lampray} is applied to a set of test problems where the result can be compared to a well-known analytical solution or to other established codes. The tests include the expansion of an HII region around a new star (Test~1), trapping of an ionization front by a dense clump of gas (Test~2), photo-evaporation of a dense clump by ionising radiation (Test~3), the chemical structure of a plane-parallel photon dominated region (Test~4), and of a spherical, externally irradiated molecular cloud (Test~5). 

The {\ctoray}-based ionization chemistry implementation has previously been tested in a code comparison on the D-type expansion of an HII region \citep{Bisbas2015MNRAS}. 
\subsection{TEST~1: HII region expansion}
The cosmological code comparison papers \citet{Iliev2006} and \citet{Iliev2009} include a series of four tests with the expansion of an HII region around a point source. We reproduce three of these here, starting with a monochromatic source in an isothermal, static, uniform density medium, then replacing the source with a $T=10^5$~K black body and including heating and cooling, and finally also including hydrodynamics. 

We use {\krome} to evolve the chemistry, with a network consisting of $\ch{H}$, $\ch{H+}$, and electrons, and the only reactions are photo-ionization with cross section given in \citet{Verner1996}, and electron recombination, where the test specifies a case B recombination rate of $\alpha_B=2.59 \times 10^{-13} (T / 10^4 \textrm{K})^{-3/4}\, \textrm{cm}^6\, \textrm{s}^{-1}$. When heating and cooling is included, this is done as follows. Photo-ionization heating is given by 
\begin{equation}
 G(H)=n_{\ch{H}}\int^\infty_{\nu_0}F_\nu h(\nu-\nu_0)\sigma_\nu(H)d\nu
\end{equation}(e.g.~\cite{Osterbrock1989}).
The dominant cooling processes are recombination cooling and free-free emission cooling. We also include collisional ionization and excitation cooling of H. All cooling rates are from \citet{Cen1992}. We employ 6 logarithmically spaced frequency bins with photon energy from 13.6 to 100~eV. 

In all three tests, the initial number density is $n_{\ch{H}}=10^{-3}$~cm$^{-3}$ of only atomic hydrogen, ionising photon rate $\dot{N}_{\gamma}=5 \times 10^{48}\, \textrm{photon}\, \textrm{s}^{-1}$, a recombination coefficient for \ch{H+ + e-} of $\alpha_B=2.59 \times 10^{-13} (T /10^4\, \textrm{K})^{-3/4}$, and simulation time $t_\mathrm{sim}=500$~Myr. 
\subsubsection{TEST~1.1: Isothermal gas}
\begin{figure}
\begin{center}
\includegraphics[width=0.48\textwidth]{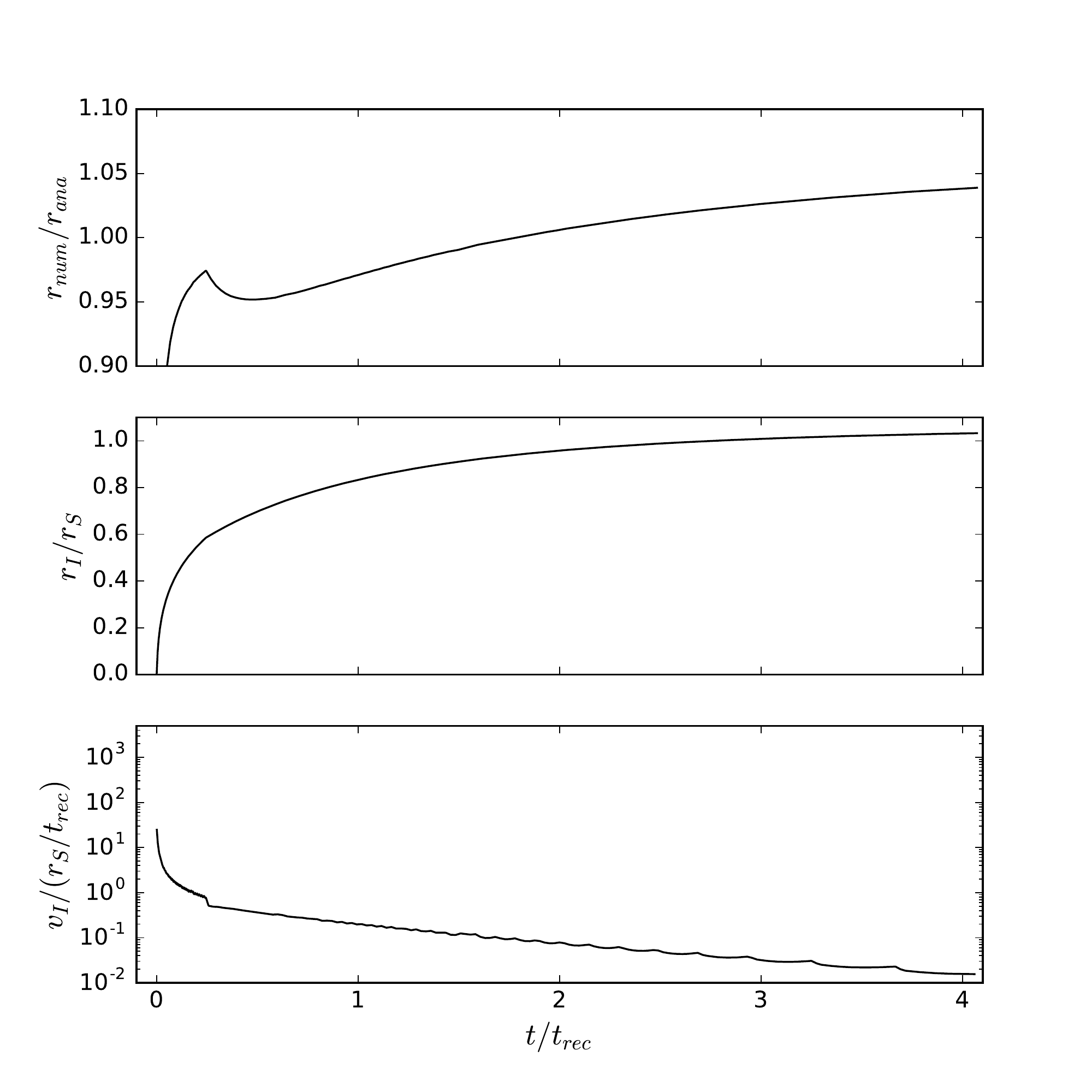}
\caption{Test~1.1: Position and velocity of the ionization front in a uniform gas at fixed temperature. }\label{fig:I06-fig7}
\end{center}
\end{figure}

\begin{figure}
\begin{center}
\includegraphics[width=0.48\textwidth]{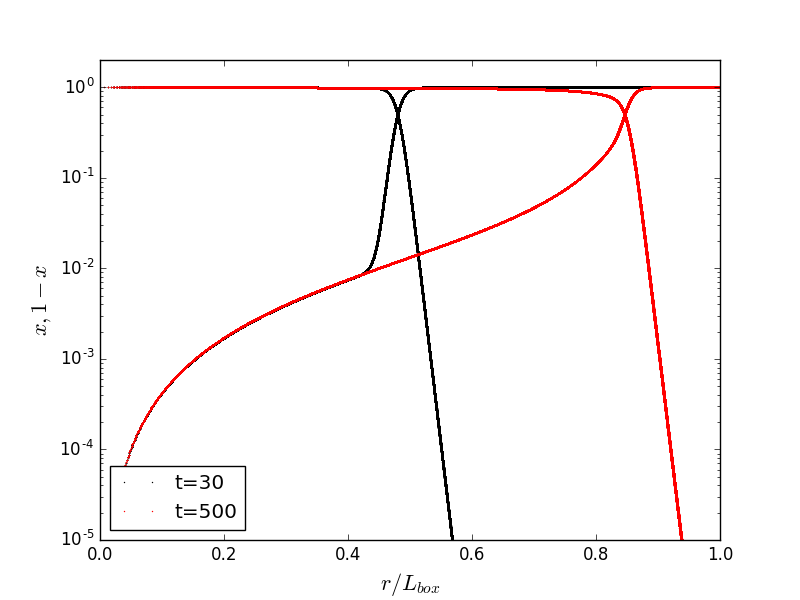}
\caption{Test~1.1: ionization fraction ($x=n_{\ch{H+}}/n_{\ch{H}_{\mathrm{tot}}}$) and neutral fraction ($x-1$). All cells are plotted, so the width of the lines show the spread at a given radius. }\label{fig:I06-fig8}
\end{center}
\end{figure}

\begin{figure}
\begin{center}
\includegraphics[width=0.48\textwidth]{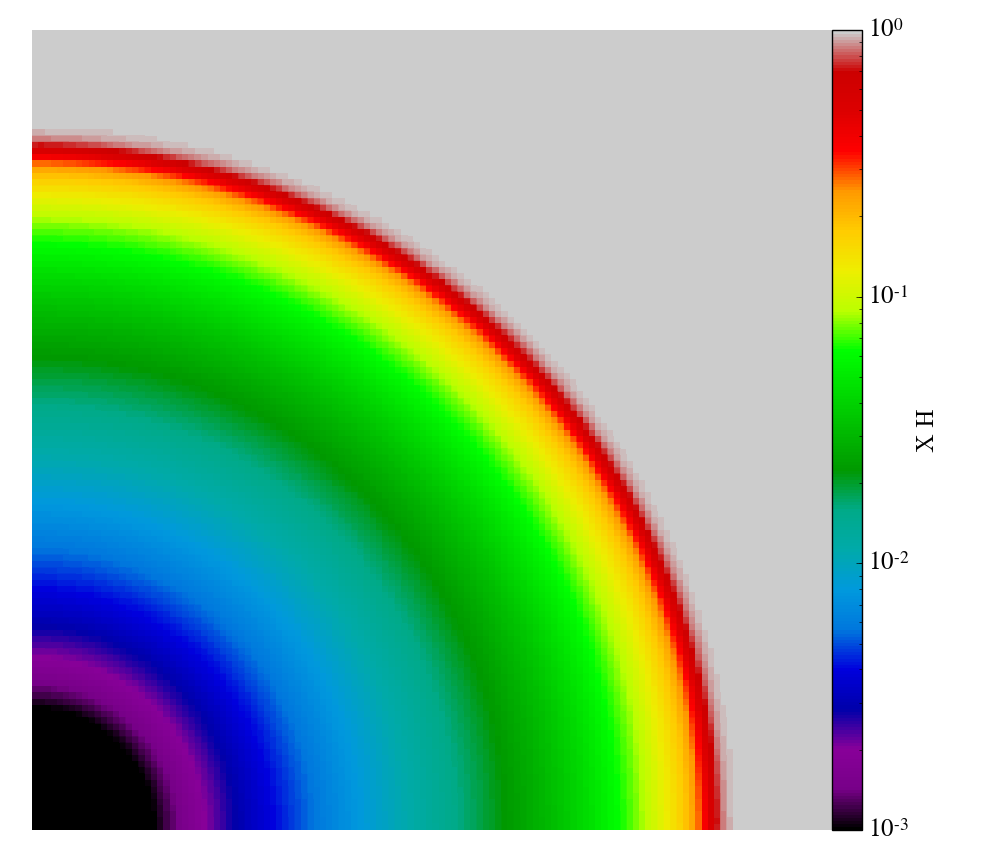}
\caption{Test~1.1: Neutral fraction, cut through volume at z=0 at time t=500~Myr. The entire 6.6 kpc box is shown. }\label{fig:I06-fig6}
\end{center}
\end{figure}

The first test reproduces \citet{Iliev2006} Test~1, which has monochromatic radiation with photon energy $13.6$~{eV}, is isothermal at $T=10^4$~K, has box length $L=6.6$~kpc, and is resolved by $128^3$ cells. 
At the final time, the ionization front is spherical (\fig{fig:I06-fig6}) with ionization structure in good agreement with the benchmark codes. The transition thickness is close to the theoretically expected $18\lambda_{\textrm{mfp}}=0.74$~kpc for monochromatic radiation, where $\lambda_{\textrm{mfp}}$ is the photon mean free path. After the very early time, the size of the front differs by less than $5\%$ from the analytically predicted radius (\fig{fig:I06-fig7}). Being explicit in time, our method requires a relatively small time step to follow the time evolution correctly. The bump around $t=0.3\,t_{\textrm{rec}}$ is due to a change in the time step there, which is set explicitly rather than by a Courant condition. Looking at spherically averaged profiles of the ionized and neutral fraction (\fig{fig:I06-fig8}), the result agrees very well with the benchmarked ray tracing codes at both $t=30$~Myr and $t=500$~Myr. 

\subsubsection{TEST~1.2: Non-isothermal gas}
\begin{figure}
\begin{center}
\includegraphics[width=0.48\textwidth]{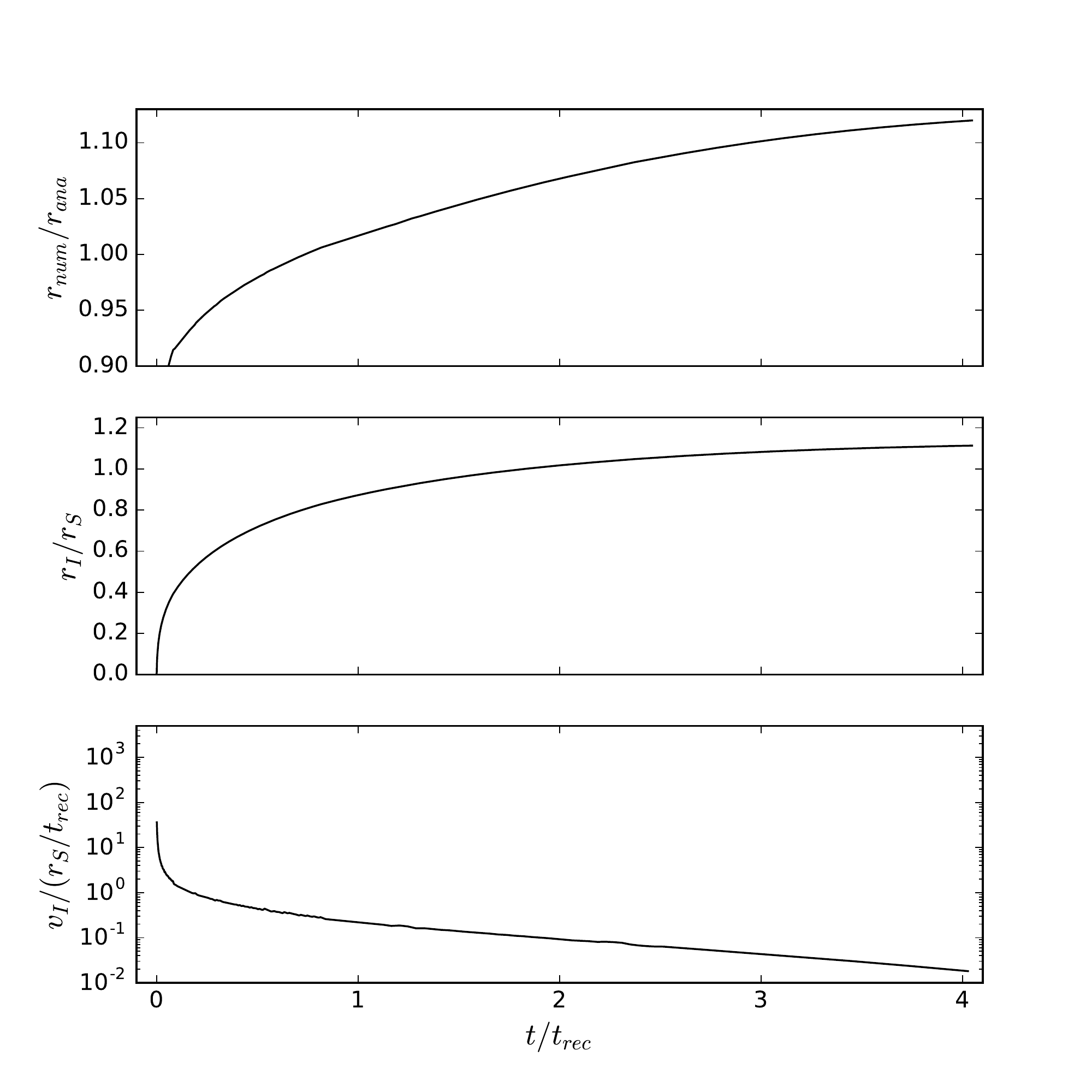}
\caption{Test~1.2: Position and velocity of the ionization front in a uniform gas with varying temperature.}\label{fig:I06-fig15}
\end{center}
\end{figure}

\begin{figure}
\begin{center}
\includegraphics[width=0.48\textwidth]{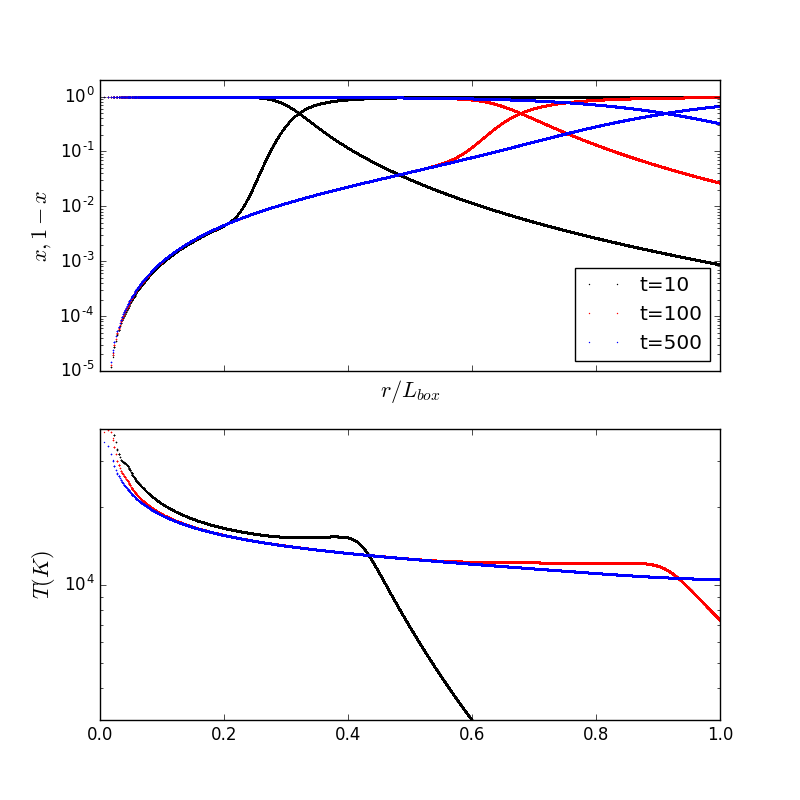}
\caption{Test~1.2: Ionized and neutral fraction (top) and gas temperature (bottom) at 10, 100 and 500 Myr. }\label{fig:I06-fig16-17}
\end{center}
\end{figure}

\begin{figure}
\begin{center}
\includegraphics[width=0.48\textwidth]{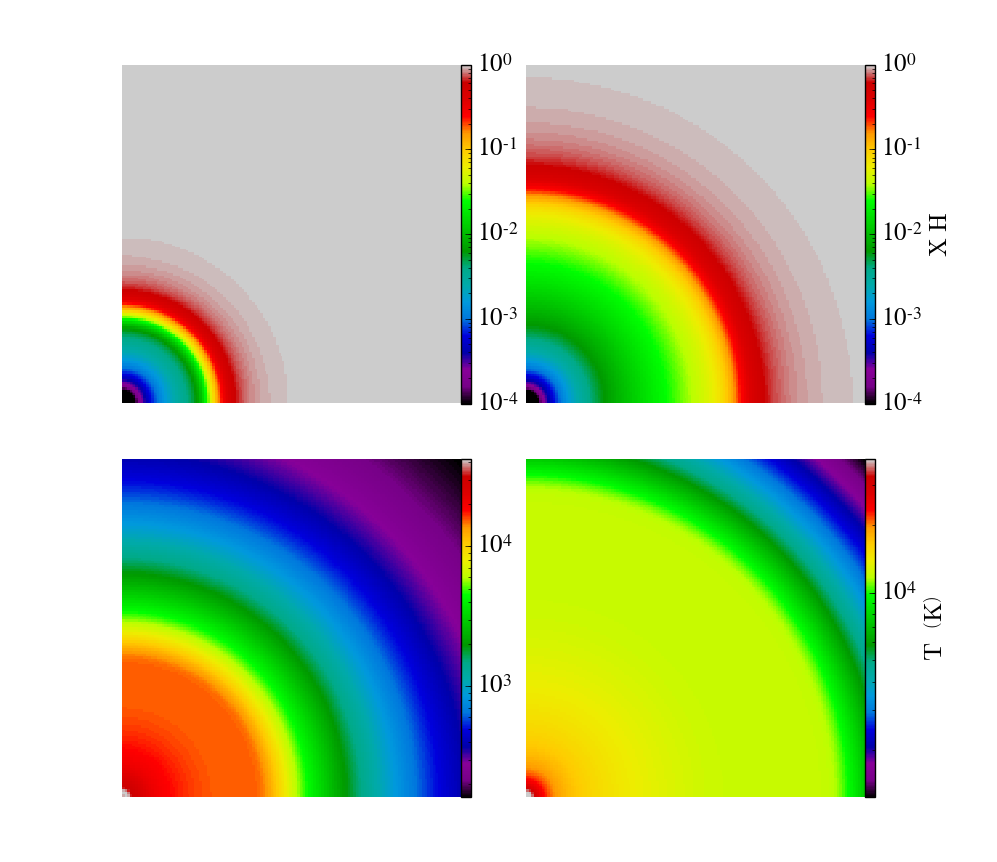}
\caption{Test~1.2: Neutral fraction (top) and temperature (bottom), cut through volume at z=0 at time t=10 (left) and t=100~Myr (right).}\label{fig:I06-fig11-14}
\end{center}
\end{figure}

The second test reproduces \citet{Iliev2006} Test~2, which has initial temperature $T=100$~K and includes heating and cooling. The box length is still $L=6.6$~kpc. Again, the ionization structure agrees well, especially with the ray tracing codes (\fig{fig:I06-fig11-14}), and the temperature structure is similar to that of especially {\rsph}, with significant pre-heating due to spectral hardening, i.e. the effect that high energy photons are predominately preserved at high optical depth since the cross section is lower at high energy. The time evolution of the ionization front (\fig{fig:I06-fig15}) is similar to that of {\ctoray}, {\rsph}, and {\ftte}, the final size being larger than the isothermal Str\"omgren radius due to a higher temperature and consequently lower recombination rate. Spherically averaged profiles (\fig{fig:I06-fig16-17}) confirms this agreement in ionized and neutral fraction at all radii, and the agreement on temperature with {\rsph}. 

\subsubsection{TEST~1.3: Hydrodynamics}
The third test reproduces \citet{Iliev2009} Test~5. It has the same parameters as Test~1.2, except that the box length is $L=15$~kpc and hydrodynamics is included. The position of the ionization front (\fig{fig:I09-fig13-17}) follows the agreement of all the benchmarked codes except \enzo{}, which is monochromatic. In density and Mach number, the double peak at $t \approx 200$~Myr due to thermal expansion from pre-heating by hard photons seen by the remaining codes is recovered, and the radial structure lies within the relatively large scatter between the codes, most closely resembling those with similar temperature structure in the previous test. Cuts through $z=0$ (\fig{fig:I09-fig2-4} and \fig{fig:I09-fig5-10}) show nicely spherical profiles that resemble those of the same codes. The test is repeated on an adaptive mesh using a 32$^3$ root grid and 2 levels of refinement with a similar result (\fig{fig:I09-fig5-10-amr}).

\begin{figure}
\begin{center}
\includegraphics[width=0.48\textwidth]{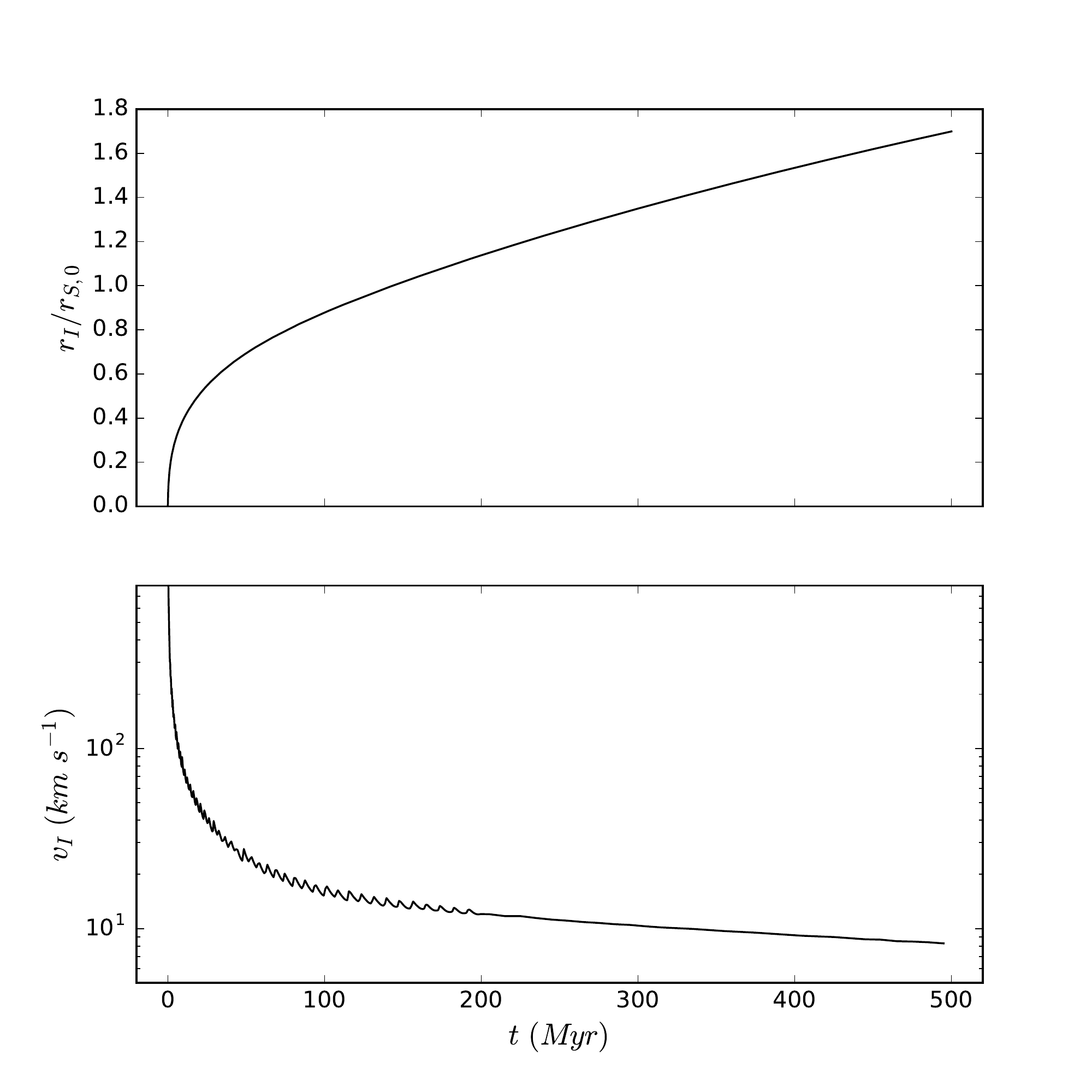}
\caption{Test~1.3: Position and velocity of ionization front with gas expansion in an initially uniform gas. }\label{fig:I09-fig18}
\end{center}
\end{figure}

\begin{figure}
\begin{center}
\includegraphics[width=0.48\textwidth]{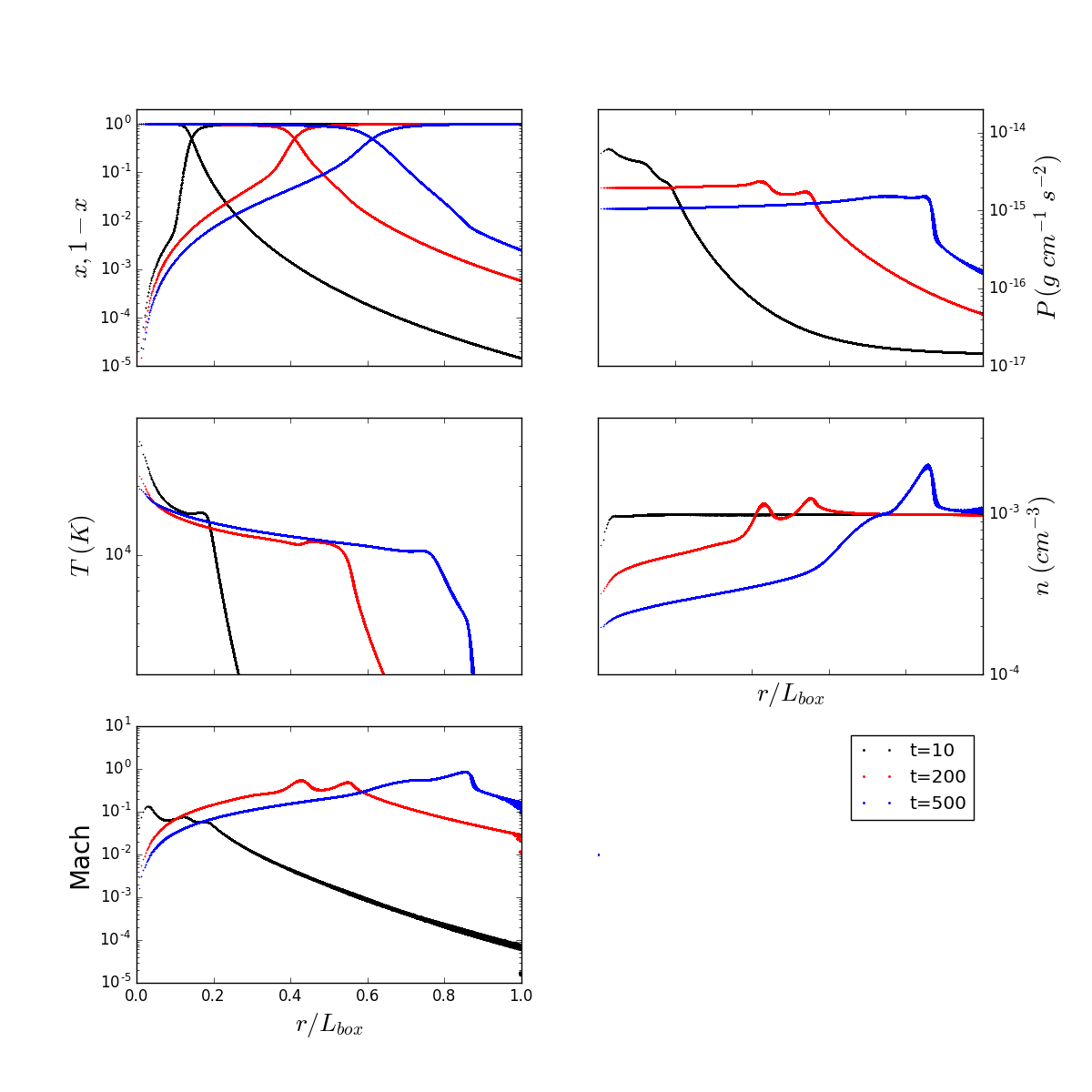}
\caption{Test~1.3: The panels show the ionization and neutral fraction, pressure, temperature, total number density and Mach number.}\label{fig:I09-fig13-17}
\end{center}
\end{figure}

\begin{figure}
\begin{center}
\includegraphics[width=0.48\textwidth]{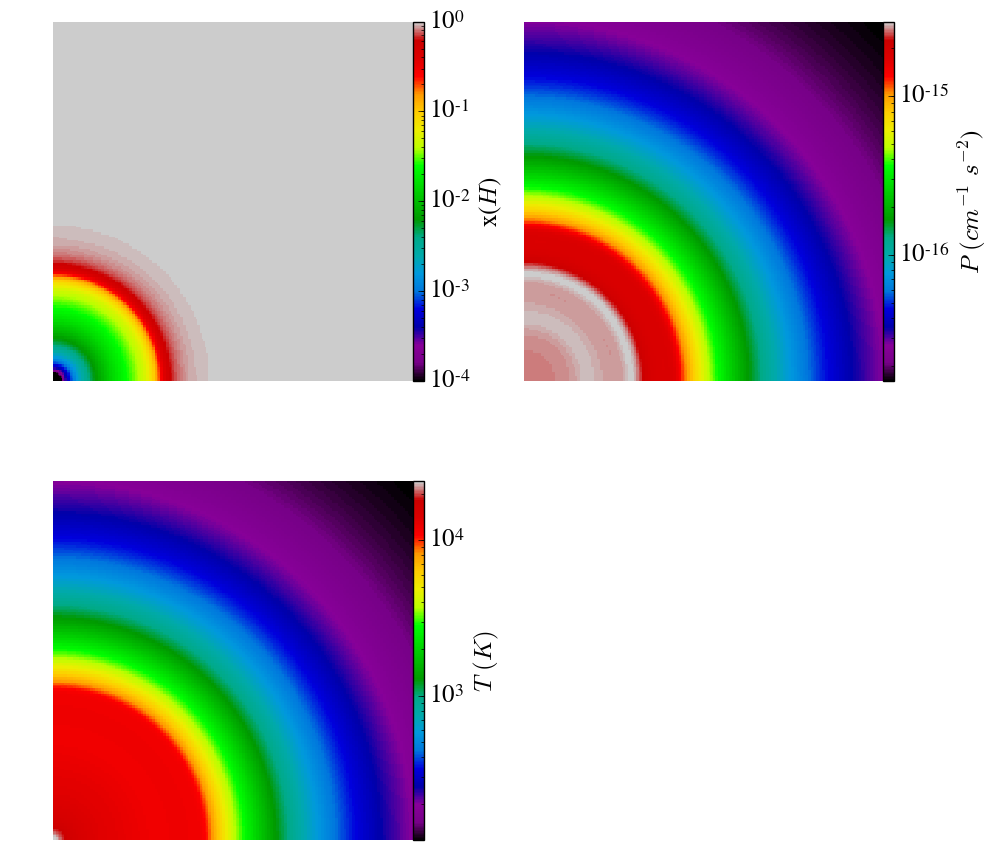}
\caption{Test~1.3: Neutral fraction, pressure and temperature, cut through volume at z=0 at time t=100~Myr}\label{fig:I09-fig2-4}
\end{center}
\end{figure}

\begin{figure}
\begin{center}
\includegraphics[width=0.48\textwidth]{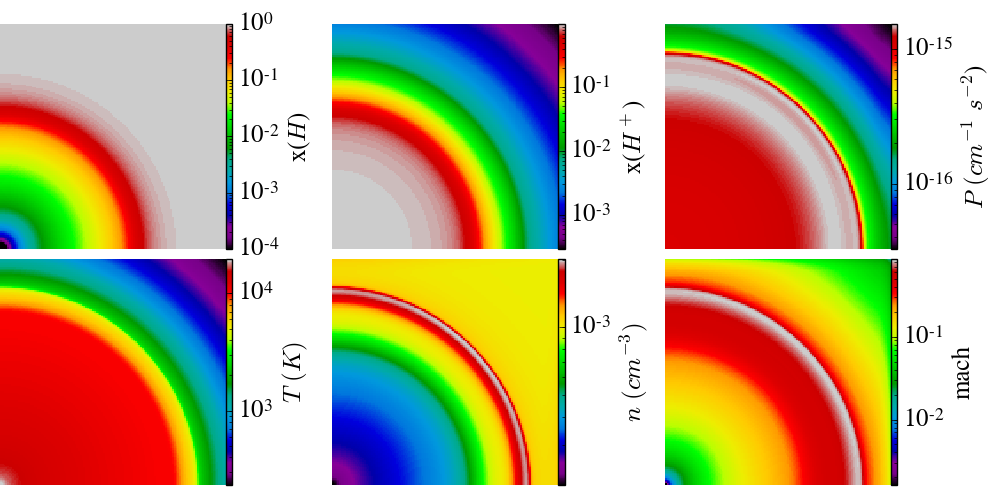}
\caption{Test~1.3: Neutral fraction, ionized fraction, pressure, temperature, total number density and Mach number, cut through volume at z=0 at time t=500~Myr}\label{fig:I09-fig5-10}
\end{center}
\end{figure}

\begin{figure}
\begin{center}
\includegraphics[width=0.48\textwidth]{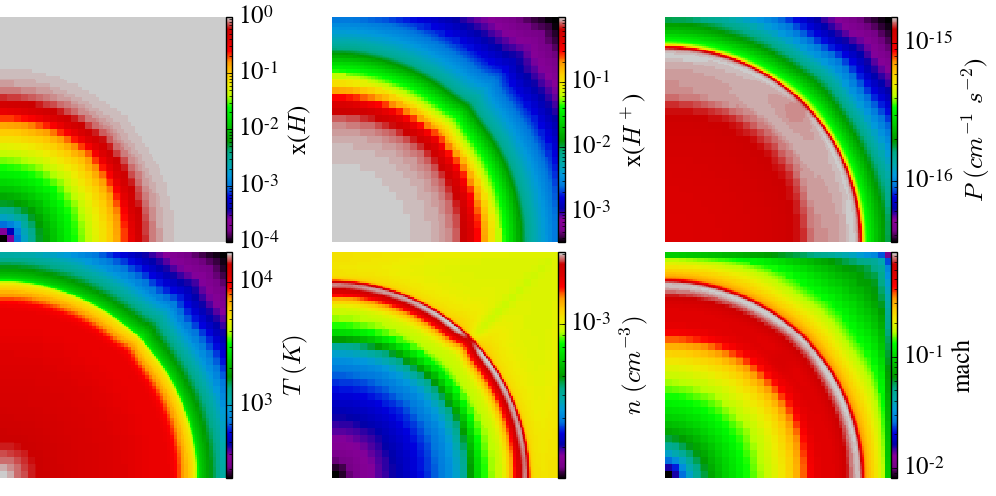}
\caption{Test~1.3: Same as \fig{fig:I09-fig5-10} except run with adaptive mesh refinement turned on using a 32$^3$ root grid and 2 levels of refinement.}\label{fig:I09-fig5-10-amr}
\end{center}
\end{figure}

\subsection{TEST~2: ionization front trapping by a dense clump}
This test is a replication of Test~3 from the code comparison paper \citet{Iliev2006}. It tests hydrogen photo-ionization chemistry and related heating and cooling processes, as well as the diffusivity of radiation transport and ability to form a shadow. The setup is briefly repeated here. The simulated volume is resolved by $128^3$~cells, is cubic with side length $L = 6.6$~kpc, filled with only atomic hydrogen of uniform number density $n_\mathrm{out}=2\times10^{-4}$~cm$^{-3}$ and initial temperature $T_\mathrm{out,init}=8000$~K. A dense clump with radius $r_\mathrm{clump}=0.8$~kpc has centre $(x_c,y_c,z_c)=(5,3.3,3.3)$~kpc, number density $n_\mathrm{clump}=200\,n_{out}$ and initial temperature $T_\mathrm{clump,init}=40$~K. A photon flux $F=10^6$~cm$^{-2}$~s$^{-1}$ with a black body spectrum with effective temperature $T_\mathrm{eff}=10^5$~K enters the box at $x=0$ and propagates along the x-axis. The chemical model is the same as in Test~1.

\begin{figure}
\begin{center}
\includegraphics[width=0.48\textwidth]{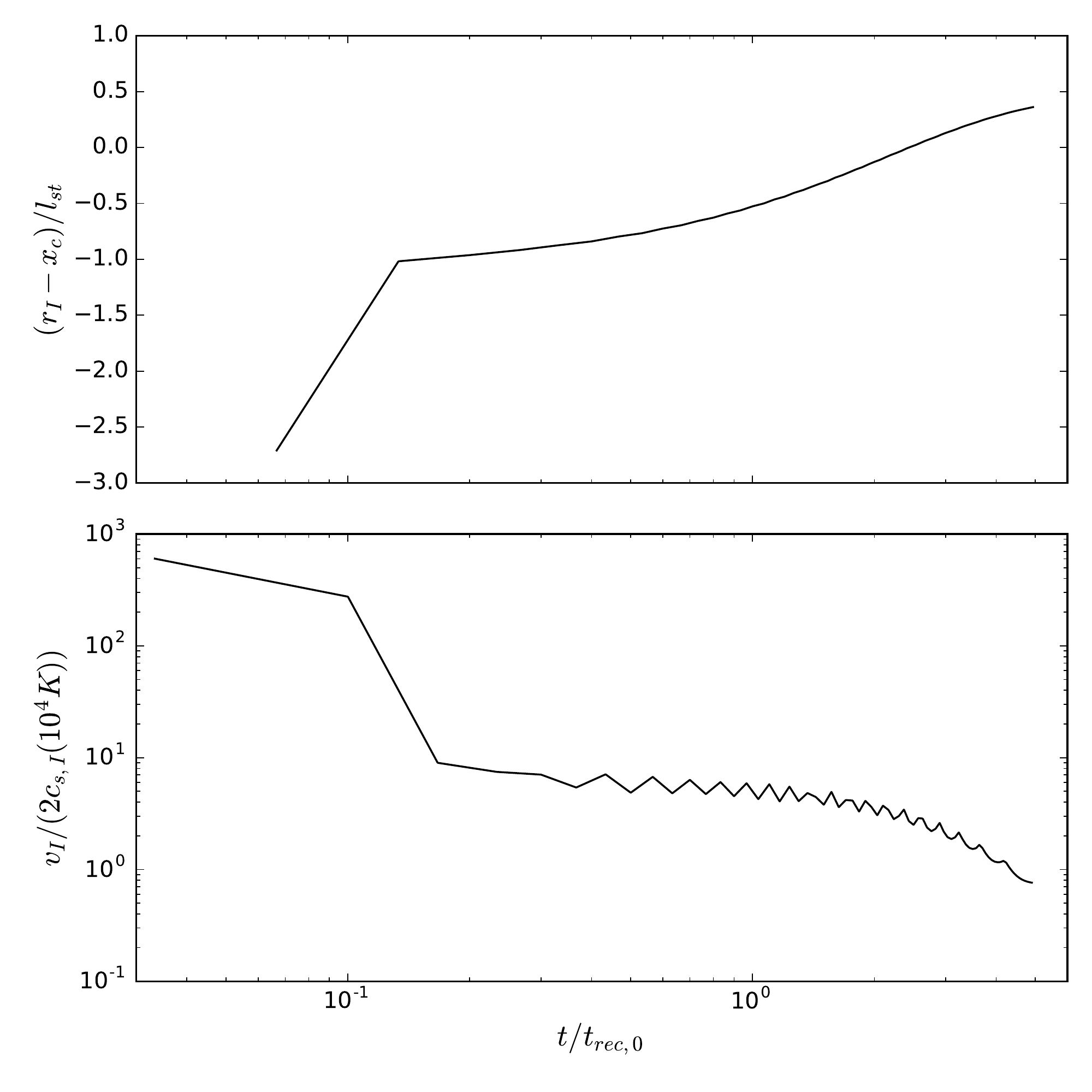}
\caption{Test~2: Position and velocity of the ionization front along the axis of symmetry through the clump in a static density field. }\label{fig:I06-fig21}
\end{center}
\end{figure}

\begin{figure}
\begin{center}
\includegraphics[width=0.48\textwidth]{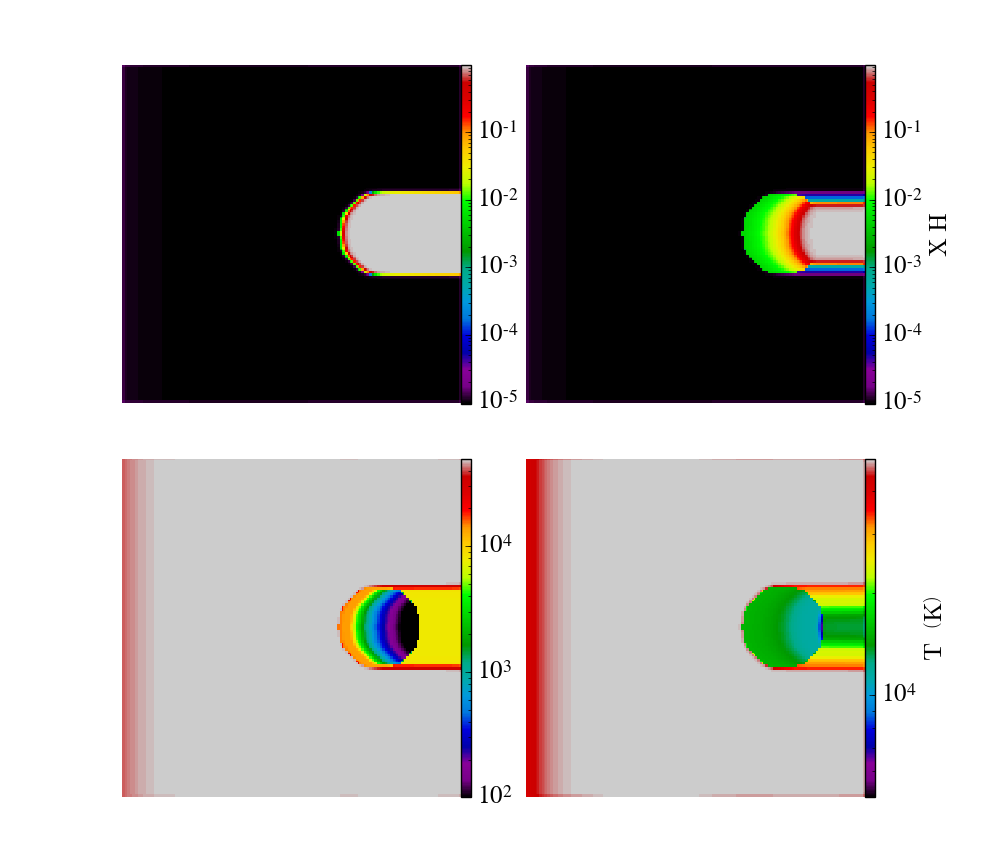}
\caption{Test~2: Slices through the central plane showing the neutral fraction and temperature at times 1, 3 and 15~Myr. }\label{fig:I06-test3-slices}
\end{center}
\end{figure}

\begin{figure}
\begin{center}
\includegraphics[width=0.48\textwidth]{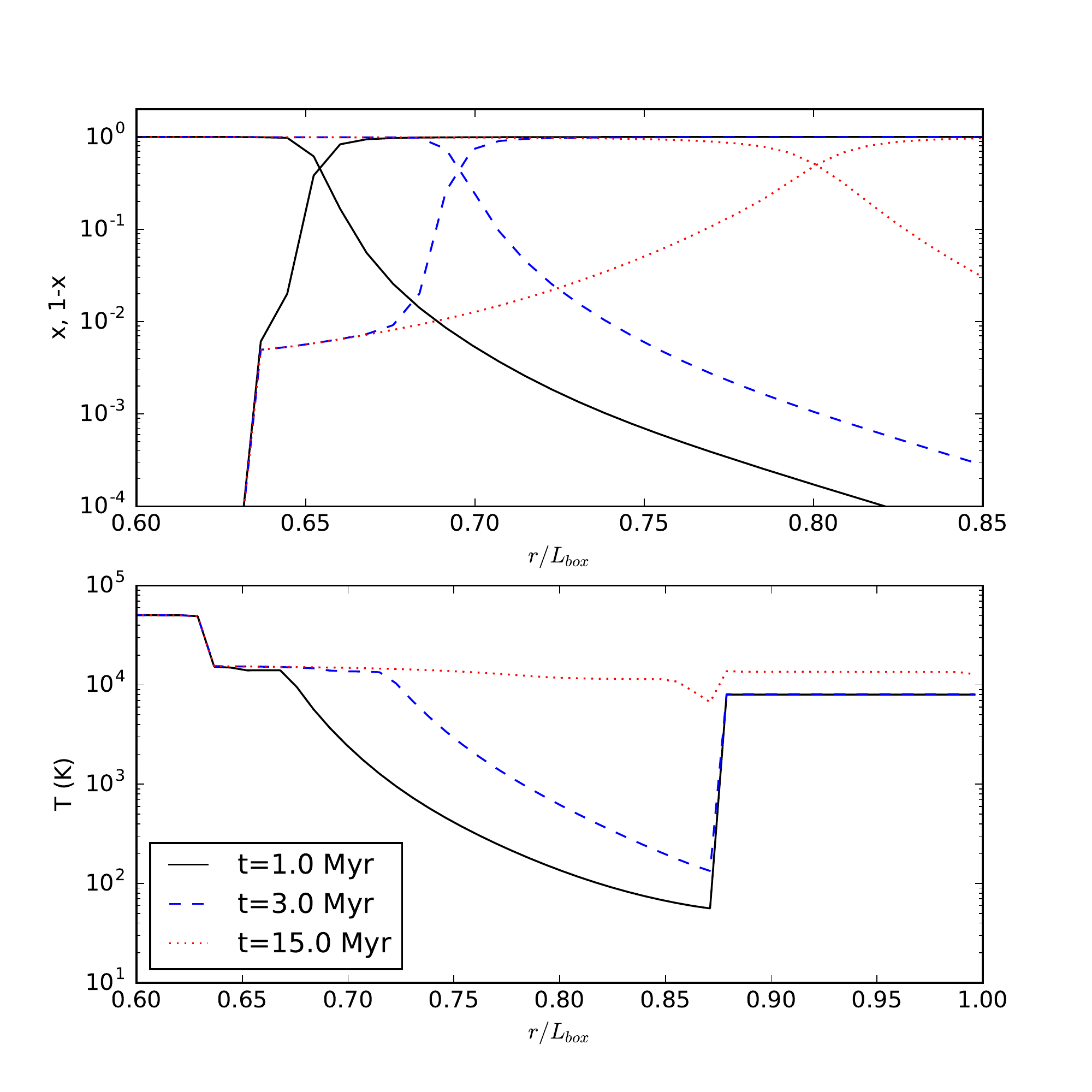}
\caption{Test~2: Neutral and ionized fraction, and temperature along the axis of symmetry through the clump. }\label{fig:I06-fig27-28}
\end{center}
\end{figure}

The time evolution of the ionization front after it reaches the clump, agrees well with the other codes, trapping the front a bit beyond the clump centre (\fig{fig:I06-fig21}). However, the position at the first two time steps deviates, because our time step of $0.2$~Myr is too large to resolve the fast propagation through the thin medium. When at 1~Myr the clump reaches the front, but is not yet trapped, the clump casts a sharp shadow (\fig{fig:I06-test3-slices}), which is both apparent in neutral fraction and temperature. Note that though similar, the color scale is not identical to the one used by \citet{Iliev2006}. The final ionization structure at 15~Myr also agrees well with the other codes except \textsc{crash}, which ionizes a smaller region and forms a stronger shadow. The temperature structure mostly resembles that of \textsc{flash-hc}, {\rsph}, and \textsc{coral}, which find relatively low self-shielding, and correspondingly higher temperatures in the shadow. The location of the ionization front agrees well with the majority of the other codes (\fig{fig:I06-fig27-28}), with a sharp transition when the clump is reached at 1~Myr, and a final extent above average of the participating codes, due to the lower recombination rate caused by the higher temperature. And like the majority of the codes, we see pre-heating of the clump, and no significant heating of the shadow at 3~Myr. Like the codes with low self-shielding, at 15~Myr the shadow has heated up, in our case by roughly a factor of~2. 

\subsection{TEST~3: Photo-evaporation of a dense clump}
\begin{figure}
\begin{center}
\includegraphics[width=0.48\textwidth]{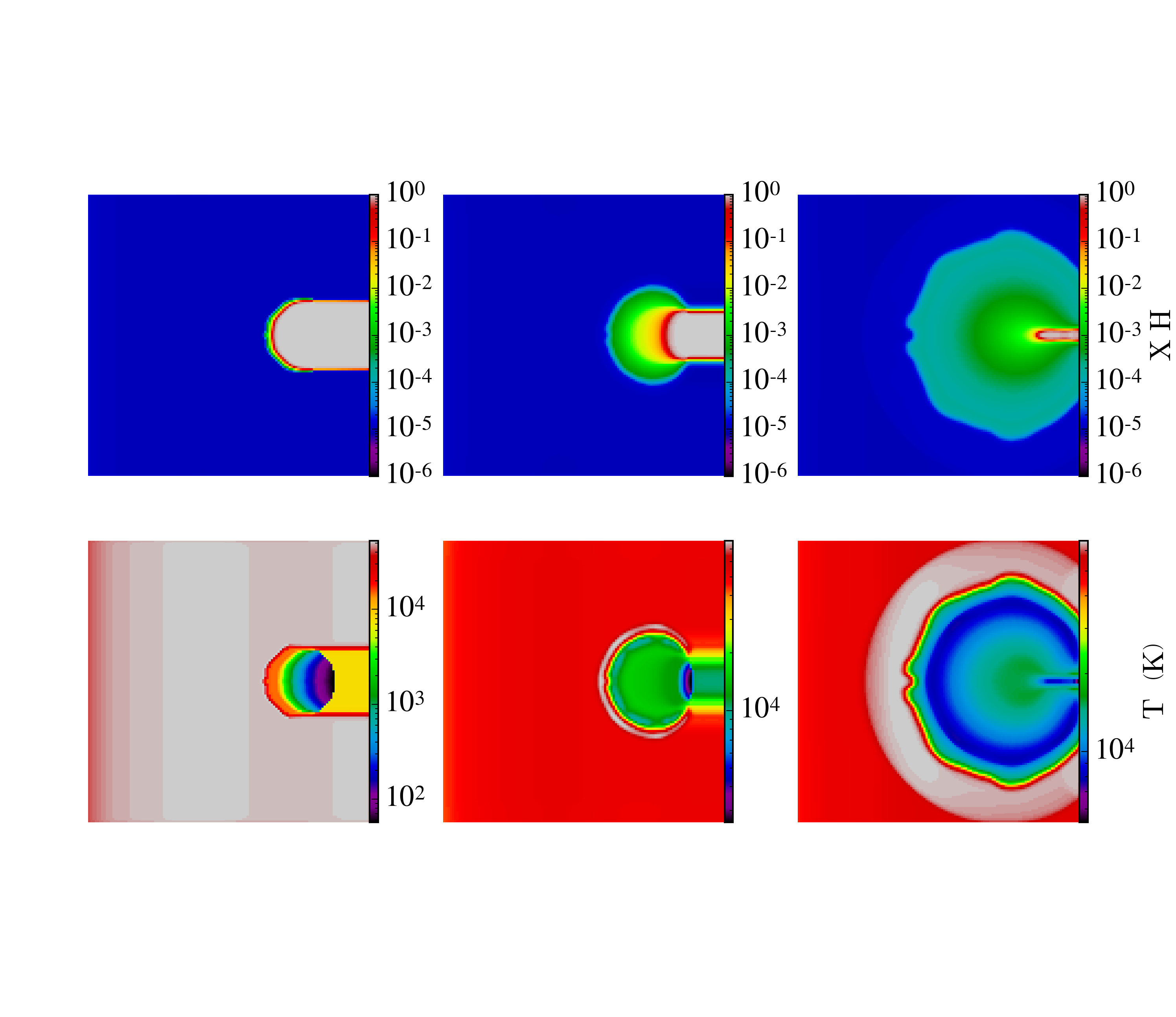}
\caption{Test~3: Slices through the central plane showing the neutral fraction and temperature at times 1, 10 and 50~Myr.}\label{fig:I09-test7-slices}
\end{center}
\end{figure}

\begin{figure}
\begin{center}
\includegraphics[width=0.48\textwidth]{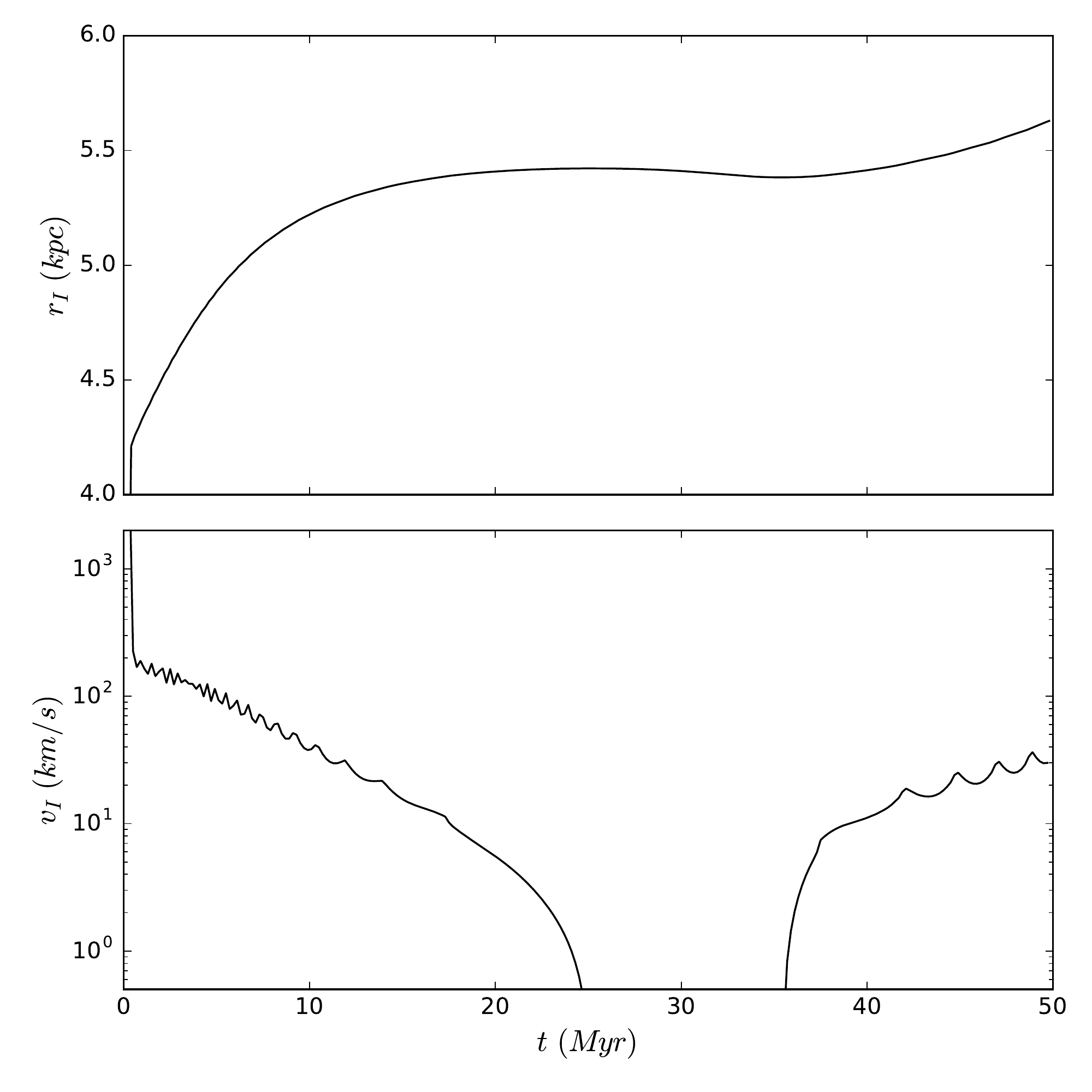}
\caption{Test~3: Position and velocity of the ionization front along the axis of symmetry through the photo-evaporating clump.}\label{fig:I09-fig30}
\end{center}
\end{figure}

\begin{figure}
\begin{center}
\includegraphics[width=0.48\textwidth]{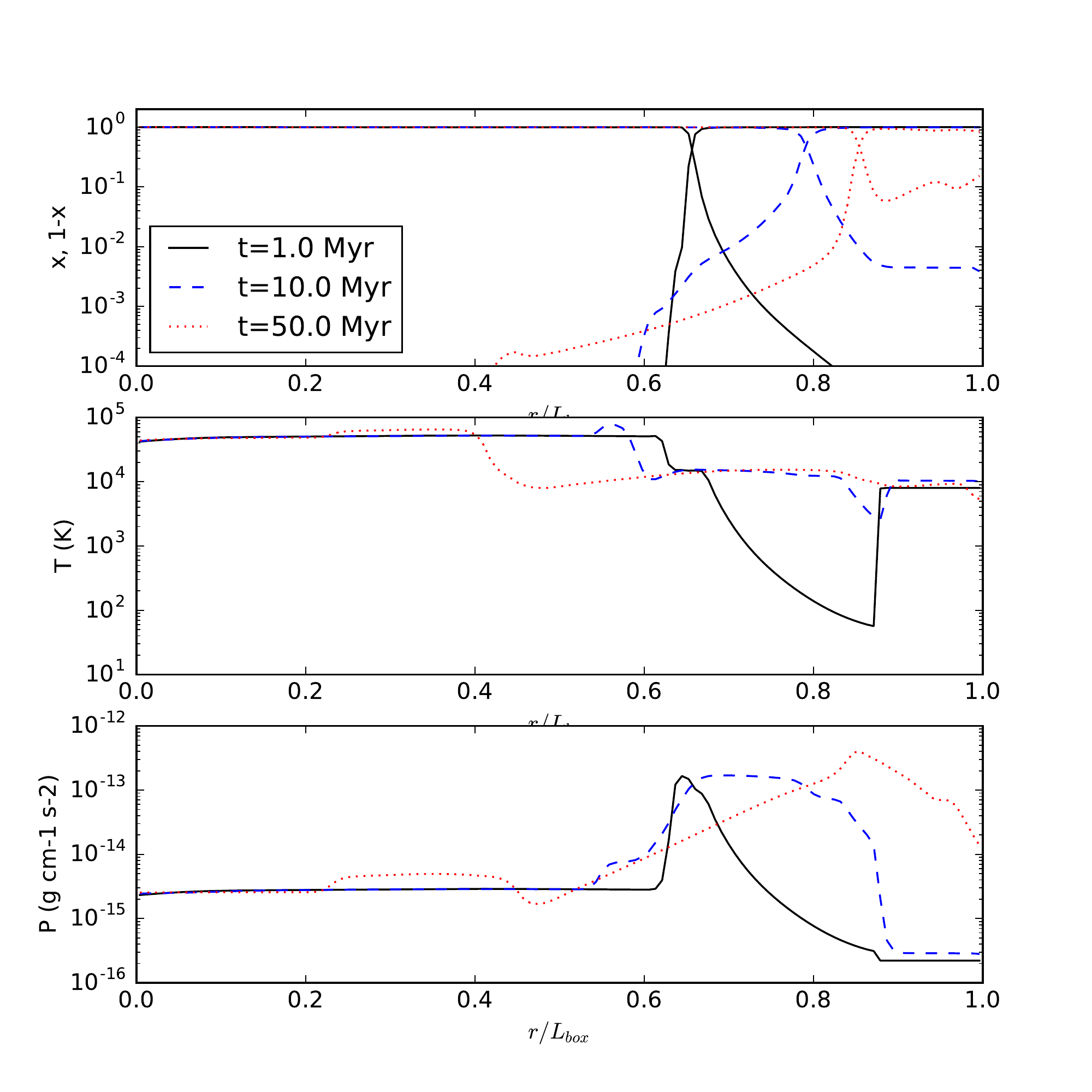}
\caption{Test~3: Neutral and ionized fraction, and temperature along the axis of symmetry through the photo-evaporating clump.}\label{fig:I09-fig42-43}
\end{center}
\end{figure}

\begin{figure}
\begin{center}
\includegraphics[width=0.48\textwidth]{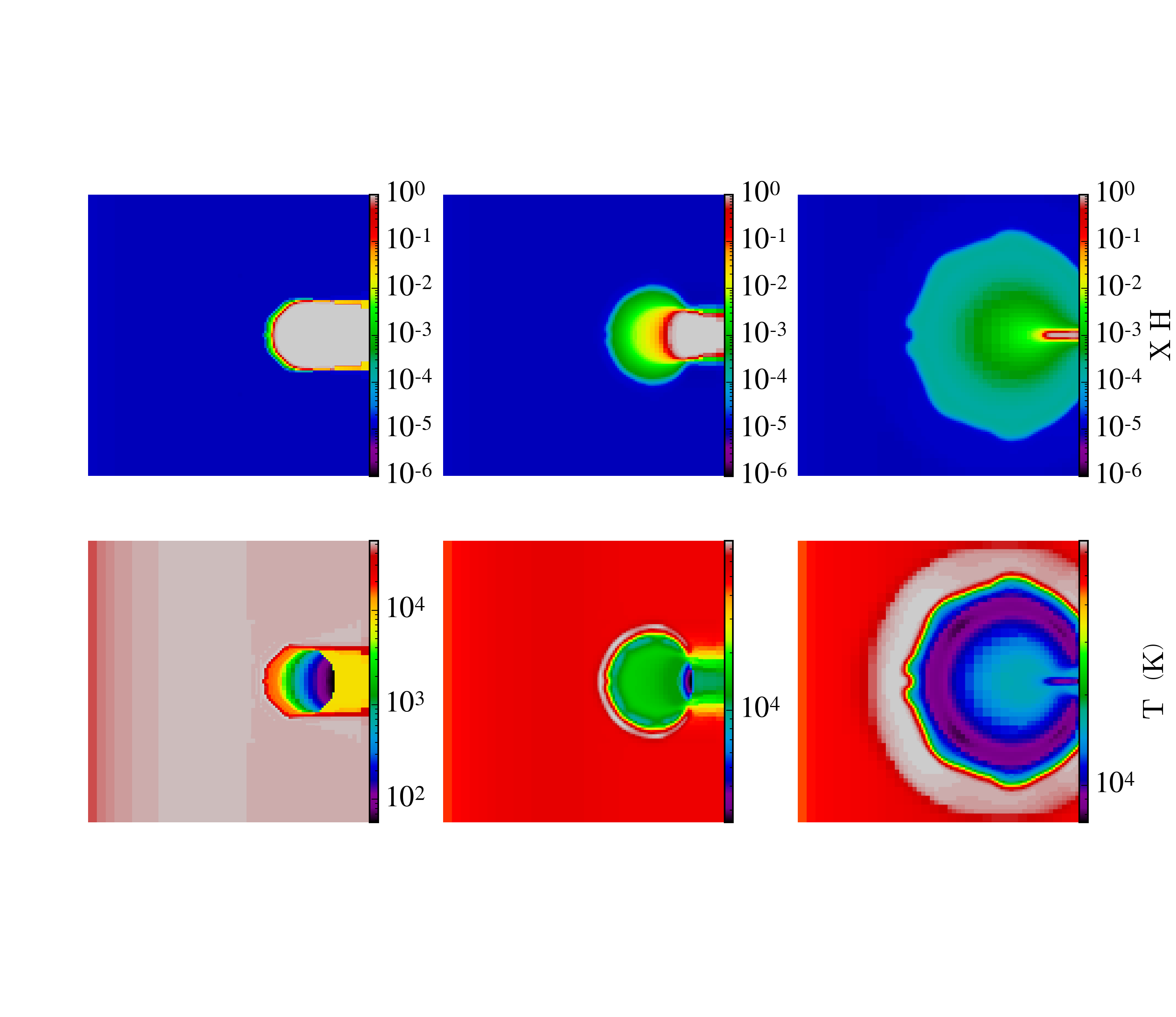}
\caption{Test~3: Same as \fig{fig:I09-test7-slices}, but with adaptive mesh refinement turned on using a 32$^3$ root grid and 2 levels of refinement.}\label{fig:I09-test7-slices-amr}
\end{center}
\end{figure}

This test corresponds to Test~7 from the code comparison paper \citet{Iliev2009}. It is identical to the above test, except hydrodynamics is included, allowing the clump to evaporate. At 1~Myr, the gas is still almost static, and the structure resembles the previous test, with virtually no flaring from the ionization front (\fig{fig:I09-test7-slices}). The evolution of position and velocity also agrees well with the majority of the other codes (\fig{fig:I09-fig30}). Like for \textsc{flash-hc}, there is a small asymmetry on the side facing the radiation at 10~Myr, which is magnified by the expansion to be clearly visible at 50~Myr (\fig{fig:I09-test7-slices}). It is due to the discretisation of the initial condition, and is absent for the other grid codes because their initial conditions have been smoothed \citep{Iliev2009}. The evolution of position, temperature, and pressure along the axis of symmetry generally agrees well with other codes (\fig{fig:I09-fig42-43}). The test is repeated on an adaptive mesh using a 32$^3$ root grid, and 2 levels of refinement with similar results (\fig{fig:I09-test7-slices-amr}).

\subsection{TEST~4: Plane-parallel PDR}\label{sec:PDR-tests}
A motivation for developing the present method is to study the effect of the interstellar radiation field on molecular clouds. Therefore we
wish to test it on the case of a PDR. The problem is so rich in micro-physical processes that no consensus solution exists, even in the
simplified case of a uniform density, plane parallel PDR \citep[see e.g.~the code comparison by ][]{Rollig2007}. Major codes use very
large chemical networks, a large number of frequency bins and track excitation states of species in detail. For 3D hydrodynamical
simulations we need to restrict ourselves to a simple network and a handful of frequency bins. 

We have tested our code using the PDR setup described in \citet{Richings2014I,Richings2014II}. They employ an H-He-C-O network with additional
metals that are important for cooling, the ISRF from \citet{Black1987ASSL}, and a modified version of the $\ch{H2}$ and $\ch{CO}$
self-shielding prescriptions by \citet{Draine1996}. Compared to our network, \citet{Richings2014II} includes additional metal coolants,
but does not include ices. The complete network we are using, including references and explicit formulas for the reaction rates, is detailed
in appendix \ref{sec:chemical-network-test-4}. 
Seven frequency bins are used, with limits in eV (5, 11.26, 13.6, 16.50, 24.6, 31.43, 54.4, 100) corresponding to a region for the dust-dominated UV part, plus the ionization threshold of \ch{C}, \ch{H}, \ch{He}, and \ch{He+}, where 16.50 eV and 31.43 eV are two additional limits added to improve the accuracy (see \sect{photo-chemistry-with-krome}).

The tests fall in two parts. The first part aims to test the network. This part is conducted as stand-alone tests where a simple 1D solver\footnote{https://bitbucket.org/troelsfrostholm/tinypdr (commit 1ba85af)} coupled to {\krome} is used together with the choice of frequency bins, and the self-shielding prescription. The second part aims to test the network, frequency bins, and self-shielding incorporated in {\ramses}, where additional issues can arise
because of the coordinate geometry and the interpolation from rays to cells.

\begin{figure}
\begin{center}
\includegraphics[width=0.48\textwidth]{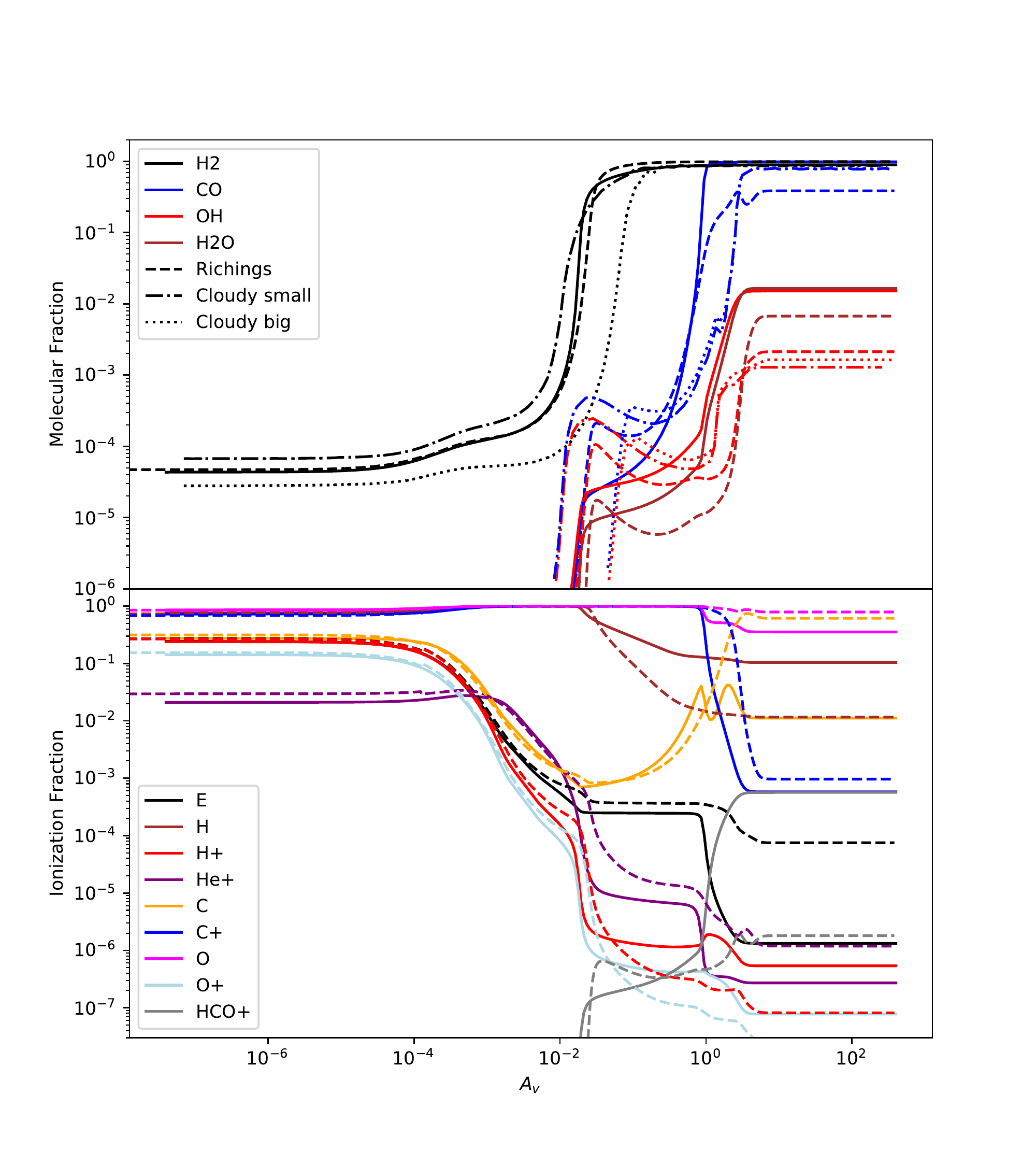}
\caption{Test~4: ionization and molecular fractions of 1D PDR test in a uniform density medium. Dashed lines are from \protect\cite{Richings2014II}
Fig.~2, and include their results compared to a small and large Cloudy PDR model. We use the same formation rate of \ch{H2} on dust grains as Richings, at a constant dust temperature of 10 K. In the top panel are shown
H2: $n_{\ch{H2}} / n_{\ch{H}_\mathrm{tot}}$, CO: $n_{\ch{CO}} / n_{\ch{C}_\mathrm{tot}}$, OH: $n_{\ch{OH}} / n_{\ch{O}_\mathrm{tot}}$ and H2O: $n_{\ch{H2O}} / n_{\ch{O}_\mathrm{tot}}$.
In the lower panel are shown E: $n_{\ch{e-}} / n_{\ch{H}_\mathrm{tot}}$, H: $n_{\ch{H}} / n_{\ch{H}_\mathrm{tot}}$,
H+: $n_{\ch{H+}} / n_{\ch{H}_\mathrm{tot}}$, He+: $n_{\ch{He+}} / n_{\ch{He}_\mathrm{tot}}$
C: $n_{\ch{C}} / n_{\ch{C}_\mathrm{tot}}$, C+: $n_{\ch{C+}} / n_{\ch{C}_\mathrm{tot}}$, O: $n_{\ch{O}} / n_{\ch{O}_\mathrm{tot}}$, O+: $n_{\ch{O+}} / n_{\ch{O}_\mathrm{tot}}$, and 
HCO+: $n_{\ch{HCO+}} / n_{\ch{C}_\mathrm{tot}}$.} 
\label{fig:R14II-fig2}
\end{center}
\end{figure}

The first test attempts to reproduce the results reported in Fig. 2 of \cite{Richings2014II}. The setup is 1D with a uniform gas with fixed density $n_{\ch{H}_{\mathrm{tot}}}=100\, \mathrm{cm}^{-3}$, gas temperature $T_{\mathrm{gas}}=100\, \mathrm{K}$, and dust temperature $T_{\mathrm{d}}=10\, \mathrm{K}$, and incident radiation corresponding to the ISRF from \cite{Black1987ASSL}. The inital composition is solar \citep{Wiersma2009MNRAS}, 
neutral, and atomic, except for hydrogen, for which initially $n_{\ch{H}}=n_{\ch{H2}}=\frac{1}{3}n_{\ch{H}_{\mathrm{tot}}}$, giving initial number densities $n_{\ch{H}}=33.3...\, \mathrm{cm}^{-3}$, $n_{\ch{H2}}=33.3...\, \mathrm{cm}^{-3}$, $n_{\ch{He}}=10\, \mathrm{cm}^{-3}$, $n_{\ch{C}}=2.46 \times 10^{-2}\, \mathrm{cm}^{-3}$, $n_{\ch{O}}=4.90 \times 10^{-2}\, \mathrm{cm}^{-3}$. The grid consists of 256 logarithmically spaced points. We use the chemical network described above, except, to ease comparison, in this test we use the same rate coefficient for \ch{H2} formation on dust grains (reaction 236 in \tab{tab:chemical_network}) as is used in \cite{Richings2014II}. The ionization transition to \ch{H+} and \ch{He+}, and the dissociation transition for \ch{H2} agrees well with Richings' result (\fig{fig:R14II-fig2}). There is roughly a 50\% deviation in the optically thin ionization degree of \ch{He+}, which is due to the coarseness of frequency bins.  The locations of the molecular transitions for \ch{CO} and \ch{OH} resemble those of \cite{Richings2014II}, but their optically thick abundances differ factors of 2 and 7 respectively. This is mainly due to a different electron balance. Our model has almost two orders of magnitude less electrons here because we leave out the heavier metals, which would otherwise donate electrons through cosmic ray ionization. Among other things, this allows for a higher abundance of \ch{H3O+}, which is the main reactant in the production of \ch{OH}, and of \ch{HCO+} which recombines to \ch{CO} and \ch{H}. 

\begin{figure}
\begin{center}
\includegraphics[width=0.48\textwidth]{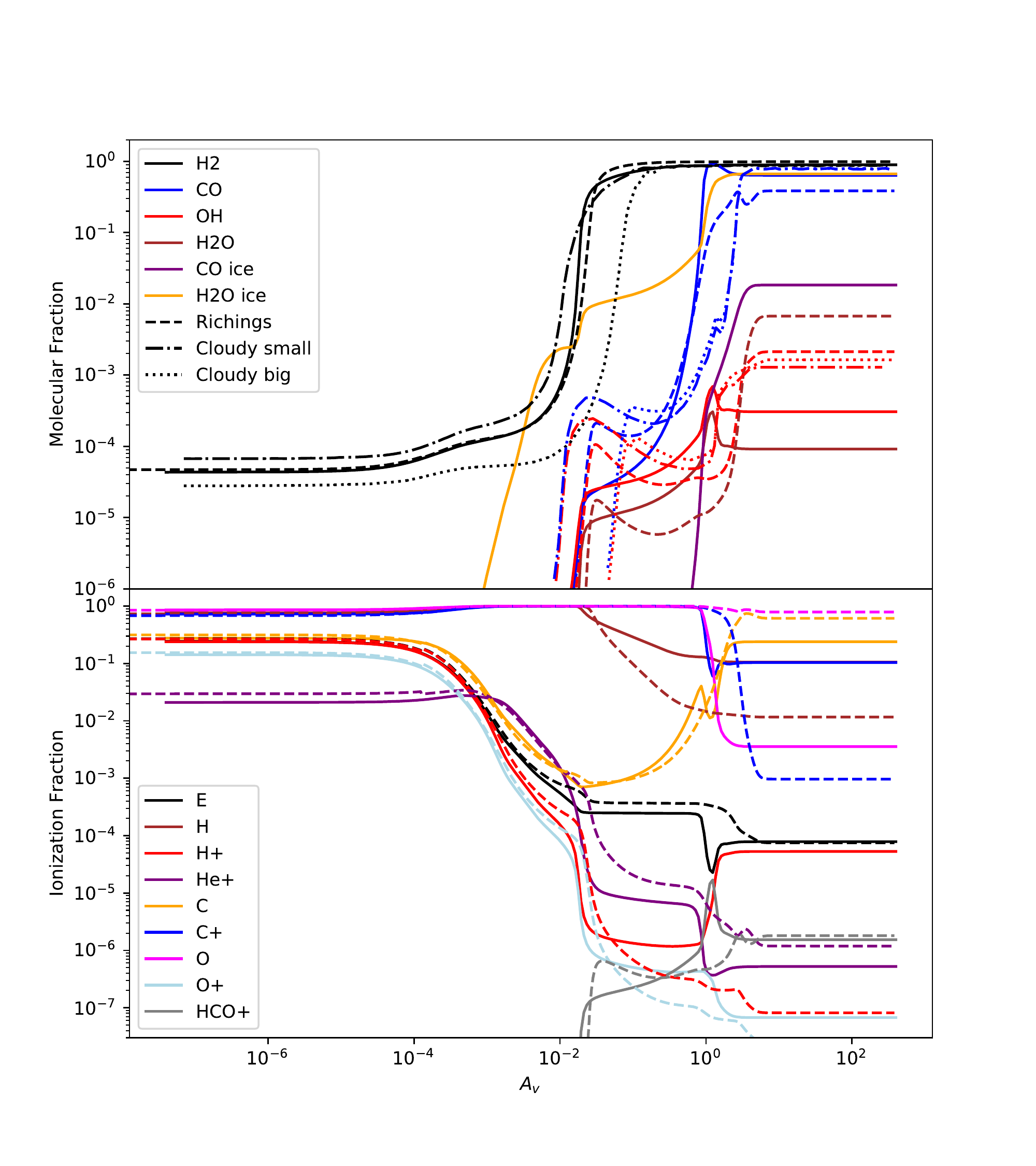}
\caption{Test~4: Same as \fig{fig:R14II-fig2}, except that \ch{H2O} and \ch{CO} ices are included in the network. In the top panel the
additional labels are CO ice: $n_{\ch{CO}_\mathrm{(ice)}}  / n_{\ch{C}_\mathrm{tot}}$ and
H2O ice: $n_{\ch{H2O}_\mathrm{(ice)}} / n_{\ch{O}_\mathrm{tot}}$. }
\label{fig:R14II-fig2-ice}
\end{center}
\end{figure}

In \fig{fig:R14II-fig2-ice} the formation of \ch{H2O} and \ch{CO} ice is also included. In this case, \ch{O} is bound in ice in the optically thick region, so the \ch{OH} abundance is lowered by a factor of around 30. The abundances of \ch{HCO+} and \ch{H3O+} are lowered by a similar amount, resulting in an increase in electron abundance on the same order and lower production of \ch{CO} and \ch{OH}. This result implies that ice formation can significantly affect observationally important molecules like \ch{OH} and \ch{HCO+}. Note that our model does not consider charged dust, which could also affect the electron density. 

\begin{figure}
\begin{center}
\includegraphics[width=0.48\textwidth]{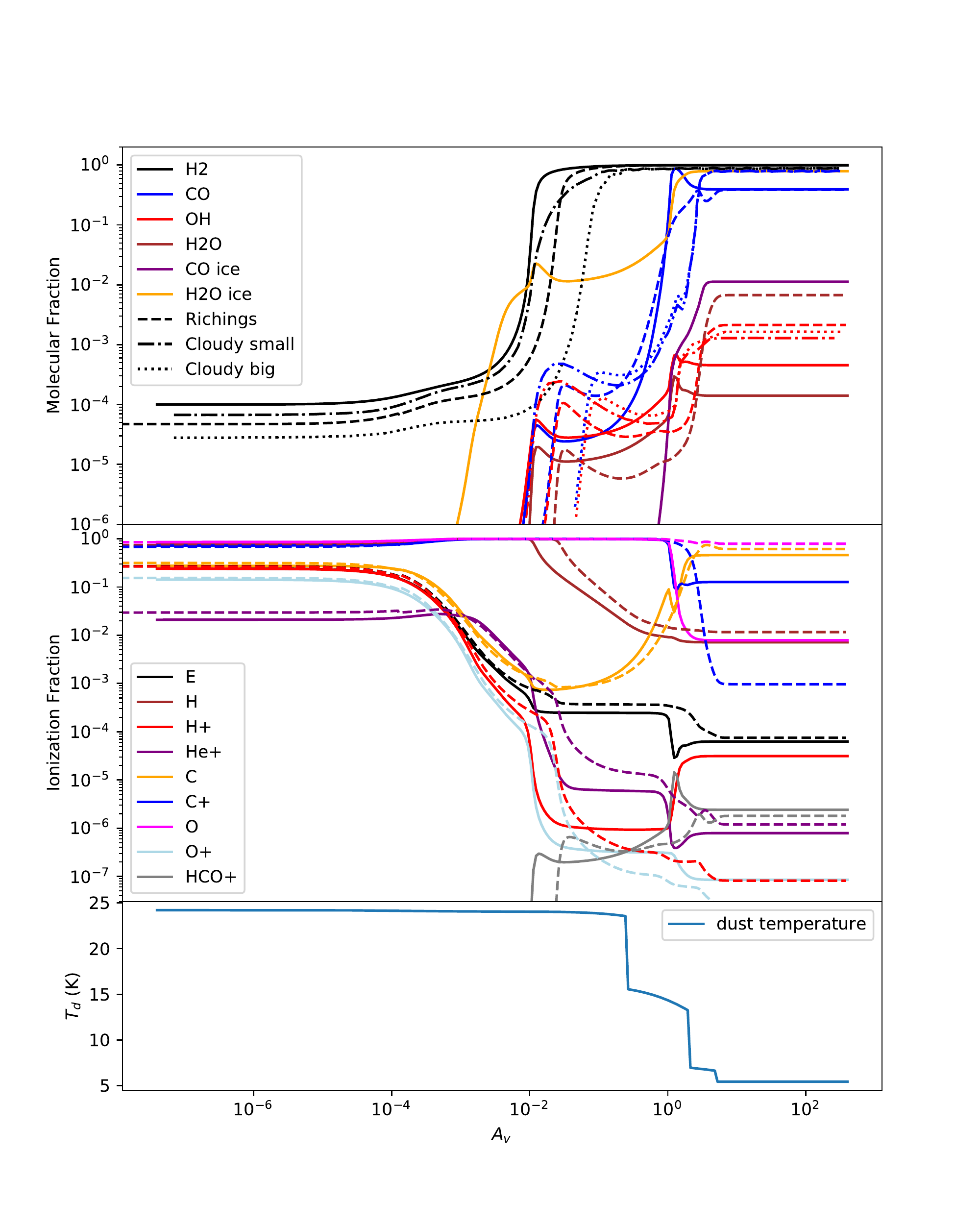}
\caption{Test~4: Same as \fig{fig:R14II-fig2-ice}, except that the dust temperature (bottom panel) and \ch{H2} formation is calculated as described in
\sect{sec:thermal-balance-dust-grains}. }\label{fig:R14II-fig2-ice-dust-table}
\end{center}
\end{figure}

As has been mentioned, the previous two runs were performed with a constant dust temperature of $10$ K, and the same rate of formation of \ch{H2} on dust grains as is used in \cite{Richings2014II}. In \fig{fig:R14II-fig2-ice-dust-table} is shown the result of using a tabulated dust temperature and \ch{H2} formation on dust treated as described in \sect{sec:thermal-balance-dust-grains}. The dust temperature now depends on intensity, and is therefore higher in the optically thin region. This in turn gives a higher \ch{H2} formation rate, which gives a higher \ch{H2} abundance here, and moves the dissociation front of \ch{H2}, as well as the fronts of \ch{CO} and \ch{OH} that are controlled by shielding from \ch{H2}, to lower {\Av}. In the optically thick region, the dust temperature becomes slightly lower, leading to a higher degree of freezeout, mainly visible in \ch{CO}. 

\begin{figure}
\begin{center}
\includegraphics[width=0.48\textwidth]{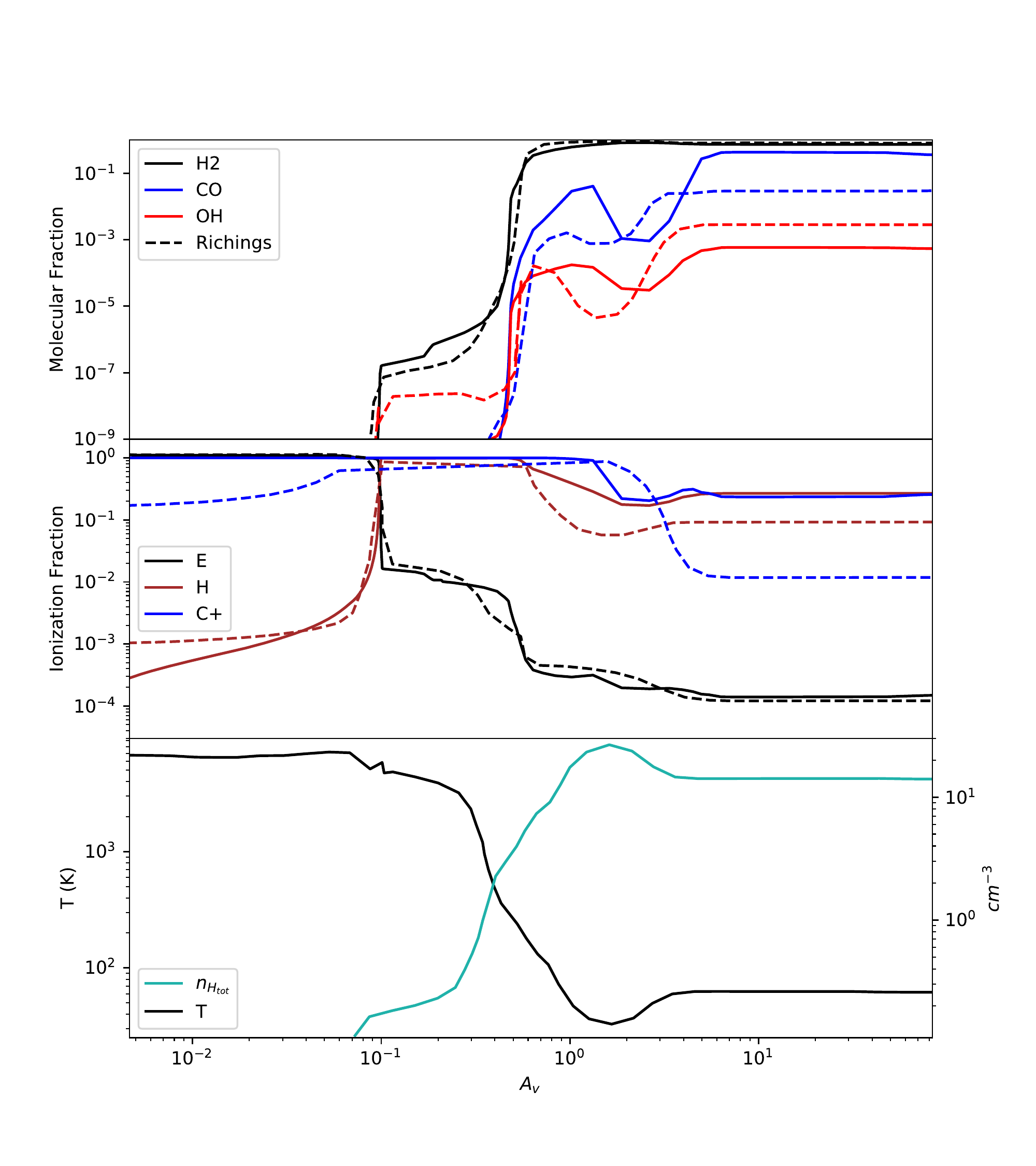}
\caption{Test~4: ionization and molecular fractions of 1D PDR test with uniform pressure, but temperature and density as given in \protect\cite{Richings2014II} Fig.~4. Species labels are as in \fig{fig:R14II-fig2}}\label{fig:R14II-fig4}
\end{center}
\end{figure}

\cite{Richings2014II} also compute a pressure and thermal equilibrium model, given in their Fig. 4. We use the resulting density and temperature profile as input to our model. In this test, to ease comparison, we again use $T_{\mathrm{d}}=10\, \mathrm{K}$ and the same rate coefficient for \ch{H2} formation on dust gains as is used in \cite{Richings2014II}. The results are compared in \fig{fig:R14II-fig4}. The location of the ionization front of \ch{H}, and the dissociation fronts of \ch{H2}, \ch{CO}, and \ch{OH} agree fairly well. However, there are some differences in the optically thick part. As compared with the previous test, the temperature here is lower ($T_{\mathrm{gas}}\approxeq 50\, \mathrm{K}$) as is the density ($n_{\ch{H}_{\mathrm{tot}}} \approxeq 14\, \mathrm{cm}^{-3}$). In contrast to the previous test, the electron density is now in agreement. Our \ch{CO} abundance is roughly a factor of four higher, but our \ch{OH} abundance is now lower, where in the previous test it was higher. 

We now turn to the tests of the implementation in {\ramses}. They are similar to the previous test in that they use constant pressure, with $n T=10^3\, \mathrm{K}\, \mathrm{cm}^{-3}$ as a model for an ISM in pressure equilibrium, except that at high densities a temperature floor is imposed at $T=10\, \textrm{K}$ for the gas temperature, to emulate an isothermal molecular cloud. The number density is 
\begin{equation}
 n\left(r\right)=\begin{cases}
n_{0} & r<r_{0}\\
n_{0}\cdot10^{\beta r} & r_{0}<r<r_{1}\\
n_{1} & r_{1}<r
\end{cases}
\end{equation}
with $n(r_1)=n_1$. We choose $n_0=10^{-2}\, \mathrm{cm}^{-3}$, $n_1=10^3\, \mathrm{cm}^{-3}$, and $\Avone=10$, and derive
\begin{equation}
 \beta=\frac{n_{0}}{N_1\ln\left(10\right)}\left(\frac{n_{1}}{n_{0}}-1\right)=5.36\cdot10^{-2}\, \mathrm{pc}^{-1}
\end{equation}
with $N_1=\frac{1}{4\times10^{-22}} \Avone$ \citep[conversion factor from][]{Krumholz2011ApJ}) and 
\begin{equation}
 r_1=\beta^{-1}\log(\frac{n_1}{n_0})=93.28\, \textrm{pc}.
\end{equation}

\begin{figure}
\begin{center}
\includegraphics[width=0.48\textwidth]{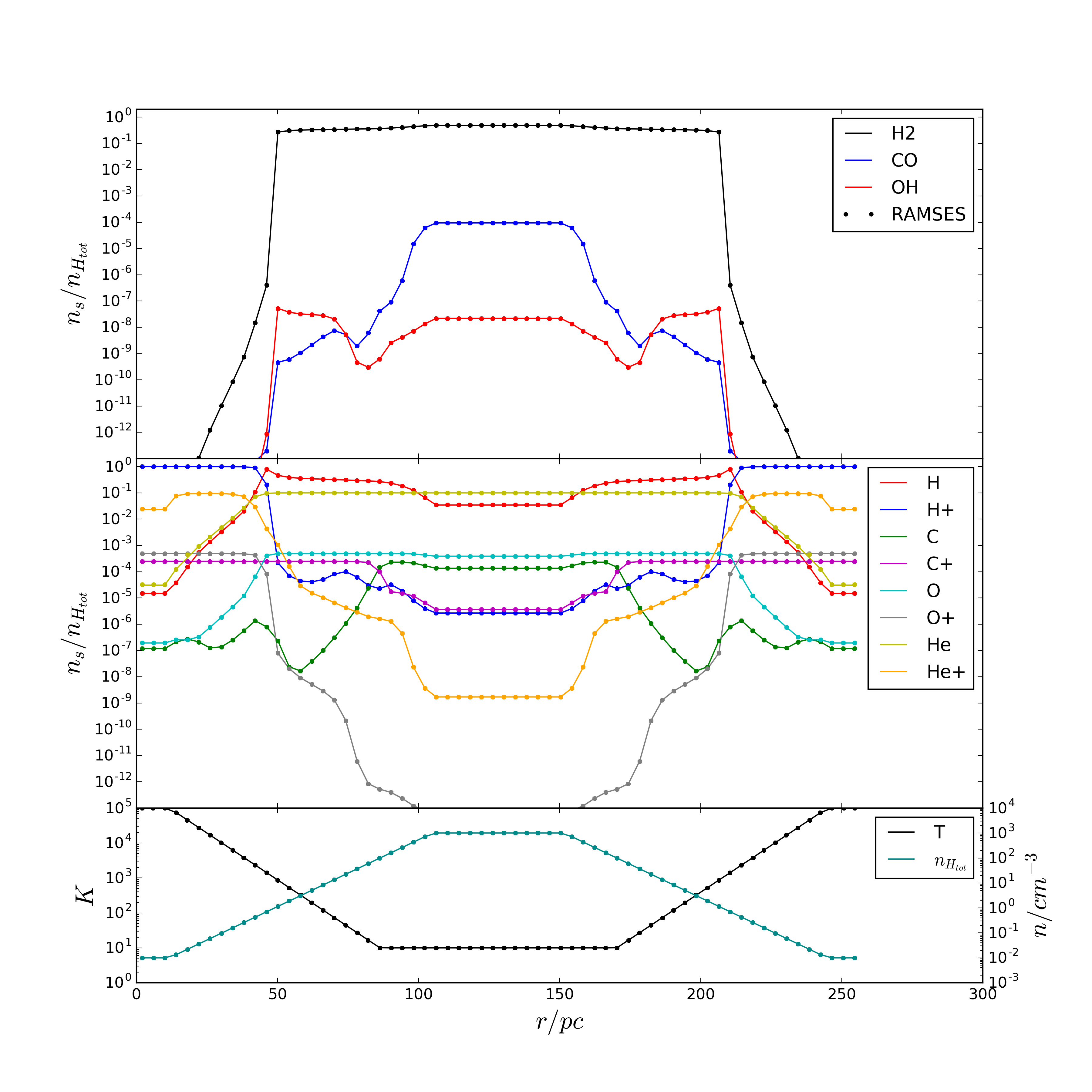}
\caption{Test~4: Comparison of 1D stand-alone code (solid line) and {\ramses} (dots) in an iso-pressure plane-parallel toy PDR model with a power-law density and temperature profile that roughly covers the range found in the ISM. }\label{fig:PDR-slab}
\end{center}
\end{figure}

We set $r_0=\frac{1}{4} r_1$. The region $r<r_0$ initially has \ch{H} ionized. In this test we further use the representation of the ISRF as a diffuse emission as described in \sect{sec:external-isrf-as-diffuse-emission}, with emissivity set such that the total energy flux is consistent with an external ISRF from \cite{Black1987ASSL}, namely to $j_{*,\nu,0}=\frac{J_{\nu,\mathrm{Black}}}{L}$, where $L$ is the side length of the simulation box. The {\ramses} run is in 3D, using plane-parallel slab geometry, a single ray direction, and periodic boundaries on the four orthogonal faces. The result is shown in \fig{fig:PDR-slab}, where the expected agreement between the stand-alone code and the implementation in {\ramses} is seen. 

\subsection{TEST~5: Spherical PDR}\label{sec:PDR-tests-spherical}
\begin{figure}
\begin{center}
\includegraphics[width=0.48\textwidth]{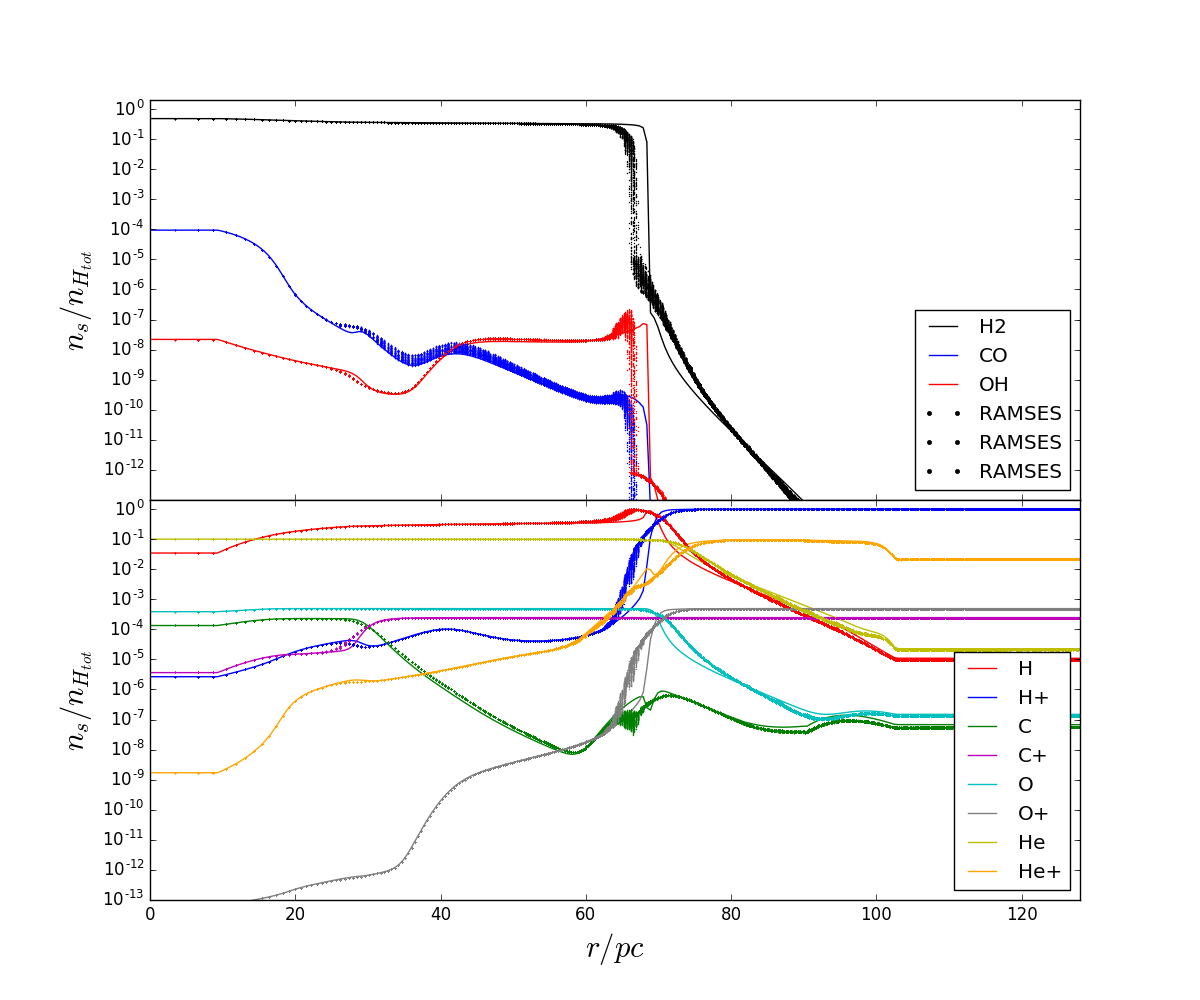}
\caption{Test~5: Comparison of 1D stand-alone code (solid lines) and {\ramses} (dots) in a spherically symmetric cloud with the same density and temperature profile as shown in \fig{fig:PDR-slab}. In the 1D solution, \eqn{eq:attenuation-spherical-cloud-isotropic} is used to account for the clouds spherical shape. }\label{fig:PDR-sphere}
\end{center}
\end{figure}

The final PDR test uses the same density and temperature profile as above, but with $r$ being the radius in a spherically symmetric cloud. 192 {\healpix} directions are used ($N_\mathrm{side}=4$), with outflow boundaries for the radiation on all faces. The emission is set such that we get one Black flux at a distance of $r_0+r_1$. This is equivalent to a total photon rate in the emitting region of 
\begin{equation}
 \dot{N}_{\mathrm{phot}}=\left(4\pi\right)^{2}\left(r_{0}+r_{1}\right)^{2}\int_{0}^{\infty}J_{\nu}^{\mathrm{Black}}/\left(h\nu\right)\mathrm{d}\nu.
\end{equation}
The resolution is $64^3$ cells. Results are in \fig{fig:PDR-sphere}. In the stand-alone reference run, the geometry of the problem must be taken into account. The mean (angle averaged) intensity $J$ due to absorption only, a distance $x$ from the centre of a spherical cloud of radius $r$, assuming an isotropic radiation field outside the cloud of $J_{\mathrm{out}}=1$ is given by
\begin{equation}
J(x)=\frac{1}{2x}\{E_3(r-x)-E_3(r+x)+rE_2(r-x)-rE_2(r+x)\}\label{eq:attenuation-spherical-cloud-isotropic}
\end{equation}
\citep[Eqn.~26]{Hulst1987}, where $x$ and $r$ are in units of optical depth and $E_n(x)=\int_1^{\infty}{\frac{e^{-x t}}{t^n}dt}$ is the $n$-th exponential integral. We use this equation to attenuate an external intensity due to \citet{Black1987ASSL}, where the optical depth is computed from the chemical composition and temperature by {\krome} as usual. For the photo-dissociation rates of \ch{H2} and \ch{CO}, where the prescription for attenuation including self-shielding due to \citet{Richings2014II} is used, we can strictly speaking not use \eqn{eq:attenuation-spherical-cloud-isotropic}, because the self-shielding attenuation does not follow an exponential decay along a given direction, and therefore does not correspond to a single well-defined optical depth. Ignoring this fact, from the shielding factors $f_s^{\ch{H2}}=\mathcal{S}_\mathrm{d}^{\ch{H2}} \mathcal{S}_\mathrm{self}^{\ch{H2}}$ and 
$f_s^{\ch{CO}}=\mathcal{S}_\mathrm{d}^{\ch{CO}} \mathcal{S}_{\mathrm{self},\ch{H2}}^{\ch{CO}}$ along the radial direction  (see \sect{sec:selfshielding}) we compute an equivalent optical depth $\tau^i=-\log{f_s^i}$, which is then transformed back to a shielding factor for the spherical cloud using \eqn{eq:attenuation-spherical-cloud-isotropic}. The deviations seen in \fig{fig:PDR-sphere} can be due to this inconsistency, to the fact that the emission in the 3D case is not perfectly isotropic, as the emitting region is bounded by a finite cube, and/or to differences in resolution. The 1D standalone test is resolved by 256 cells. 

\begin{figure}
\begin{center}
\includegraphics[width=0.48\textwidth]{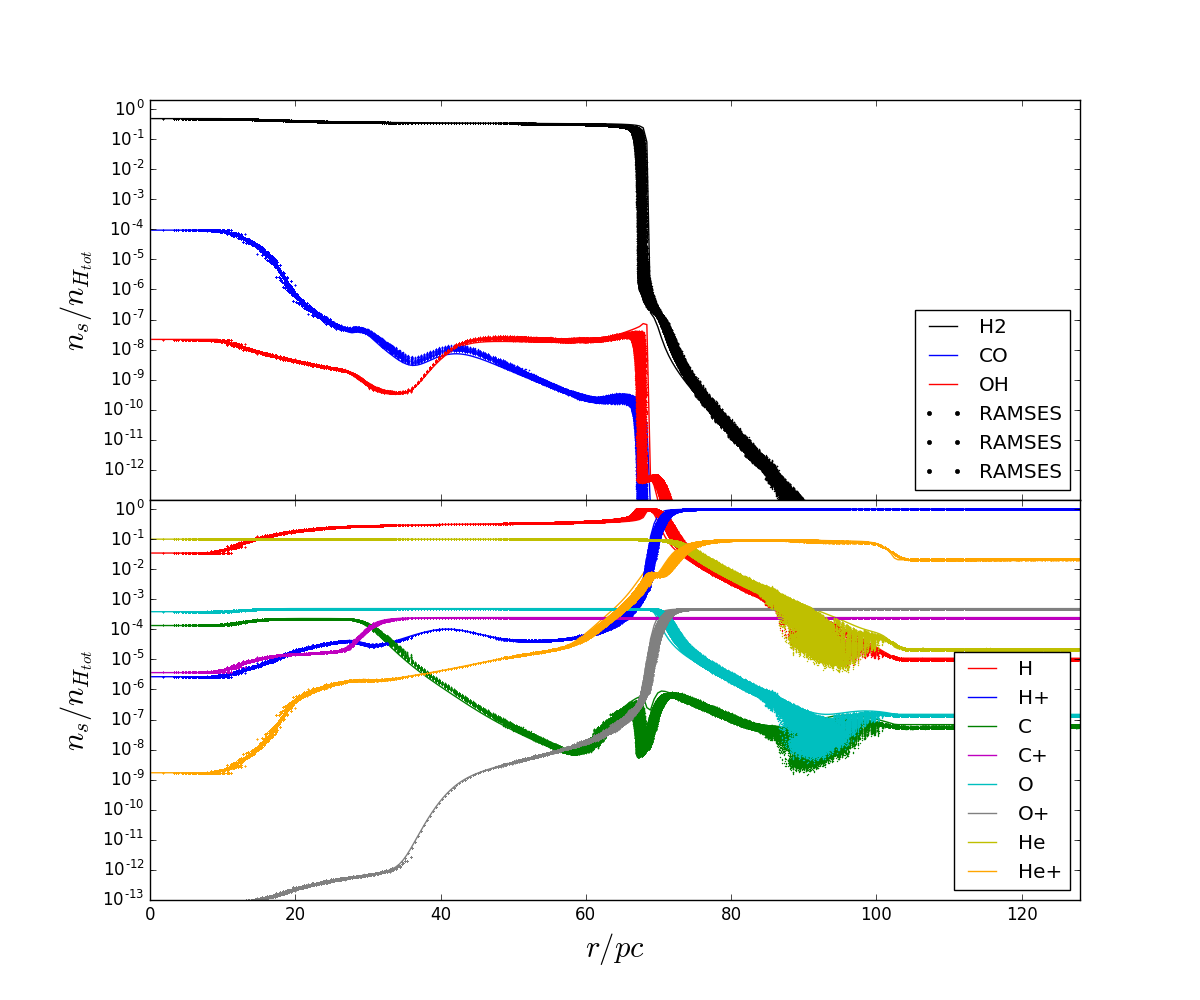}
\caption{Test~5: Same as \fig{fig:PDR-sphere}, but with two extra adaptive mesh levels. Refinement is applied where the relative gradient in \ch{H}, \ch{H+} or \ch{H2} exceeds 0.2, and the abundance of that species is above 1\% by mass.}\label{fig:PDR-sphere-amr}
\end{center}
\end{figure}

\begin{figure}
\begin{center}
\includegraphics[width=0.48\textwidth]{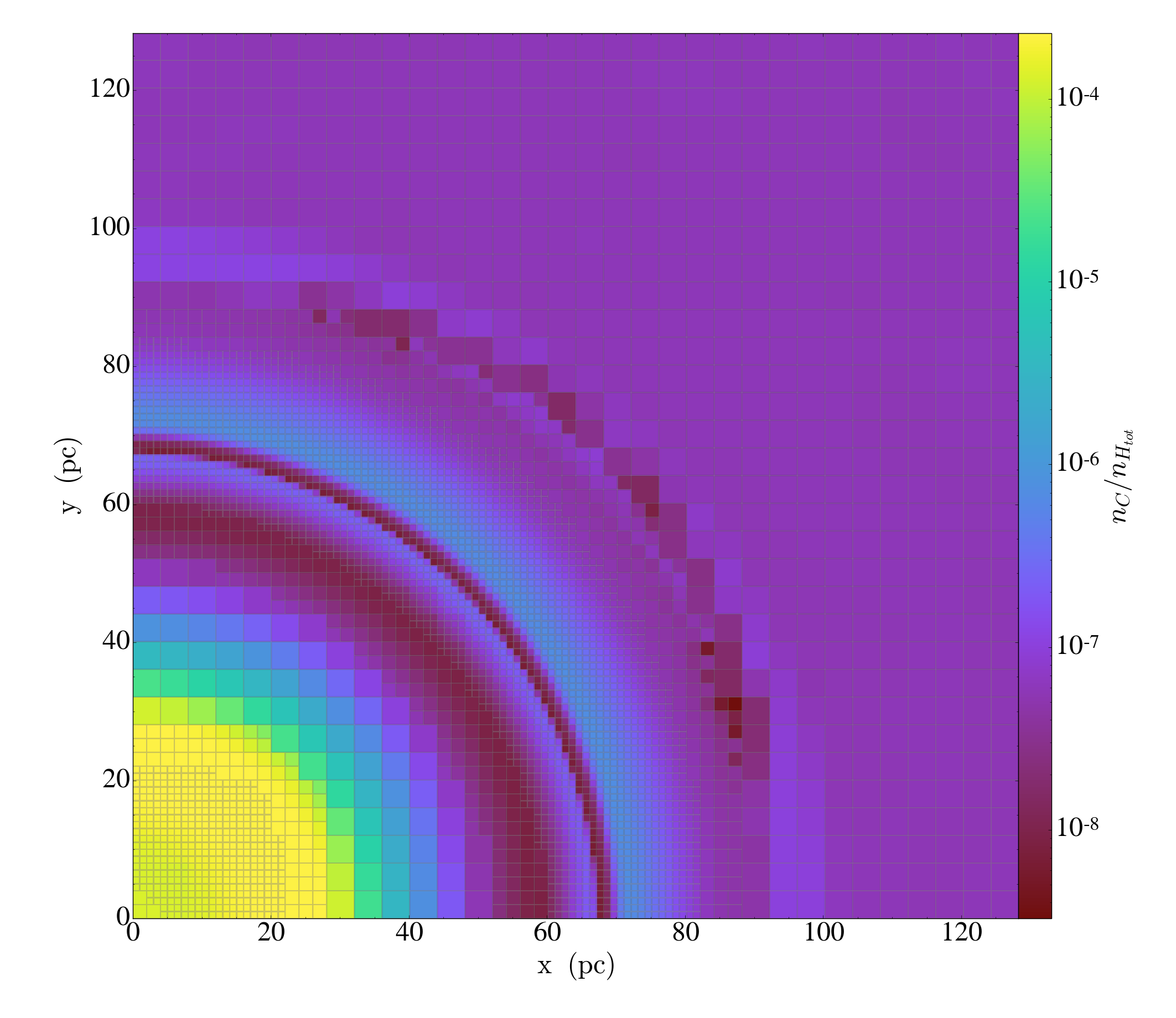}
\caption{Test~5: Slice through the centre of the cloud, showing neutral \ch{C} normalised to total number density of \ch{H}. 
The noise at the edge of the cloud is at a very low abundance, and is a consequence of the specific AMR refinement
criteria.}\label{fig:PDR-sphere-amr-slice}
\end{center}
\end{figure}

The test is repeated with two additional levels of active mesh refinement (\fig{fig:PDR-sphere-amr}). Refinement is activated where the relative gradient in the abundance of \ch{H}, \ch{H+}, or \ch{H2} exceeds 0.2, and the corresponding abundance is above 1\% by mass.  In this case, the shapes of the different transitions are better resolved. In particular, the small bumps around 65 pc in \ch{He+} and \ch{C} are recovered. But the construction and removal of refined mesh following the front as it expands introduces a small amount of asymmetry, which results in the scatter in the plot, with a width of roughly a single coarse cell. As seen in \fig{fig:PDR-sphere-amr-slice}, the actual transition is perfectly smooth, and the scatter is due to an ever so slight non-spherical form of the ionization fronts. The noise around $r=90$ pc is located at the outer refinement boundary, where the abundance of \ch{H} has dropped below 1\%.

\section{Performance and scaling}\label{sec:performance}
\begin{figure}
\begin{center}
\includegraphics[width=0.48\textwidth]{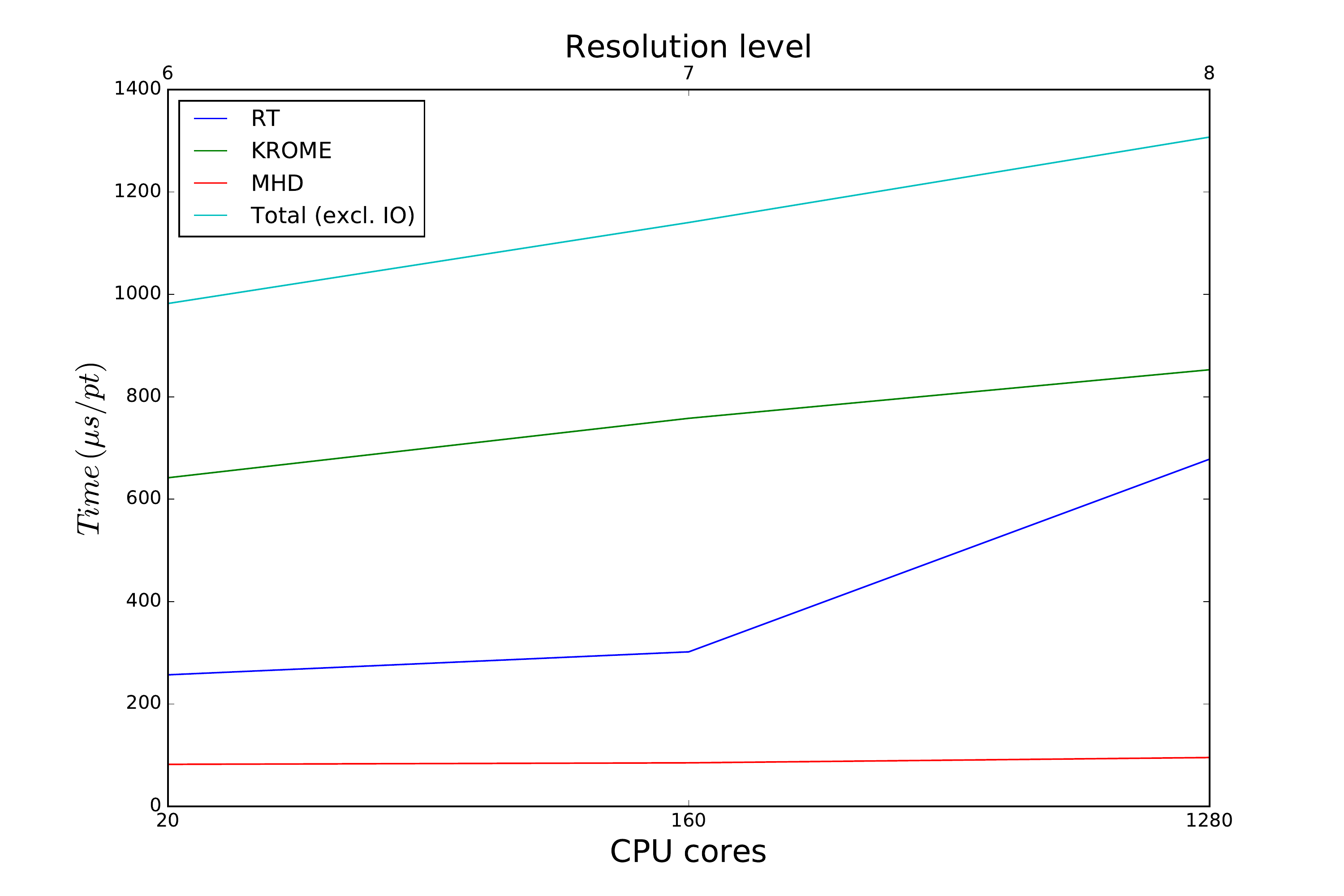}
\caption{Weak scaling (case A) using 5 {\openmp} threads per {\mpi} rank. }\label{fig:scaling-MC-5-threads}
\end{center}
\end{figure}

\begin{figure}
\begin{center}
\includegraphics[width=0.48\textwidth]{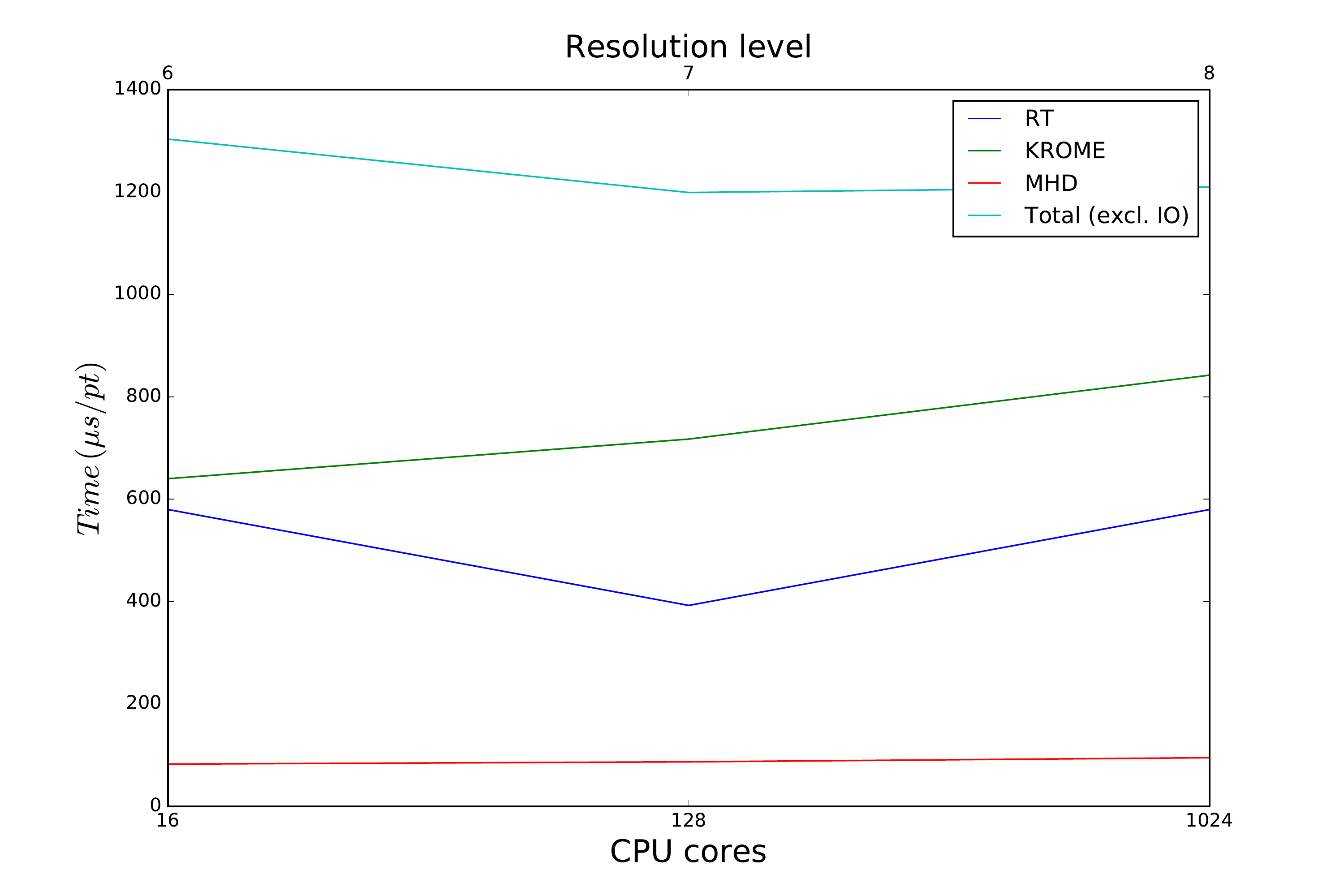}
\caption{Weak scaling (case B) using 8 {\openmp} threads per {\mpi} rank. }\label{fig:scaling-MC-8-threads}
\end{center}
\end{figure}

The weak-scaling properties of the photo-chemistry solver with {\krome} are measured on the spherical cloud setup of Test~5, except using 48 {\healpix} directions instead of 192. The test is performed on the local HPC cluster at University of Copenhagen using nodes with 2x 10-core Xeon E5-2680v2 running at 2.8~GHz with 64 GB DDR3-1866 MHz RAM per node interconnected with FDR infiniband. Two sets of runs are performed -- one with 5 {\openmp} threads per {\mpi} rank (A) and one with 8 threads per rank (B). In both cases we exploit hyper threading and run 2 threads per core. . Resolution and computational resources are given in table \ref{table:scaling-resolution-and-resources}. The results are given in \fig{fig:scaling-MC-5-threads} and \fig{fig:scaling-MC-8-threads}, where running time is in core $\mu s$ per cell update averaged over 5 time steps. 

\begin{table}
\begin{tabular}{|c|c|c|c|c|}
\hline 
Case & level & cells & {\mpi} ranks & {\openmp} threads\tabularnewline
\hline 
\hline 
A & 6 & $64^{3}$ & 8 & 5\tabularnewline
\hline 
 & 7 & $128^{3}$ & 64 & 5\tabularnewline
\hline 
 & 8 & $256^{3}$ & 512 & 5\tabularnewline
\hline 
\hline 
B & 6 & $64^{3}$ & 4 & 8\tabularnewline
\hline 
 & 7 & $128^{3}$ & 32 & 8\tabularnewline
\hline 
 & 8 & $256^{3}$ & 256 & 8\tabularnewline
\hline 
\end{tabular}

\caption{Resolution and resources for the weak scaling test.}\label{table:scaling-resolution-and-resources}
\end{table}

\begin{figure}
\begin{center}
\includegraphics[width=0.48\textwidth]{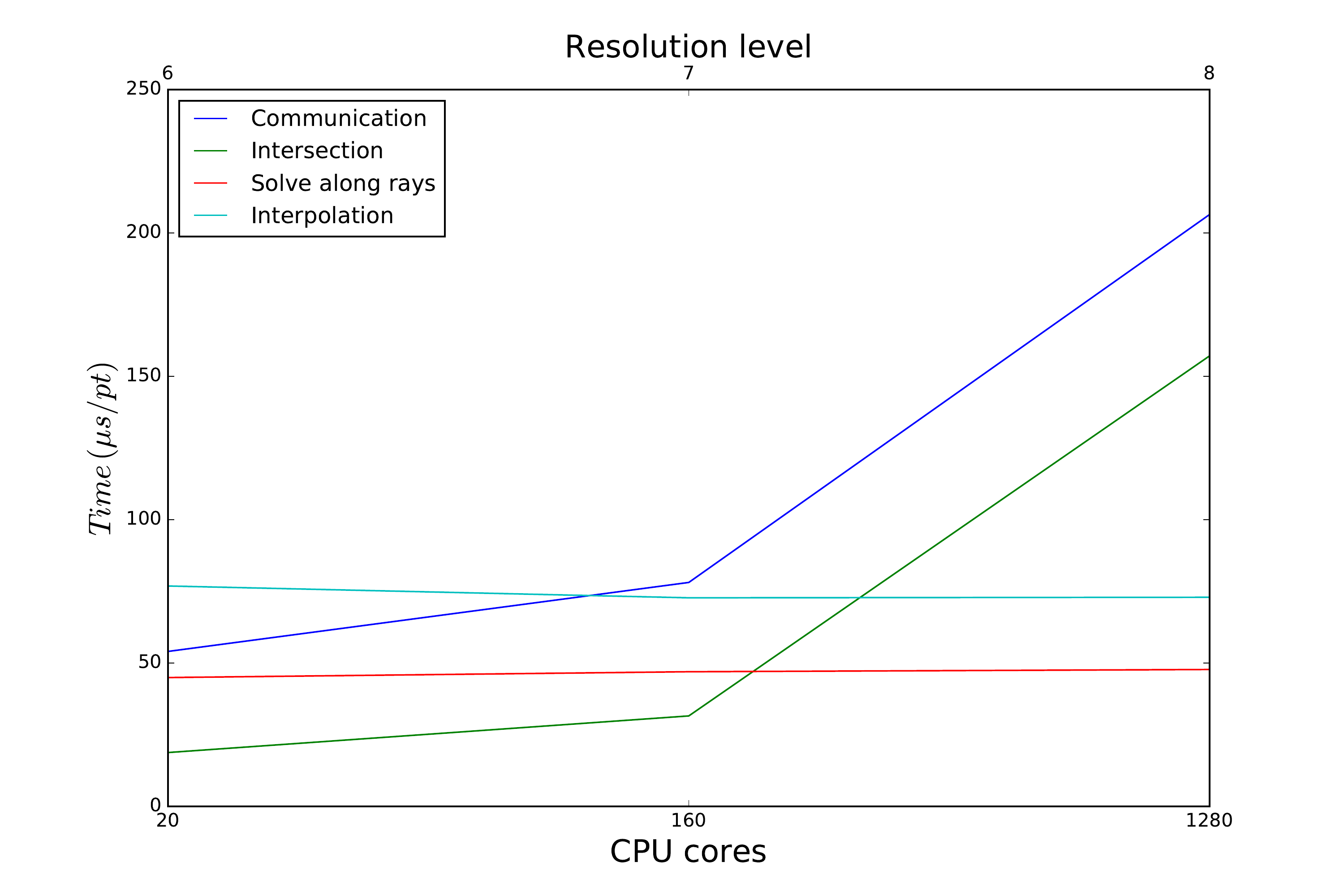}
\caption{Weak scaling (case A) -- timing of radiative transfer steps. }\label{fig:scaling-MC-5-threads-lampray}
\end{center}
\end{figure}

While the rest of {\ramses} used in Copenhagen and the coupling to {\krome} is {\openmp} parallelised, this is not yet the case for the radiative transfer. It therefore naturally runs longer per core in case B where more {\openmp} threads are available. In case B with twice as many cells per {\mpi} rank, and which goes up to half the number of ranks as case A, the running time is roughly constant. At $256^3$ cells on 512 ranks, the time in RT doubles. As can be seen from \fig{fig:scaling-MC-5-threads-lampray}, it is because the cost of communication in RT more than doubles at this point, and becomes dominant. 

\begin{figure}
\begin{center}
\includegraphics[width=0.48\textwidth]{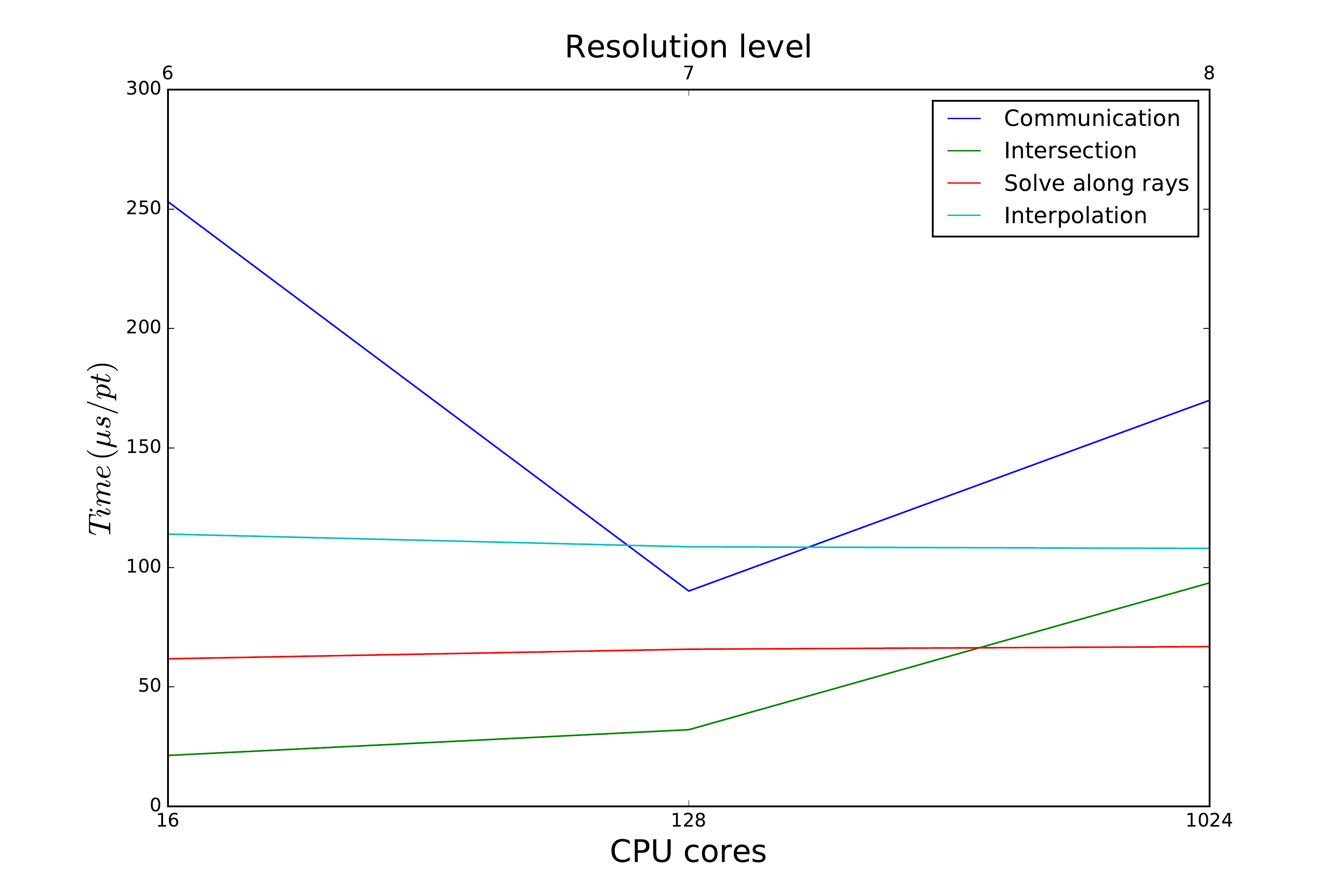}
\caption{Weak scaling (case B) -- timing of radiative transfer steps. }\label{fig:scaling-MC-8-threads-lampray}
\end{center}
\end{figure}

\section{Summary and future work}\label{sec:summary}

We have presented a new method we call {\lampray}, and its implementation into the {\ramses} code, for solving the time-independent radiative transfer problem on an adaptive octree mesh by means of tracing rays, that each cover the entire length of the domain. The diffuse radiation field is represented by rays that uniformly cover leaf cells at each refinement level for each of a set of directions chosen by the {\healpix} scheme. Because rays in highly refined cells extend into coarser cells, the coarse cells will be covered by several rays in a given direction. The radiation field due to point sources is represented by a fixed set of rays that intersects the source, also with directions chosen by the {\healpix} scheme. 

The {\lampray} code is parallelised for computer clusters with {\mpi} by changing the domain decomposition of the already {\mpi} parallel {\ramses} code in such a way that points along any given ray is placed optimally in the same memory space. The method is shown to scale to at least 1280 CPUs in a weak scaling test. It is shown to perform well enough to support radiation hydrodynamics with photo-ionization and -dissociation chemistry using a simplified H-He-C-O chemical network, spending between 1000 and 1300 core $\mu$s per point. 

The radiation field is either coupled to hydrogen ionization chemistry using an implicit, photon-conserving method similar to what is used in {\ctoray}, or to arbitrary photo-ionization and -dissociation chemistry through a time-explicit method using the non-equilibrium chemistry code {\krome}. The former has previously participated in the STARBENCH code comparison on the D-type expansion of an HII region \citep{Bisbas2015MNRAS}. The latter has been tested here in the following circumstances. 

Using a minimal hydrogen ionization chemistry network, most of the benchmark tests in \citet{Iliev2006,Iliev2009} have been reproduced, with good results. These include 1.1) the expansion of an HII region in static, isothermal gas, 1.2) the previous, except with heating and cooling included, 1.3) the previous, except with hydrodynamics included, 2) the trapping of an ionization front by a dense clump in a static gas, 3) and with active hydrodynamics, letting the clump evaporate. Because the method is time-explicit, the early fast propagation of the ionization front is not well resolved, and there is some deviation in ionization front position and velocity in the earliest part of the evolution in all the tests. There is generally some disagreement among the benchmarked codes regarding the level of shielding and the temperature structure. {\lampray} tends to agree with those that predict a relatively low level of shielding, and high temperatures, namely \textsc{flash-hc}, {\rsph}, and \textsc{coral}. Test~3 has been repeated with active mesh refinement, producing almost identical results. 

A new model for the H-He-C-O chemistry in photon dominated regions, for use in 3D hydrodynamic simulations, has been presented. It is in part based on the network presented in \citet{Glover2010MNRAS}, but updated with newer reaction rates and broader temperature coverage, a better treatment of dust, explicit photo chemistry including cross sections, and inclusion of \ch{H2O} and \ch{CO} ices. It is compared with the model in \citet{Richings2014I,Richings2014II}. It includes 240 reactions, among which are photo-ionization, photo-dissociation, recombinations on dust grains, and formation of \ch{CO} and \ch{H2O} ices. It reproduces the ionization transitions for \ch{H}, \ch{He}, and \ch{C}, and dissociation transition for \ch{H2}, while for \ch{CO} and \ch{OH}, only the transition location is reproduced, while the shape differs. Electron and \ch{OH} density in the optically thick part differs due to the omission of metals that would donate electrons if included. 

A new method for including an interstellar radiation field in periodic boundary gas simulations has been presented, in which the interstellar field is modelled as an emissivity in the diffuse ionized medium. Using this, the above chemical model has been tested on a 3D hydrostatic model of a spherical molecular cloud. This results in stable, spherically symmetric ionization and molecular transitions that agree well with a corresponding 1D simulation presented here. The test has been repeated with active mesh refinement, which allows certain transition features to be better resolved, but also introduces a small degree of asymmetry into the solution, that slightly deforms the spherical transitions. 

We envision a number of future improvements. As mentioned, coarse cells in the adaptive mesh are covered by more rays than necessary. Alleviating this problem would require fundamental changes to the method. One solution for the diffuse field would be to use the fact that one can decompose the solution to the radiative transfer equation in a set of regions into an intrinsic solution in each region, and a solution at the region boundaries \citep{Heinemann2006A&A}. This could be exploited to find the intrinsic solution one level of refinement at a time, in the regions that are at that level. These solutions would then finally be recomposed in the ray domain, using interpolation at refinement boundaries. The downside of this method is that the solution for all ray directions, at least at the refinement boundaries, must be held in memory at once, multiplying memory consumption roughly by the number of directions. It also requires an additional communication of the intensity at refinement boundaries. For point sources, a common technique is to split rays as the distance to the source increases \citep{Abel2002MNRAS}. 

As described, photo-ionization chemistry using the coupling with {\krome} is time-explicit, and therefore not well suited for tracking fast ionization fronts. By modifying the ODE solver in {\krome} to also compute time-averaged ionization fractions over the time step, an implicit method like that of {\ctoray} could be implemented.

In {\ramses}, stars may be represented as accreting point particles. The accretion history has previously been used as input to the stellar structure model {\mesa} \citep{Jensen2018MNRAS,Kuffmeier2018MNRAS}, showing a significant influence on the luminosity of young stellar objects. In the future, we plan to use the stellar properties of such models to determine the spectrum of point sources, by evolving {\mesa} in parallel with {\ramses}. 

\section*{Acknowledgements}
We thank Remo Collet and \AA ke Nordlund for useful discussions and for their contributions to the initial version of the radiative transfer framework
in the version of {\ramses} used at the Centre for Star and Planet Formation. We thank Alexander James Richings for kindly sharing his simulation data, to which we compare our model in the present paper. 
We thank the developers of the python-based analysing tool YT \url{http://yt-project.org} \citep{2011ApJS..192....9T} that has been used to produce
some of the figures.
The research leading to these results has received funding from the Danish Council for Independent Research through a Sapere Aude Starting Grant
to TH. The Centre for Star and Planet Formation is financed by the Danish National Research Foundation (DNRF97).
We acknowledge PRACE for awarding us access to the computing resource MARCONI based in Italy at CINECA for carrying out part of the
development and tests.
Computing nodes at the University of Copenhagen HPC centre, funded with a research grant (VKR023406) from Villum Fonden, were used for
carrying out part of the simulations and the post-processing.

\bibliographystyle{mnras}
\bibliography{references.bib,network.bib}

\begin{thebibliography}{}
\makeatletter
\relax
\def\mn@urlcharsother{\let\do\@makeother \do\$\do\&\do\#\do\^\do\_\do\%\do\~}
\def\mn@doi{\begingroup\mn@urlcharsother \@ifnextchar [ {\mn@doi@}
  {\mn@doi@[]}}
\def\mn@doi@[#1]#2{\def\@tempa{#1}\ifx\@tempa\@empty \href
  {http://dx.doi.org/#2} {doi:#2}\else \href {http://dx.doi.org/#2} {#1}\fi
  \endgroup}
\def\mn@eprint#1#2{\mn@eprint@#1:#2::\@nil}
\def\mn@eprint@arXiv#1{\href {http://arxiv.org/abs/#1} {{\tt arXiv:#1}}}
\def\mn@eprint@dblp#1{\href {http://dblp.uni-trier.de/rec/bibtex/#1.xml}
  {dblp:#1}}
\def\mn@eprint@#1:#2:#3:#4\@nil{\def\@tempa {#1}\def\@tempb {#2}\def\@tempc
  {#3}\ifx \@tempc \@empty \let \@tempc \@tempb \let \@tempb \@tempa \fi \ifx
  \@tempb \@empty \def\@tempb {arXiv}\fi \@ifundefined
  {mn@eprint@\@tempb}{\@tempb:\@tempc}{\expandafter \expandafter \csname
  mn@eprint@\@tempb\endcsname \expandafter{\@tempc}}}

\bibitem[\protect\citeauthoryear{{Abel} \& {Wandelt}}{{Abel} \&
  {Wandelt}}{2002}]{Abel2002MNRAS}
{Abel} T.,  {Wandelt} B.~D.,  2002, \mn@doi [\mnras]
  {10.1046/j.1365-8711.2002.05206.x}, \href
  {http://adsabs.harvard.edu/abs/2002MNRAS.330L..53A} {330, L53}

\bibitem[\protect\citeauthoryear{{Abel}, {Anninos}, {Zhang}  \&
  {Norman}}{{Abel} et~al.}{1997}]{Abel1997NewAstronomy}
{Abel} T.,  {Anninos} P.,  {Zhang} Y.,   {Norman} M.~L.,  1997, \mn@doi [\na]
  {10.1016/S1384-1076(97)00010-9}, \href
  {http://adsabs.harvard.edu/abs/1997NewA....2..181A} {2, 181}

\bibitem[\protect\citeauthoryear{{Abel}, {Norman}  \& {Madau}}{{Abel}
  et~al.}{1999}]{Abel1999ApJ}
{Abel} T.,  {Norman} M.~L.,   {Madau} P.,  1999, \mn@doi [\apj]
  {10.1086/307739}, \href {http://adsabs.harvard.edu/abs/1999ApJ...523...66A}
  {523, 66}

\bibitem[\protect\citeauthoryear{{Adams} \& {Smith}}{{Adams} \&
  {Smith}}{1977}]{1977CPL....47..383A}
{Adams} N.~G.,  {Smith} D.,  1977, \mn@doi [Chemical Physics Letters]
  {10.1016/0009-2614(77)80043-2}, \href
  {http://adsabs.harvard.edu/abs/1977CPL....47..383A} {47, 383}

\bibitem[\protect\citeauthoryear{{Adams}, {Smith}  \& {Millar}}{{Adams}
  et~al.}{1984}]{1984MNRAS.211..857A}
{Adams} N.~G.,  {Smith} D.,   {Millar} T.~J.,  1984, \mn@doi [\mnras]
  {10.1093/mnras/211.4.857}, \href
  {http://adsabs.harvard.edu/abs/1984MNRAS.211..857A} {211, 857}

\bibitem[\protect\citeauthoryear{{Aldrovandi} \& {Pequignot}}{{Aldrovandi} \&
  {Pequignot}}{1973}]{Aldrovandi1973A&A}
{Aldrovandi} S.~M.~V.,  {Pequignot} D.,  1973, \aap, \href
  {http://adsabs.harvard.edu/abs/1973A%26A....25..137A} {25, 137}

\bibitem[\protect\citeauthoryear{{Alge}, {Adams}  \& {Smith}}{{Alge}
  et~al.}{1983}]{1983JPhB...16.1433A}
{Alge} E.,  {Adams} N.~G.,   {Smith} D.,  1983, \mn@doi [Journal of Physics B
  Atomic Molecular Physics] {10.1088/0022-3700/16/8/017}, \href
  {http://adsabs.harvard.edu/abs/1983JPhB...16.1433A} {16, 1433}

\bibitem[\protect\citeauthoryear{{Andreazza} \& {Singh}}{{Andreazza} \&
  {Singh}}{1997}]{1997MNRAS.287..287A}
{Andreazza} C.~M.,  {Singh} P.~D.,  1997, \mn@doi [\mnras]
  {10.1093/mnras/287.2.287}, \href
  {http://adsabs.harvard.edu/abs/1997MNRAS.287..287A} {287, 287}

\bibitem[\protect\citeauthoryear{Anicich}{Anicich}{2003}]{Anicich2003}
Anicich V.~G.,  2003, JPL Publication 03-19, 1-1194

\bibitem[\protect\citeauthoryear{{Aubert} \& {Teyssier}}{{Aubert} \&
  {Teyssier}}{2008}]{Aubert2008MNRAS}
{Aubert} D.,  {Teyssier} R.,  2008, \mn@doi [\mnras]
  {10.1111/j.1365-2966.2008.13223.x}, \href
  {http://adsabs.harvard.edu/abs/2008MNRAS.387..295A} {387, 295}

\bibitem[\protect\citeauthoryear{Azatyan, Aleksandrov  \& Troshin}{Azatyan
  et~al.}{}]{osti_4089545}
Azatyan V.,  Aleksandrov E.,   Troshin A., , Kinet. Catal. (USSR) (Engl.
  Transl.), v. 16, no. 2, pp. 261-265

\bibitem[\protect\citeauthoryear{{Barinovs} \& {van Hemert}}{{Barinovs} \& {van
  Hemert}}{2006}]{2006ApJ...636..923B}
{Barinovs} {\u G}.,  {van Hemert} M.~C.,  2006, \mn@doi [\apj]
  {10.1086/498080}, \href {http://adsabs.harvard.edu/abs/2006ApJ...636..923B}
  {636, 923}

\bibitem[\protect\citeauthoryear{{Barreira}, {Llinares}, {Bose}  \&
  {Li}}{{Barreira} et~al.}{2016}]{Barreira2016JCAP}
{Barreira} A.,  {Llinares} C.,  {Bose} S.,   {Li} B.,  2016, \mn@doi [\jcap]
  {10.1088/1475-7516/2016/05/001}, \href
  {http://adsabs.harvard.edu/abs/2016JCAP...05..001B} {5, 001}

\bibitem[\protect\citeauthoryear{{Baulch} et~al.,}{{Baulch}
  et~al.}{1992}]{Baulch1992JPCRD}
{Baulch} D.~L.,  et~al., 1992, \mn@doi [Journal of Physical and Chemical
  Reference Data] {10.1063/1.555908}, \href
  {http://adsabs.harvard.edu/abs/1992JPCRD..21..411B} {21, 411}

\bibitem[\protect\citeauthoryear{{Baulch} et~al.,}{{Baulch}
  et~al.}{2005}]{Baulch2005JPCRD}
{Baulch} D.~L.,  et~al., 2005, \mn@doi [Journal of Physical and Chemical
  Reference Data] {10.1063/1.1748524}, \href
  {http://adsabs.harvard.edu/abs/2005JPCRD..34..757B} {34, 757}

\bibitem[\protect\citeauthoryear{{Bisbas} et~al.,}{{Bisbas}
  et~al.}{2015}]{Bisbas2015MNRAS}
{Bisbas} T.~G.,  et~al., 2015, \mn@doi [\mnras] {10.1093/mnras/stv1659}, \href
  {http://adsabs.harvard.edu/abs/2015MNRAS.453.1324B} {453, 1324}

\bibitem[\protect\citeauthoryear{{Black}}{{Black}}{1987}]{Black1987ASSL}
{Black} J.~H.,  1987, in {Hollenbach} D.~J.,  {Thronson} Jr. H.~A.,  eds,
  Astrophysics and Space Science Library Vol. 134, Interstellar Processes. pp
  731--744, \mn@doi{10.1007/978-94-009-3861-8_27}

\bibitem[\protect\citeauthoryear{{Buntemeyer}, {Banerjee}, {Peters}, {Klassen}
  \& {Pudritz}}{{Buntemeyer} et~al.}{2016}]{Buntemeyer2016NewA}
{Buntemeyer} L.,  {Banerjee} R.,  {Peters} T.,  {Klassen} M.,   {Pudritz}
  R.~E.,  2016, \mn@doi [\na] {10.1016/j.newast.2015.07.002}, \href
  {http://adsabs.harvard.edu/abs/2016NewA...43...49B} {43, 49}

\bibitem[\protect\citeauthoryear{{Butler}, {Tan}, {Teyssier}, {Rosdahl}, {Van
  Loo}  \& {Nickerson}}{{Butler} et~al.}{2017}]{Butler2017}
{Butler} M.~J.,  {Tan} J.~C.,  {Teyssier} R.,  {Rosdahl} J.,  {Van Loo} S.,
  {Nickerson} S.,  2017, \mn@doi [\apj] {10.3847/1538-4357/aa7054}, \href
  {http://adsabs.harvard.edu/abs/2017ApJ...841...82B} {841, 82}

\bibitem[\protect\citeauthoryear{{Cazaux} \& {Spaans}}{{Cazaux} \&
  {Spaans}}{2009}]{Cazaux2009A&A}
{Cazaux} S.,  {Spaans} M.,  2009, \mn@doi [\aap] {10.1051/0004-6361:200811302},
  \href {http://adsabs.harvard.edu/abs/2009A%26A...496..365C} {496, 365}

\bibitem[\protect\citeauthoryear{{Cen}}{{Cen}}{1992a}]{Cen1992}
{Cen} R.,  1992a, \mn@doi [\apjs] {10.1086/191630}, \href
  {http://adsabs.harvard.edu/abs/1992ApJS...78..341C} {78, 341}

\bibitem[\protect\citeauthoryear{{Cen}}{{Cen}}{1992b}]{Cen1992ApJS}
{Cen} R.,  1992b, \mn@doi [\apjs] {10.1086/191630}, \href
  {http://adsabs.harvard.edu/abs/1992ApJS...78..341C} {78, 341}

\bibitem[\protect\citeauthoryear{{Clark}, {Glover}  \& {Klessen}}{{Clark}
  et~al.}{2012}]{Clark2012MNRAS}
{Clark} P.~C.,  {Glover} S.~C.~O.,   {Klessen} R.~S.,  2012, \mn@doi [\mnras]
  {10.1111/j.1365-2966.2011.20087.x}, \href
  {http://adsabs.harvard.edu/abs/2012MNRAS.420..745C} {420, 745}

\bibitem[\protect\citeauthoryear{Cohen \& Westberg}{Cohen \&
  Westberg}{1979}]{cohen1979evaluation}
Cohen N.,  Westberg K.,  1979, Journal of Physical Chemistry, 83, 46

\bibitem[\protect\citeauthoryear{{Commer{\c c}on}, {Teyssier}, {Audit},
  {Hennebelle}  \& {Chabrier}}{{Commer{\c c}on}
  et~al.}{2011}]{Commercon2011A&A}
{Commer{\c c}on} B.,  {Teyssier} R.,  {Audit} E.,  {Hennebelle} P.,
  {Chabrier} G.,  2011, \mn@doi [\aap] {10.1051/0004-6361/201015880}, \href
  {http://adsabs.harvard.edu/abs/2011A%26A...529A..35C} {529, A35}

\bibitem[\protect\citeauthoryear{{Commer{\c c}on}, {Debout}  \&
  {Teyssier}}{{Commer{\c c}on} et~al.}{2014}]{Commercon2014}
{Commer{\c c}on} B.,  {Debout} V.,   {Teyssier} R.,  2014, \mn@doi [\aap]
  {10.1051/0004-6361/201322858}, \href
  {http://adsabs.harvard.edu/abs/2014A%26A...563A..11C} {563, A11}

\bibitem[\protect\citeauthoryear{{Coppola}, {Longo}, {Capitelli}, {Palla}  \&
  {Galli}}{{Coppola} et~al.}{2011}]{Coppola2011ApJS}
{Coppola} C.~M.,  {Longo} S.,  {Capitelli} M.,  {Palla} F.,   {Galli} D.,
  2011, \mn@doi [\apjs] {10.1088/0067-0049/193/1/7}, \href
  {http://adsabs.harvard.edu/abs/2011ApJS..193....7C} {193, 7}

\bibitem[\protect\citeauthoryear{{Corrigan}}{{Corrigan}}{1965}]{Corrigan1965JChPh}
{Corrigan} S.~J.~B.,  1965, \mn@doi [\jcp] {10.1063/1.1696701}, \href
  {http://adsabs.harvard.edu/abs/1965JChPh..43.4381C} {43, 4381}

\bibitem[\protect\citeauthoryear{{Dale}, {Ercolano}  \& {Bonnell}}{{Dale}
  et~al.}{2013}]{Dale2013}
{Dale} J.~E.,  {Ercolano} B.,   {Bonnell} I.~A.,  2013, \mn@doi [\mnras]
  {10.1093/mnras/sts592}, \href
  {http://adsabs.harvard.edu/abs/2013MNRAS.430..234D} {430, 234}

\bibitem[\protect\citeauthoryear{{Dalgarno} \& {Lepp}}{{Dalgarno} \&
  {Lepp}}{1987}]{Dalgarno1987IAUS}
{Dalgarno} A.,  {Lepp} S.,  1987, in {Vardya} M.~S.,  {Tarafdar} S.~P.,  eds,
  IAU Symposium Vol. 120, Astrochemistry. pp 109--118

\bibitem[\protect\citeauthoryear{Dean, Davidson  \& Hanson}{Dean
  et~al.}{1991}]{Dean1991}
Dean A.~J.,  Davidson D.~F.,   Hanson R.~K.,  1991, \mn@doi [The Journal of
  Physical Chemistry] {10.1021/j100154a037}, 95, 183

\bibitem[\protect\citeauthoryear{{Draine}}{{Draine}}{1978}]{Draine1978ApJS}
{Draine} B.~T.,  1978, \mn@doi [\apjs] {10.1086/190513}, \href
  {http://adsabs.harvard.edu/abs/1978ApJS...36..595D} {36, 595}

\bibitem[\protect\citeauthoryear{{Draine}}{{Draine}}{2009}]{Draine2009}
{Draine} B.~T.,  2009, in {Henning} T.,  {Gr{\"u}n} E.,   {Steinacker} J.,
  eds,  Astronomical Society of the Pacific Conference Series Vol. 414, Cosmic
  Dust - Near and Far. p.~453 (\mn@eprint {arXiv} {0903.1658})

\bibitem[\protect\citeauthoryear{{Draine} \& {Bertoldi}}{{Draine} \&
  {Bertoldi}}{1996}]{Draine1996}
{Draine} B.~T.,  {Bertoldi} F.,  1996, \mn@doi [\apj] {10.1086/177689}, \href
  {http://adsabs.harvard.edu/abs/1996ApJ...468..269D} {468, 269}

\bibitem[\protect\citeauthoryear{{Dullemond} \& {Monnier}}{{Dullemond} \&
  {Monnier}}{2010}]{Dullemond2010}
{Dullemond} C.~P.,  {Monnier} J.~D.,  2010, \mn@doi [\araa]
  {10.1146/annurev-astro-081309-130932}, \href
  {http://adsabs.harvard.edu/abs/2010ARA%26A..48..205D} {48, 205}

\bibitem[\protect\citeauthoryear{{Dunn}}{{Dunn}}{1968}]{Dunn1968}
{Dunn} G.~H.,  1968, \mn@doi [Physical Review] {10.1103/PhysRev.172.1}, \href
  {http://adsabs.harvard.edu/abs/1968PhRv..172....1D} {172, 1}

\bibitem[\protect\citeauthoryear{{Fairbairn}}{{Fairbairn}}{1969}]{1969RSPSA.312..207F}
{Fairbairn} A.~R.,  1969, \mn@doi [Proceedings of the Royal Society of London
  Series A] {10.1098/rspa.1969.0149}, \href
  {http://adsabs.harvard.edu/abs/1969RSPSA.312..207F} {312, 207}

\bibitem[\protect\citeauthoryear{{Feautrier}}{{Feautrier}}{1964}]{Feautrier1964SAOSR}
{Feautrier} P.,  1964, SAO Special Report, \href
  {http://adsabs.harvard.edu/abs/1964SAOSR.167...80F} {167, 80}

\bibitem[\protect\citeauthoryear{{Federer}, {Villinger}, {Howorka},
  {Lindinger}, {Tosis}, {Bassi}  \& {Ferguson}}{{Federer}
  et~al.}{1984}]{1984PhRvL..52.2084F}
{Federer} W.,  {Villinger} H.,  {Howorka} F.,  {Lindinger} W.,  {Tosis} P.,
  {Bassi} D.,   {Ferguson} E.,  1984, \mn@doi [Physical Review Letters]
  {10.1103/PhysRevLett.52.2084}, \href
  {http://adsabs.harvard.edu/abs/1984PhRvL..52.2084F} {52, 2084}

\bibitem[\protect\citeauthoryear{{Ferland}, {Peterson}, {Horne}, {Welsh}  \&
  {Nahar}}{{Ferland} et~al.}{1992}]{Ferland1992ApJ}
{Ferland} G.~J.,  {Peterson} B.~M.,  {Horne} K.,  {Welsh} W.~F.,   {Nahar}
  S.~N.,  1992, \mn@doi [\apj] {10.1086/171063}, \href
  {http://adsabs.harvard.edu/abs/1992ApJ...387...95F} {387, 95}

\bibitem[\protect\citeauthoryear{{Ferland}, {Korista}, {Verner}, {Ferguson},
  {Kingdon}  \& {Verner}}{{Ferland} et~al.}{1998}]{Ferland1998}
{Ferland} G.~J.,  {Korista} K.~T.,  {Verner} D.~A.,  {Ferguson} J.~W.,
  {Kingdon} J.~B.,   {Verner} E.~M.,  1998, \mn@doi [\pasp] {10.1086/316190},
  \href {http://adsabs.harvard.edu/abs/1998PASP..110..761F} {110, 761}

\bibitem[\protect\citeauthoryear{{Ferland} et~al.,}{{Ferland}
  et~al.}{2013}]{Ferland2013}
{Ferland} G.~J.,  et~al., 2013, \rmxaa, \href
  {http://adsabs.harvard.edu/abs/2013RMxAA..49..137F} {49, 137}

\bibitem[\protect\citeauthoryear{{Forrey}}{{Forrey}}{2013}]{Forrey2013ApJ}
{Forrey} R.~C.,  2013, \mn@doi [\apjl] {10.1088/2041-8205/773/2/L25}, \href
  {http://adsabs.harvard.edu/abs/2013ApJ...773L..25F} {773, L25}

\bibitem[\protect\citeauthoryear{Frostholm}{Frostholm}{2014}]{Frostholm2014}
Frostholm T.,  2014, Master's thesis, University of Copenhagen, Denmark

\bibitem[\protect\citeauthoryear{{Geppert} et~al.,}{{Geppert}
  et~al.}{2005}]{2005JPhCS...4...26G}
{Geppert} W.~D.,  et~al., 2005, in {Wolf} A.,  {Lammich} L.,   {Schmelcher} P.,
   eds,  Journal of Physics Conference Series Vol. 4, Journal of Physics
  Conference Series. pp 26--31, \mn@doi{10.1088/1742-6596/4/1/004}

\bibitem[\protect\citeauthoryear{Gerlich \& Horning}{Gerlich \&
  Horning}{1992}]{gerlich1992experimental}
Gerlich D.,  Horning S.,  1992, Chemical Reviews, 92, 1509

\bibitem[\protect\citeauthoryear{{Glover} \& {Abel}}{{Glover} \&
  {Abel}}{2008}]{Glover2008MNRAS}
{Glover} S.~C.~O.,  {Abel} T.,  2008, \mn@doi [\mnras]
  {10.1111/j.1365-2966.2008.13224.x}, \href
  {http://adsabs.harvard.edu/abs/2008MNRAS.388.1627G} {388, 1627}

\bibitem[\protect\citeauthoryear{{Glover} \& {Jappsen}}{{Glover} \&
  {Jappsen}}{2007}]{Glover2007}
{Glover} S.~C.~O.,  {Jappsen} A.-K.,  2007, \mn@doi [\apj] {10.1086/519445},
  \href {http://adsabs.harvard.edu/abs/2007ApJ...666....1G} {666, 1}

\bibitem[\protect\citeauthoryear{{Glover} \& {Mac Low}}{{Glover} \& {Mac
  Low}}{2007}]{Glover2007a}
{Glover} S.~C.~O.,  {Mac Low} M.-M.,  2007, \mn@doi [\apjs] {10.1086/512238},
  \href {http://adsabs.harvard.edu/abs/2007ApJS..169..239G} {169, 239}

\bibitem[\protect\citeauthoryear{{Glover} \& {Savin}}{{Glover} \&
  {Savin}}{2009}]{2009MNRAS.393..911G}
{Glover} S.~C.~O.,  {Savin} D.~W.,  2009, \mn@doi [\mnras]
  {10.1111/j.1365-2966.2008.14156.x}, \href
  {http://adsabs.harvard.edu/abs/2009MNRAS.393..911G} {393, 911}

\bibitem[\protect\citeauthoryear{{Glover}, {Federrath}, {Mac Low}  \&
  {Klessen}}{{Glover} et~al.}{2010}]{Glover2010MNRAS}
{Glover} S.~C.~O.,  {Federrath} C.,  {Mac Low} M.-M.,   {Klessen} R.~S.,  2010,
  \mn@doi [\mnras] {10.1111/j.1365-2966.2009.15718.x}, \href
  {http://adsabs.harvard.edu/abs/2010MNRAS.404....2G} {404, 2}

\bibitem[\protect\citeauthoryear{{Gonz{\'a}lez}, {Vaytet}, {Commer{\c c}on}  \&
  {Masson}}{{Gonz{\'a}lez} et~al.}{2015}]{Gonzalez2015A&A}
{Gonz{\'a}lez} M.,  {Vaytet} N.,  {Commer{\c c}on} B.,   {Masson} J.,  2015,
  \mn@doi [\aap] {10.1051/0004-6361/201525971}, \href
  {http://adsabs.harvard.edu/abs/2015A%26A...578A..12G} {578, A12}

\bibitem[\protect\citeauthoryear{{G{\'o}rski}, {Hivon}, {Banday}, {Wandelt},
  {Hansen}, {Reinecke}  \& {Bartelmann}}{{G{\'o}rski}
  et~al.}{2005}]{Gorski2005}
{G{\'o}rski} K.~M.,  {Hivon} E.,  {Banday} A.~J.,  {Wandelt} B.~D.,  {Hansen}
  F.~K.,  {Reinecke} M.,   {Bartelmann} M.,  2005, \mn@doi [\apj]
  {10.1086/427976}, \href {http://adsabs.harvard.edu/abs/2005ApJ...622..759G}
  {622, 759}

\bibitem[\protect\citeauthoryear{{Grassi}, {Krstic}, {Merlin}, {Buonomo},
  {Piovan}  \& {Chiosi}}{{Grassi} et~al.}{2011}]{Grassi2011A&A}
{Grassi} T.,  {Krstic} P.,  {Merlin} E.,  {Buonomo} U.,  {Piovan} L.,
  {Chiosi} C.,  2011, \mn@doi [\aap] {10.1051/0004-6361/200913779}, \href
  {http://adsabs.harvard.edu/abs/2011A%26A...533A.123G} {533, A123}

\bibitem[\protect\citeauthoryear{{Grassi}, {Bovino}, {Schleicher}, {Prieto},
  {Seifried}, {Simoncini}  \& {Gianturco}}{{Grassi}
  et~al.}{2014}]{Grassi2014MNRAS}
{Grassi} T.,  {Bovino} S.,  {Schleicher} D.~R.~G.,  {Prieto} J.,  {Seifried}
  D.,  {Simoncini} E.,   {Gianturco} F.~A.,  2014, \mn@doi [\mnras]
  {10.1093/mnras/stu114}, \href
  {http://adsabs.harvard.edu/abs/2014MNRAS.439.2386G} {439, 2386}

\bibitem[\protect\citeauthoryear{{Grassi}, {Bovino}, {Haugb{\o}lle}  \&
  {Schleicher}}{{Grassi} et~al.}{2017}]{Grassi2017MNRAS}
{Grassi} T.,  {Bovino} S.,  {Haugb{\o}lle} T.,   {Schleicher} D.~R.~G.,  2017,
  \mn@doi [\mnras] {10.1093/mnras/stw2871}, \href
  {http://adsabs.harvard.edu/abs/2017MNRAS.466.1259G} {466, 1259}

\bibitem[\protect\citeauthoryear{{Guberman}}{{Guberman}}{1995}]{1995JChPh.102.1699G}
{Guberman} S.~L.,  1995, \mn@doi [\jcp] {10.1063/1.468902}, \href
  {http://adsabs.harvard.edu/abs/1995JChPh.102.1699G} {102, 1699}

\bibitem[\protect\citeauthoryear{{Harada}, {Herbst}  \& {Wakelam}}{{Harada}
  et~al.}{2010}]{Harada2010ApJ}
{Harada} N.,  {Herbst} E.,   {Wakelam} V.,  2010, \mn@doi [\apj]
  {10.1088/0004-637X/721/2/1570}, \href
  {http://adsabs.harvard.edu/abs/2010ApJ...721.1570H} {721, 1570}

\bibitem[\protect\citeauthoryear{Harding, Guadagnini  \& Schatz}{Harding
  et~al.}{1993}]{Harding1993}
Harding L.~B.,  Guadagnini R.,   Schatz G.~C.,  1993, \mn@doi [The Journal of
  Physical Chemistry] {10.1021/j100123a005}, 97, 5472

\bibitem[\protect\citeauthoryear{{Hasegawa} \& {Herbst}}{{Hasegawa} \&
  {Herbst}}{1993}]{Hasegawa1993}
{Hasegawa} T.~I.,  {Herbst} E.,  1993, \mn@doi [\mnras]
  {10.1093/mnras/261.1.83}, \href
  {http://adsabs.harvard.edu/abs/1993MNRAS.261...83H} {261, 83}

\bibitem[\protect\citeauthoryear{{Haworth}, {Glover}, {Koepferl}, {Bisbas}  \&
  {Dale}}{{Haworth} et~al.}{2018}]{Haworth2018}
{Haworth} T.~J.,  {Glover} S.~C.~O.,  {Koepferl} C.~M.,  {Bisbas} T.~G.,
  {Dale} J.~E.,  2018, \mn@doi [\nar] {10.1016/j.newar.2018.06.001}, \href
  {http://adsabs.harvard.edu/abs/2018NewAR..82....1H} {82, 1}

\bibitem[\protect\citeauthoryear{{Heinemann}, {Dobler}, {Nordlund}  \&
  {Brandenburg}}{{Heinemann} et~al.}{2006}]{Heinemann2006A&A}
{Heinemann} T.,  {Dobler} W.,  {Nordlund} {\AA}.,   {Brandenburg} A.,  2006,
  \mn@doi [\aap] {10.1051/0004-6361:20053120}, \href
  {http://adsabs.harvard.edu/abs/2006A%26A...448..731H} {448, 731}

\bibitem[\protect\citeauthoryear{{Hocuk} \& {Cazaux}}{{Hocuk} \&
  {Cazaux}}{2015}]{Hocuk2015}
{Hocuk} S.,  {Cazaux} S.,  2015, \mn@doi [\aap] {10.1051/0004-6361/201424503},
  \href {http://adsabs.harvard.edu/abs/2015A%26A...576A..49H} {576, A49}

\bibitem[\protect\citeauthoryear{{Hollenbach} \& {McKee}}{{Hollenbach} \&
  {McKee}}{1979}]{Hollenbach1979}
{Hollenbach} D.,  {McKee} C.~F.,  1979, \mn@doi [\apjs] {10.1086/190631}, \href
  {http://adsabs.harvard.edu/abs/1979ApJS...41..555H} {41, 555}

\bibitem[\protect\citeauthoryear{{Hollenbach}, {Kaufman}, {Bergin}  \&
  {Melnick}}{{Hollenbach} et~al.}{2009}]{Hollenbach2009}
{Hollenbach} D.,  {Kaufman} M.~J.,  {Bergin} E.~A.,   {Melnick} G.~J.,  2009,
  \mn@doi [\apj] {10.1088/0004-637X/690/2/1497}, \href
  {http://adsabs.harvard.edu/abs/2009ApJ...690.1497H} {690, 1497}

\bibitem[\protect\citeauthoryear{{Howard}, {Pudritz}  \& {Harris}}{{Howard}
  et~al.}{2017}]{Howard2017}
{Howard} C.~S.,  {Pudritz} R.~E.,   {Harris} W.~E.,  2017, \mn@doi [\mnras]
  {10.1093/mnras/stx1363}, \href
  {http://adsabs.harvard.edu/abs/2017MNRAS.470.3346H} {470, 3346}

\bibitem[\protect\citeauthoryear{{Huebner}, {Keady}  \& {Lyon}}{{Huebner}
  et~al.}{1992}]{Huebner1992ApSS}
{Huebner} W.~F.,  {Keady} J.~J.,   {Lyon} S.~P.,  1992, \mn@doi [\apss]
  {10.1007/BF00644558}, \href
  {http://adsabs.harvard.edu/abs/1992Ap%26SS.195....1H} {195, 1}

\bibitem[\protect\citeauthoryear{{Hummer} \& {Rybicki}}{{Hummer} \&
  {Rybicki}}{1971}]{1971MNRAS.152....1H}
{Hummer} D.~G.,  {Rybicki} G.~B.,  1971, \mn@doi [\mnras]
  {10.1093/mnras/152.1.1}, \href
  {http://adsabs.harvard.edu/abs/1971MNRAS.152....1H} {152, 1}

\bibitem[\protect\citeauthoryear{{Iliev} et~al.,}{{Iliev}
  et~al.}{2006}]{Iliev2006}
{Iliev} I.~T.,  et~al., 2006, \mn@doi [\mnras]
  {10.1111/j.1365-2966.2006.10775.x}, \href
  {http://adsabs.harvard.edu/abs/2006MNRAS.371.1057I} {371, 1057}

\bibitem[\protect\citeauthoryear{{Iliev} et~al.,}{{Iliev}
  et~al.}{2009}]{Iliev2009}
{Iliev} I.~T.,  et~al., 2009, \mn@doi [\mnras]
  {10.1111/j.1365-2966.2009.15558.x}, \href
  {http://adsabs.harvard.edu/abs/2009MNRAS.400.1283I} {400, 1283}

\bibitem[\protect\citeauthoryear{{Janev}, {Langer}, {Post}  \& {Evans}}{{Janev}
  et~al.}{1987}]{Janev1987Springer}
{Janev} R.~K.,  {Langer} W.~D.,  {Post} Jr. D.~E.,   {Evans} Jr. K.,  eds,
  1987, {Elementary processes in hydrogen-helium plasmas: Cross sections and
  reaction rate coefficients,}  Springer Series on Atoms and Plasmas Vol. 4.
Springer-Verlag

\bibitem[\protect\citeauthoryear{{Jensen} \& {Haugb{\o}lle}}{{Jensen} \&
  {Haugb{\o}lle}}{2018}]{Jensen2018MNRAS}
{Jensen} S.~S.,  {Haugb{\o}lle} T.,  2018, \mn@doi [\mnras]
  {10.1093/mnras/stx2844}, \href
  {http://adsabs.harvard.edu/abs/2018MNRAS.474.1176J} {474, 1176}

\bibitem[\protect\citeauthoryear{{Jensen}, {Bilodeau}, {Safvan}, {Seiersen},
  {Andersen}, {Pedersen}  \& {Heber}}{{Jensen}
  et~al.}{2000}]{2000ApJ...543..764J}
{Jensen} M.~J.,  {Bilodeau} R.~C.,  {Safvan} C.~P.,  {Seiersen} K.,  {Andersen}
  L.~H.,  {Pedersen} H.~B.,   {Heber} O.,  2000, \mn@doi [\apj]
  {10.1086/317137}, \href {http://adsabs.harvard.edu/abs/2000ApJ...543..764J}
  {543, 764}

\bibitem[\protect\citeauthoryear{{Jones}, {Birkinshaw}  \& {Twiddy}}{{Jones}
  et~al.}{1981}]{1981CPL....77..484J}
{Jones} J.~D.~C.,  {Birkinshaw} K.,   {Twiddy} N.~D.,  1981, \mn@doi [Chemical
  Physics Letters] {10.1016/0009-2614(81)85191-3}, \href
  {http://adsabs.harvard.edu/abs/1981CPL....77..484J} {77, 484}

\bibitem[\protect\citeauthoryear{{Karpas}, {Anicich}  \& {Huntress}}{{Karpas}
  et~al.}{1979}]{Karpas1979JChPh}
{Karpas} Z.,  {Anicich} V.,   {Huntress} W.~T.,  1979, \mn@doi [\jcp]
  {10.1063/1.437823}, \href {http://adsabs.harvard.edu/abs/1979JChPh..70.2877K}
  {70, 2877}

\bibitem[\protect\citeauthoryear{{Khokhlov}}{{Khokhlov}}{1998}]{Khokhlov1998}
{Khokhlov} A.,  1998, \mn@doi [Journal of Computational Physics]
  {10.1006/jcph.1998.9998}, \href
  {http://adsabs.harvard.edu/abs/1998JCoPh.143..519K} {143, 519}

\bibitem[\protect\citeauthoryear{{Kimura}, {Dalgarno}, {Chantranupong}, {Li},
  {Hirsch}  \& {Buenker}}{{Kimura} et~al.}{1993}]{Kimura1993}
{Kimura} M.,  {Dalgarno} A.,  {Chantranupong} L.,  {Li} Y.,  {Hirsch} G.,
  {Buenker} R.~J.,  1993, \mn@doi [\apj] {10.1086/173361}, \href
  {http://adsabs.harvard.edu/abs/1993ApJ...417..812K} {417, 812}

\bibitem[\protect\citeauthoryear{{Kreckel}, {Bruhns}, {{\v C}{\'{\i}}{\v z}ek},
  {Glover}, {Miller}, {Urbain}  \& {Savin}}{{Kreckel}
  et~al.}{2010}]{Kreckel2010Sci}
{Kreckel} H.,  {Bruhns} H.,  {{\v C}{\'{\i}}{\v z}ek} M.,  {Glover} S.~C.~O.,
  {Miller} K.~A.,  {Urbain} X.,   {Savin} D.~W.,  2010, \mn@doi [Science]
  {10.1126/science.1187191}, \href
  {http://adsabs.harvard.edu/abs/2010Sci...329...69K} {329, 69}

\bibitem[\protect\citeauthoryear{{Krumholz}, {Leroy}  \& {McKee}}{{Krumholz}
  et~al.}{2011}]{Krumholz2011ApJ}
{Krumholz} M.~R.,  {Leroy} A.~K.,   {McKee} C.~F.,  2011, \mn@doi [\apj]
  {10.1088/0004-637X/731/1/25}, \href
  {http://adsabs.harvard.edu/abs/2011ApJ...731...25K} {731, 25}

\bibitem[\protect\citeauthoryear{{Krumholz} et~al.,}{{Krumholz}
  et~al.}{2014}]{KrumholzPPVI}
{Krumholz} M.~R.,  et~al., 2014, \mn@doi [Protostars and Planets VI]
  {10.2458/azu_uapress_9780816531240-ch011}, \href
  {http://adsabs.harvard.edu/abs/2014prpl.conf..243K} {pp 243--266}

\bibitem[\protect\citeauthoryear{{Kuffmeier}, {Haugb{\o}lle}  \&
  {Nordlund}}{{Kuffmeier} et~al.}{2017}]{Kuffmeier2017}
{Kuffmeier} M.,  {Haugb{\o}lle} T.,   {Nordlund} {\AA}.,  2017, \mn@doi [\apj]
  {10.3847/1538-4357/aa7c64}, \href
  {http://adsabs.harvard.edu/abs/2017ApJ...846....7K} {846, 7}

\bibitem[\protect\citeauthoryear{{Kuffmeier}, {Frimann}, {Jensen}  \&
  {Haugb{\o}lle}}{{Kuffmeier} et~al.}{2018}]{Kuffmeier2018MNRAS}
{Kuffmeier} M.,  {Frimann} S.,  {Jensen} S.~S.,   {Haugb{\o}lle} T.,  2018,
  \mn@doi [\mnras] {10.1093/mnras/sty024}, \href
  {http://adsabs.harvard.edu/abs/2018MNRAS.475.2642K} {475, 2642}

\bibitem[\protect\citeauthoryear{{Larson} et~al.,}{{Larson}
  et~al.}{1998}]{1998ApJ...505..459L}
{Larson} {\AA}.,  et~al., 1998, \mn@doi [\apj] {10.1086/306164}, \href
  {http://adsabs.harvard.edu/abs/1998ApJ...505..459L} {505, 459}

\bibitem[\protect\citeauthoryear{{Le Teuff}, {Millar}  \& {Markwick}}{{Le
  Teuff} et~al.}{2000}]{LeTeuff2000}
{Le Teuff} Y.~H.,  {Millar} T.~J.,   {Markwick} A.~J.,  2000, \mn@doi [\aaps]
  {10.1051/aas:2000265}, \href
  {http://adsabs.harvard.edu/abs/2000A%26AS..146..157L} {146, 157}

\bibitem[\protect\citeauthoryear{{Leger}, {Jura}  \& {Omont}}{{Leger}
  et~al.}{1985}]{Leger1985}
{Leger} A.,  {Jura} M.,   {Omont} A.,  1985, \aap, \href
  {http://adsabs.harvard.edu/abs/1985A%26A...144..147L} {144, 147}

\bibitem[\protect\citeauthoryear{{Lepp} \& {Shull}}{{Lepp} \&
  {Shull}}{1983}]{Lepp1983ApJ}
{Lepp} S.,  {Shull} J.~M.,  1983, \mn@doi [\apj] {10.1086/161149}, \href
  {http://adsabs.harvard.edu/abs/1983ApJ...270..578L} {270, 578}

\bibitem[\protect\citeauthoryear{Linder, Janev  \& Botero}{Linder
  et~al.}{1995}]{linder1995reactive}
Linder F.,  Janev R.,   Botero J.,  1995, in , Atomic and Molecular Processes
  in Fusion Edge Plasmas.
Springer, pp 397--431

\bibitem[\protect\citeauthoryear{{Loison}, {Wakelam}, {Hickson}, {Bergeat}  \&
  {Mereau}}{{Loison} et~al.}{2014}]{Loison2014MNRAS}
{Loison} J.-C.,  {Wakelam} V.,  {Hickson} K.~M.,  {Bergeat} A.,   {Mereau} R.,
  2014, \mn@doi [\mnras] {10.1093/mnras/stt1956}, \href
  {http://adsabs.harvard.edu/abs/2014MNRAS.437..930L} {437, 930}

\bibitem[\protect\citeauthoryear{{Mathis}, {Rumpl}  \& {Nordsieck}}{{Mathis}
  et~al.}{1977}]{Mathis1977}
{Mathis} J.~S.,  {Rumpl} W.,   {Nordsieck} K.~H.,  1977, \mn@doi [\apj]
  {10.1086/155591}, \href {http://adsabs.harvard.edu/abs/1977ApJ...217..425M}
  {217, 425}

\bibitem[\protect\citeauthoryear{{McCall} et~al.,}{{McCall}
  et~al.}{2004}]{2004PhRvA..70e2716M}
{McCall} B.~J.,  et~al., 2004, \mn@doi [\pra] {10.1103/PhysRevA.70.052716},
  \href {http://adsabs.harvard.edu/abs/2004PhRvA..70e2716M} {70, 052716}

\bibitem[\protect\citeauthoryear{{McElroy}, {Walsh}, {Markwick}, {Cordiner},
  {Smith}  \& {Millar}}{{McElroy} et~al.}{2013}]{McElroy2013A&A}
{McElroy} D.,  {Walsh} C.,  {Markwick} A.~J.,  {Cordiner} M.~A.,  {Smith} K.,
  {Millar} T.~J.,  2013, \mn@doi [\aap] {10.1051/0004-6361/201220465}, \href
  {http://adsabs.harvard.edu/abs/2013A%26A...550A..36M} {550, A36}

\bibitem[\protect\citeauthoryear{{McEwan}, {Scott}, {Adams}, {Babcock},
  {Terzieva}  \& {Herbst}}{{McEwan} et~al.}{1999}]{1999ApJ...513..287M}
{McEwan} M.~J.,  {Scott} G.~B.~I.,  {Adams} N.~G.,  {Babcock} L.~M.,
  {Terzieva} R.,   {Herbst} E.,  1999, \mn@doi [\apj] {10.1086/306861}, \href
  {http://adsabs.harvard.edu/abs/1999ApJ...513..287M} {513, 287}

\bibitem[\protect\citeauthoryear{{Mellema}, {Iliev}, {Alvarez}  \&
  {Shapiro}}{{Mellema} et~al.}{2006}]{Mellema2006NewA}
{Mellema} G.,  {Iliev} I.~T.,  {Alvarez} M.~A.,   {Shapiro} P.~R.,  2006,
  \mn@doi [\na] {10.1016/j.newast.2005.09.004}, \href
  {http://adsabs.harvard.edu/abs/2006NewA...11..374M} {11, 374}

\bibitem[\protect\citeauthoryear{{Millar}, {Farquhar}  \& {Willacy}}{{Millar}
  et~al.}{1997}]{Millar1997}
{Millar} T.~J.,  {Farquhar} P.~R.~A.,   {Willacy} K.,  1997, \mn@doi [\aaps]
  {10.1051/aas:1997118}, \href
  {http://adsabs.harvard.edu/abs/1997A%26AS..121..139M} {121, 139}

\bibitem[\protect\citeauthoryear{{Mitchell}}{{Mitchell}}{1990}]{1990PhR...186..215M}
{Mitchell} J.~B.~A.,  1990, \physrep, \href
  {http://adsabs.harvard.edu/abs/1990PhR...186..215M} {186, 215}

\bibitem[\protect\citeauthoryear{{Mitchell} \& {Deveau}}{{Mitchell} \&
  {Deveau}}{1983}]{Mitchell1983ApJ}
{Mitchell} G.~F.,  {Deveau} T.~J.,  1983, \mn@doi [\apj] {10.1086/160812},
  \href {http://adsabs.harvard.edu/abs/1983ApJ...266..646M} {266, 646}

\bibitem[\protect\citeauthoryear{Murrell \& Rodriguez}{Murrell \&
  Rodriguez}{1986}]{Murrell1986}
Murrell J.,  Rodriguez J.,  1986, \mn@doi [Journal of Molecular Structure:
  THEOCHEM] {https://doi.org/10.1016/0166-1280(86)87044-0}, 139, 267

\bibitem[\protect\citeauthoryear{{Nahar}}{{Nahar}}{1999}]{Nahar1999}
{Nahar} S.~N.,  1999, \mn@doi [\apjs] {10.1086/313173}, \href
  {http://adsabs.harvard.edu/abs/1999ApJS..120..131N} {120, 131}

\bibitem[\protect\citeauthoryear{{Nahar} \& {Pradhan}}{{Nahar} \&
  {Pradhan}}{1997}]{Nahar1997}
{Nahar} S.~N.,  {Pradhan} A.~K.,  1997, \mn@doi [\apjs] {10.1086/313013}, \href
  {http://adsabs.harvard.edu/abs/1997ApJS..111..339N} {111, 339}

\bibitem[\protect\citeauthoryear{{Nelson} \& {Langer}}{{Nelson} \&
  {Langer}}{1997}]{Nelson1997}
{Nelson} R.~P.,  {Langer} W.~D.,  1997, \mn@doi [\apj] {10.1086/304167}, \href
  {http://adsabs.harvard.edu/abs/1997ApJ...482..796N} {482, 796}

\bibitem[\protect\citeauthoryear{{O'Connor}, {Urbain}, {St{\"u}tzel}, {Miller},
  {de Ruette}, {Garrido}  \& {Savin}}{{O'Connor}
  et~al.}{2015}]{OConnor2015ApJS}
{O'Connor} A.~P.,  {Urbain} X.,  {St{\"u}tzel} J.,  {Miller} K.~A.,  {de
  Ruette} N.,  {Garrido} M.,   {Savin} D.~W.,  2015, \mn@doi [\apjs]
  {10.1088/0067-0049/219/1/6}, \href
  {http://adsabs.harvard.edu/abs/2015ApJS..219....6O} {219, 6}

\bibitem[\protect\citeauthoryear{{Omukai}}{{Omukai}}{2000}]{Omukai2000}
{Omukai} K.,  2000, \mn@doi [\apj] {10.1086/308776}, \href
  {http://adsabs.harvard.edu/abs/2000ApJ...534..809O} {534, 809}

\bibitem[\protect\citeauthoryear{{Osterbrock}}{{Osterbrock}}{1989}]{Osterbrock1989}
{Osterbrock} D.~E.,  1989, {Astrophysics of gaseous nebulae and active galactic
  nuclei}

\bibitem[\protect\citeauthoryear{{Padoan}, {Pan}, {Haugb{\o}lle}  \&
  {Nordlund}}{{Padoan} et~al.}{2016}]{Padoan2016}
{Padoan} P.,  {Pan} L.,  {Haugb{\o}lle} T.,   {Nordlund} {\AA}.,  2016, \mn@doi
  [\apj] {10.3847/0004-637X/822/1/11}, \href
  {http://adsabs.harvard.edu/abs/2016ApJ...822...11P} {822, 11}

\bibitem[\protect\citeauthoryear{{Peart} \& {Hayton}}{{Peart} \&
  {Hayton}}{1994}]{1994JPhB...27.2551P}
{Peart} B.,  {Hayton} D.~A.,  1994, \mn@doi [Journal of Physics B Atomic
  Molecular Physics] {10.1088/0953-4075/27/12/013}, \href
  {http://adsabs.harvard.edu/abs/1994JPhB...27.2551P} {27, 2551}

\bibitem[\protect\citeauthoryear{{Peters}, {Banerjee}, {Klessen}, {Mac Low},
  {Galv{\'a}n-Madrid}  \& {Keto}}{{Peters} et~al.}{2010}]{Peters2010}
{Peters} T.,  {Banerjee} R.,  {Klessen} R.~S.,  {Mac Low} M.-M.,
  {Galv{\'a}n-Madrid} R.,   {Keto} E.~R.,  2010, \mn@doi [\apj]
  {10.1088/0004-637X/711/2/1017}, \href
  {http://adsabs.harvard.edu/abs/2010ApJ...711.1017P} {711, 1017}

\bibitem[\protect\citeauthoryear{{Peters} et~al.,}{{Peters}
  et~al.}{2017}]{Peters2017}
{Peters} T.,  et~al., 2017, \mn@doi [\mnras] {10.1093/mnras/stw3216}, \href
  {http://adsabs.harvard.edu/abs/2017MNRAS.466.3293P} {466, 3293}

\bibitem[\protect\citeauthoryear{{Petuchowski}, {Dwek}, {Allen}  \&
  {Nuth}}{{Petuchowski} et~al.}{1989}]{Petuchowski1989}
{Petuchowski} S.~J.,  {Dwek} E.,  {Allen} Jr. J.~E.,   {Nuth} III J.~A.,  1989,
  \mn@doi [\apj] {10.1086/167601}, \href
  {http://adsabs.harvard.edu/abs/1989ApJ...342..406P} {342, 406}

\bibitem[\protect\citeauthoryear{{Poulaert}, {Brouillard}, {Claeys}, {McGowan}
  \& {Van Wassenhove}}{{Poulaert} et~al.}{1978}]{Poulaert1978}
{Poulaert} G.,  {Brouillard} F.,  {Claeys} W.,  {McGowan} J.~W.,   {Van
  Wassenhove} G.,  1978, \mn@doi [Journal of Physics B Atomic Molecular
  Physics] {10.1088/0022-3700/11/21/006}, \href
  {http://adsabs.harvard.edu/abs/1978JPhB...11L.671P} {11, L671}

\bibitem[\protect\citeauthoryear{{Prasad} \& {Huntress}}{{Prasad} \&
  {Huntress}}{1980}]{Prasad1980ApJS}
{Prasad} S.~S.,  {Huntress} Jr. W.~T.,  1980, \mn@doi [\apjs] {10.1086/190665},
  \href {http://adsabs.harvard.edu/abs/1980ApJS...43....1P} {43, 1}

\bibitem[\protect\citeauthoryear{Rakshit \& Warneck}{Rakshit \&
  Warneck}{1980}]{F29807601084}
Rakshit A.~B.,  Warneck P.,  1980, \mn@doi [J. Chem. Soc.{,} Faraday Trans. 2]
  {10.1039/F29807601084}, 76, 1084

\bibitem[\protect\citeauthoryear{{Ramaker} \& {Peek}}{{Ramaker} \&
  {Peek}}{1976}]{Ramaker1976PhRvA}
{Ramaker} D.~E.,  {Peek} J.~M.,  1976, \mn@doi [\pra] {10.1103/PhysRevA.13.58},
  \href {http://adsabs.harvard.edu/abs/1976PhRvA..13...58R} {13, 58}

\bibitem[\protect\citeauthoryear{{Reboussin}, {Wakelam}, {Guilloteau}  \&
  {Hersant}}{{Reboussin} et~al.}{2014}]{Reboussin2014}
{Reboussin} L.,  {Wakelam} V.,  {Guilloteau} S.,   {Hersant} F.,  2014, \mn@doi
  [\mnras] {10.1093/mnras/stu462}, \href
  {http://adsabs.harvard.edu/abs/2014MNRAS.440.3557R} {440, 3557}

\bibitem[\protect\citeauthoryear{{Richings}, {Schaye}  \&
  {Oppenheimer}}{{Richings} et~al.}{2014a}]{Richings2014I}
{Richings} A.~J.,  {Schaye} J.,   {Oppenheimer} B.~D.,  2014a, \mn@doi [\mnras]
  {10.1093/mnras/stu525}, \href
  {http://adsabs.harvard.edu/abs/2014MNRAS.440.3349R} {440, 3349}

\bibitem[\protect\citeauthoryear{{Richings}, {Schaye}  \&
  {Oppenheimer}}{{Richings} et~al.}{2014b}]{Richings2014II}
{Richings} A.~J.,  {Schaye} J.,   {Oppenheimer} B.~D.,  2014b, \mn@doi [\mnras]
  {10.1093/mnras/stu1046}, \href
  {http://adsabs.harvard.edu/abs/2014MNRAS.442.2780R} {442, 2780}

\bibitem[\protect\citeauthoryear{{Rijkhorst}, {Plewa}, {Dubey}  \&
  {Mellema}}{{Rijkhorst} et~al.}{2006}]{Rijkhorst2006A&A}
{Rijkhorst} E.-J.,  {Plewa} T.,  {Dubey} A.,   {Mellema} G.,  2006, \mn@doi
  [\aap] {10.1051/0004-6361:20053401}, \href
  {http://adsabs.harvard.edu/abs/2006A%26A...452..907R} {452, 907}

\bibitem[\protect\citeauthoryear{{Roberge}, {Jones}, {Lepp}  \&
  {Dalgarno}}{{Roberge} et~al.}{1991}]{Roberge1991}
{Roberge} W.~G.,  {Jones} D.,  {Lepp} S.,   {Dalgarno} A.,  1991, \mn@doi
  [\apjs] {10.1086/191604}, \href
  {http://adsabs.harvard.edu/abs/1991ApJS...77..287R} {77, 287}

\bibitem[\protect\citeauthoryear{{R{\"o}llig} et~al.,}{{R{\"o}llig}
  et~al.}{2007}]{Rollig2007}
{R{\"o}llig} M.,  et~al., 2007, \mn@doi [\aap] {10.1051/0004-6361:20065918},
  \href {http://adsabs.harvard.edu/abs/2007A%26A...467..187R} {467, 187}

\bibitem[\protect\citeauthoryear{{Rosdahl} \& {Teyssier}}{{Rosdahl} \&
  {Teyssier}}{2015}]{Rosdahl2015}
{Rosdahl} J.,  {Teyssier} R.,  2015, \mn@doi [\mnras] {10.1093/mnras/stv567},
  \href {http://adsabs.harvard.edu/abs/2015MNRAS.449.4380R} {449, 4380}

\bibitem[\protect\citeauthoryear{{Rosdahl}, {Blaizot}, {Aubert}, {Stranex}  \&
  {Teyssier}}{{Rosdahl} et~al.}{2013}]{Rosdahl2013}
{Rosdahl} J.,  {Blaizot} J.,  {Aubert} D.,  {Stranex} T.,   {Teyssier} R.,
  2013, \mn@doi [\mnras] {10.1093/mnras/stt1722}, \href
  {http://adsabs.harvard.edu/abs/2013MNRAS.436.2188R} {436, 2188}

\bibitem[\protect\citeauthoryear{{Ros{\'e}n} et~al.,}{{Ros{\'e}n}
  et~al.}{1998}]{1998PhRvA..57.4462R}
{Ros{\'e}n} S.,  et~al., 1998, \mn@doi [\pra] {10.1103/PhysRevA.57.4462}, \href
  {http://adsabs.harvard.edu/abs/1998PhRvA..57.4462R} {57, 4462}

\bibitem[\protect\citeauthoryear{{Ros{\'e}n} et~al.,}{{Ros{\'e}n}
  et~al.}{2000}]{2000FaDi..115..295R}
{Ros{\'e}n} S.,  et~al., 2000, \mn@doi [Faraday Discussions]
  {10.1039/a909314a}, \href {http://adsabs.harvard.edu/abs/2000FaDi..115..295R}
  {115, 295}

\bibitem[\protect\citeauthoryear{{Safranek-Shrader}, {Krumholz}, {Kim},
  {Ostriker}, {Klein}, {Li}, {McKee}  \& {Stone}}{{Safranek-Shrader}
  et~al.}{2017}]{Safranek-Shrader2017MNRAS}
{Safranek-Shrader} C.,  {Krumholz} M.~R.,  {Kim} C.-G.,  {Ostriker} E.~C.,
  {Klein} R.~I.,  {Li} S.,  {McKee} C.~F.,   {Stone} J.~M.,  2017, \mn@doi
  [\mnras] {10.1093/mnras/stw2647}, \href
  {http://adsabs.harvard.edu/abs/2017MNRAS.465..885S} {465, 885}

\bibitem[\protect\citeauthoryear{{Savin}, {Krsti{\'c}}, {Haiman}  \&
  {Stancil}}{{Savin} et~al.}{2004}]{Savin2004ApJ}
{Savin} D.~W.,  {Krsti{\'c}} P.~S.,  {Haiman} Z.,   {Stancil} P.~C.,  2004,
  \mn@doi [\apjl] {10.1086/421108}, \href
  {http://adsabs.harvard.edu/abs/2004ApJ...606L.167S} {606, L167}

\bibitem[\protect\citeauthoryear{{Schneider}, {Dulieu}, {Giusti-Suzor}  \&
  {Roueff}}{{Schneider} et~al.}{1994}]{Schneider1994ApJ}
{Schneider} I.~F.,  {Dulieu} O.,  {Giusti-Suzor} A.,   {Roueff} E.,  1994,
  \mn@doi [\apj] {10.1086/173948}, \href
  {http://adsabs.harvard.edu/abs/1994ApJ...424..983S} {424, 983}

\bibitem[\protect\citeauthoryear{{Sidhu}, {Miller}  \& {Tennyson}}{{Sidhu}
  et~al.}{1992}]{1992A&A...255..453S}
{Sidhu} K.~S.,  {Miller} S.,   {Tennyson} J.,  1992, \aap, \href
  {http://adsabs.harvard.edu/abs/1992A%26A...255..453S} {255, 453}

\bibitem[\protect\citeauthoryear{{Singh}, {Sanzovo}, {Borin}  \&
  {Ornellas}}{{Singh} et~al.}{1999}]{1999MNRAS.303..235S}
{Singh} P.~D.,  {Sanzovo} G.~C.,  {Borin} A.~C.,   {Ornellas} F.~R.,  1999,
  \mnras, \href {http://adsabs.harvard.edu/abs/1999MNRAS.303..235S} {303, 235}

\bibitem[\protect\citeauthoryear{{Slack}}{{Slack}}{1976}]{1976JChPh..64..228S}
{Slack} M.~W.,  1976, \mn@doi [\jcp] {10.1063/1.431955}, \href
  {http://adsabs.harvard.edu/abs/1976JChPh..64..228S} {64, 228}

\bibitem[\protect\citeauthoryear{{Smith} \& {Adams}}{{Smith} \&
  {Adams}}{1977}]{1977IJMSI..23..123S}
{Smith} D.,  {Adams} N.~G.,  1977, \mn@doi [International Journal of Mass
  Spectrometry and Ion Processes] {10.1016/0020-7381(77)80094-6}, \href
  {http://adsabs.harvard.edu/abs/1977IJMSI..23..123S} {23, 123}

\bibitem[\protect\citeauthoryear{{Smith}, {Spanel}  \& {Mayhew}}{{Smith}
  et~al.}{1992}]{1992IJMSI.117..457S}
{Smith} D.,  {Spanel} P.,   {Mayhew} C.~A.,  1992, \mn@doi [International
  Journal of Mass Spectrometry and Ion Processes]
  {10.1016/0168-1176(92)80108-D}, \href
  {http://adsabs.harvard.edu/abs/1992IJMSI.117..457S} {117, 457}

\bibitem[\protect\citeauthoryear{{Smith}, {Schlemmer}, {von Richthofen}  \&
  {Gerlich}}{{Smith} et~al.}{2002}]{Smith2002ApJ}
{Smith} M.~A.,  {Schlemmer} S.,  {von Richthofen} J.,   {Gerlich} D.,  2002,
  \mn@doi [\apjl] {10.1086/344404}, \href
  {http://adsabs.harvard.edu/abs/2002ApJ...578L..87S} {578, L87}

\bibitem[\protect\citeauthoryear{{Smith}, {Herbst}  \& {Chang}}{{Smith}
  et~al.}{2004}]{2004MNRAS.350..323S}
{Smith} I.~W.~M.,  {Herbst} E.,   {Chang} Q.,  2004, \mn@doi [\mnras]
  {10.1111/j.1365-2966.2004.07656.x}, \href
  {http://adsabs.harvard.edu/abs/2004MNRAS.350..323S} {350, 323}

\bibitem[\protect\citeauthoryear{{Stancil} \& {Dalgarno}}{{Stancil} \&
  {Dalgarno}}{1998}]{1998FaDi..109...61S}
{Stancil} P.~C.,  {Dalgarno} A.,  1998, \mn@doi [Faraday Discussions]
  {10.1039/a800074c}, \href {http://adsabs.harvard.edu/abs/1998FaDi..109...61S}
  {109, 61}

\bibitem[\protect\citeauthoryear{{Stancil} et~al.,}{{Stancil}
  et~al.}{1998}]{Stancil1998}
{Stancil} P.~C.,  et~al., 1998, \mn@doi [\apj] {10.1086/305937}, \href
  {http://adsabs.harvard.edu/abs/1998ApJ...502.1006S} {502, 1006}

\bibitem[\protect\citeauthoryear{{Stancil}, {Schultz}, {Kimura}, {Gu}, {Hirsch}
   \& {Buenker}}{{Stancil} et~al.}{1999}]{Stancil1999}
{Stancil} P.~C.,  {Schultz} D.~R.,  {Kimura} M.,  {Gu} J.-P.,  {Hirsch} G.,
  {Buenker} R.~J.,  1999, \mn@doi [\aaps] {10.1051/aas:1999419}, \href
  {http://adsabs.harvard.edu/abs/1999A%26AS..140..225S} {140, 225}

\bibitem[\protect\citeauthoryear{{Stenrup}, {Larson}  \& {Elander}}{{Stenrup}
  et~al.}{2009}]{Stenrup2009PhRvA}
{Stenrup} M.,  {Larson} {\AA}.,   {Elander} N.,  2009, \mn@doi [\pra]
  {10.1103/PhysRevA.79.012713}, \href
  {http://adsabs.harvard.edu/abs/2009PhRvA..79a2713S} {79, 012713}

\bibitem[\protect\citeauthoryear{{Sternberg} \& {Dalgarno}}{{Sternberg} \&
  {Dalgarno}}{1995}]{Sternberg1995}
{Sternberg} A.,  {Dalgarno} A.,  1995, \mn@doi [\apjs] {10.1086/192198}, \href
  {http://adsabs.harvard.edu/abs/1995ApJS...99..565S} {99, 565}

\bibitem[\protect\citeauthoryear{Takagi, Kosugi  \& Dourneuf}{Takagi
  et~al.}{1991}]{0953-4075-24-3-026}
Takagi H.,  Kosugi N.,   Dourneuf M.~L.,  1991, Journal of Physics B: Atomic,
  Molecular and Optical Physics, 24, 711

\bibitem[\protect\citeauthoryear{{Tan}, {Beltr{\'a}n}, {Caselli}, {Fontani},
  {Fuente}, {Krumholz}, {McKee}  \& {Stolte}}{{Tan} et~al.}{2014}]{TanPPVI}
{Tan} J.~C.,  {Beltr{\'a}n} M.~T.,  {Caselli} P.,  {Fontani} F.,  {Fuente} A.,
  {Krumholz} M.~R.,  {McKee} C.~F.,   {Stolte} A.,  2014, \mn@doi [Protostars
  and Planets VI] {10.2458/azu_uapress_9780816531240-ch007}, \href
  {http://adsabs.harvard.edu/abs/2014prpl.conf..149T} {pp 149--172}

\bibitem[\protect\citeauthoryear{{Teyssier}}{{Teyssier}}{2002}]{Teyssier2002}
{Teyssier} R.,  2002, \mn@doi [\aap] {10.1051/0004-6361:20011817}, \href
  {http://adsabs.harvard.edu/abs/2002A%26A...385..337T} {385, 337}

\bibitem[\protect\citeauthoryear{{Tielens}}{{Tielens}}{2010}]{TielensBook}
{Tielens} A.~G.~G.~M.,  2010, {The Physics and Chemistry of the Interstellar
  Medium}.
Cambridge University Press

\bibitem[\protect\citeauthoryear{{Tsang} \& {Hampson}}{{Tsang} \&
  {Hampson}}{1986a}]{Tsang1986}
{Tsang} W.,  {Hampson} R.~F.,  1986a, \mn@doi [Journal of Physical and Chemical
  Reference Data] {10.1063/1.555759}, \href
  {http://adsabs.harvard.edu/abs/1986JPCRD..15.1087T} {15, 1087}

\bibitem[\protect\citeauthoryear{{Tsang} \& {Hampson}}{{Tsang} \&
  {Hampson}}{1986b}]{1986JPCRD..15.1087T}
{Tsang} W.,  {Hampson} R.~F.,  1986b, \mn@doi [Journal of Physical and Chemical
  Reference Data] {10.1063/1.555759}, \href
  {http://adsabs.harvard.edu/abs/1986JPCRD..15.1087T} {15, 1087}

\bibitem[\protect\citeauthoryear{{Turk}, {Smith}, {Oishi}, {Skory}, {Skillman},
  {Abel}  \& {Norman}}{{Turk} et~al.}{2011}]{2011ApJS..192....9T}
{Turk} M.~J.,  {Smith} B.~D.,  {Oishi} J.~S.,  {Skory} S.,  {Skillman} S.~W.,
  {Abel} T.,   {Norman} M.~L.,  2011, \mn@doi [The Astrophysical Journal
  Supplement Series] {10.1088/0067-0049/192/1/9}, \href
  {http://adsabs.harvard.edu/abs/2011ApJS..192....9T} {192, 9}

\bibitem[\protect\citeauthoryear{{Valdivia} \& {Hennebelle}}{{Valdivia} \&
  {Hennebelle}}{2014}]{Valdivia2014}
{Valdivia} V.,  {Hennebelle} P.,  2014, \mn@doi [\aap]
  {10.1051/0004-6361/201423720}, \href
  {http://adsabs.harvard.edu/abs/2014A%26A...571A..46V} {571, A46}

\bibitem[\protect\citeauthoryear{{Vaytet} \& {Haugb{\o}lle}}{{Vaytet} \&
  {Haugb{\o}lle}}{2017}]{Vaytet2017}
{Vaytet} N.,  {Haugb{\o}lle} T.,  2017, \mn@doi [\aap]
  {10.1051/0004-6361/201628194}, \href
  {http://adsabs.harvard.edu/abs/2017A%26A...598A.116V} {598, A116}

\bibitem[\protect\citeauthoryear{{Verner} \& {Ferland}}{{Verner} \&
  {Ferland}}{1996}]{Verner1996}
{Verner} D.~A.,  {Ferland} G.~J.,  1996, \mn@doi [\apjs] {10.1086/192284},
  \href {http://adsabs.harvard.edu/abs/1996ApJS..103..467V} {103, 467}

\bibitem[\protect\citeauthoryear{{Verner}, {Ferland}, {Korista}  \&
  {Yakovlev}}{{Verner} et~al.}{1996}]{Verner1996ApJ}
{Verner} D.~A.,  {Ferland} G.~J.,  {Korista} K.~T.,   {Yakovlev} D.~G.,  1996,
  \mn@doi [\apj] {10.1086/177435}, \href
  {http://adsabs.harvard.edu/abs/1996ApJ...465..487V} {465, 487}

\bibitem[\protect\citeauthoryear{{Viggiano}, {Albritton}, {Fehsenfeld},
  {Adams}, {Smith}  \& {Howorka}}{{Viggiano}
  et~al.}{1980}]{1980ApJ...236..492V}
{Viggiano} A.~A.,  {Albritton} D.~L.,  {Fehsenfeld} F.~C.,  {Adams} N.~G.,
  {Smith} D.,   {Howorka} F.,  1980, \mn@doi [\apj] {10.1086/157766}, \href
  {http://adsabs.harvard.edu/abs/1980ApJ...236..492V} {236, 492}

\bibitem[\protect\citeauthoryear{{Visser}, {van Dishoeck}  \& {Black}}{{Visser}
  et~al.}{2009}]{Visser2009A&A}
{Visser} R.,  {van Dishoeck} E.~F.,   {Black} J.~H.,  2009, \mn@doi [\aap]
  {10.1051/0004-6361/200912129}, \href
  {http://adsabs.harvard.edu/abs/2009A%26A...503..323V} {503, 323}

\bibitem[\protect\citeauthoryear{{Voronov}}{{Voronov}}{1997}]{Voronov1997}
{Voronov} G.~S.,  1997, \mn@doi [Atomic Data and Nuclear Data Tables]
  {10.1006/adnd.1997.0732}, \href
  {http://adsabs.harvard.edu/abs/1997ADNDT..65....1V} {65, 1}

\bibitem[\protect\citeauthoryear{{Wakelam} et~al.,}{{Wakelam}
  et~al.}{2012}]{Wakelam2012ApJS}
{Wakelam} V.,  et~al., 2012, \mn@doi [\apjs] {10.1088/0067-0049/199/1/21},
  \href {http://adsabs.harvard.edu/abs/2012ApJS..199...21W} {199, 21}

\bibitem[\protect\citeauthoryear{{Walch} et~al.,}{{Walch}
  et~al.}{2015}]{Walch2015}
{Walch} S.,  et~al., 2015, \mn@doi [\mnras] {10.1093/mnras/stv1975}, \href
  {http://adsabs.harvard.edu/abs/2015MNRAS.454..238W} {454, 238}

\bibitem[\protect\citeauthoryear{Walkauskas \& Kaufman}{Walkauskas \&
  Kaufman}{1975}]{WALKAUSKAS1975691}
Walkauskas L.,  Kaufman F.,  1975, \mn@doi [Symposium (International) on
  Combustion] {https://doi.org/10.1016/S0082-0784(75)80339-0}, 15, 691

\bibitem[\protect\citeauthoryear{Warnatz}{Warnatz}{1984}]{Warnatz1984}
Warnatz J.,  1984, Rate Coefficients in the C/H/O System.
Springer US, New York, NY, pp 197--360, \mn@doi{10.1007/978-1-4684-0186-8\_5},
  \url {https://doi.org/10.1007/978-1-4684-0186-8\_5}

\bibitem[\protect\citeauthoryear{{Wiersma}, {Schaye}  \& {Smith}}{{Wiersma}
  et~al.}{2009}]{Wiersma2009MNRAS}
{Wiersma} R.~P.~C.,  {Schaye} J.,   {Smith} B.~D.,  2009, \mn@doi [\mnras]
  {10.1111/j.1365-2966.2008.14191.x}, \href
  {http://adsabs.harvard.edu/abs/2009MNRAS.393...99W} {393, 99}

\bibitem[\protect\citeauthoryear{{Wise} \& {Abel}}{{Wise} \&
  {Abel}}{2011}]{Wise2011MNRAS}
{Wise} J.~H.,  {Abel} T.,  2011, \mn@doi [\mnras]
  {10.1111/j.1365-2966.2011.18646.x}, \href
  {http://adsabs.harvard.edu/abs/2011MNRAS.414.3458W} {414, 3458}

\bibitem[\protect\citeauthoryear{{Woodall}, {Ag{\'u}ndez}, {Markwick-Kemper}
  \& {Millar}}{{Woodall} et~al.}{2007}]{2007A&A...466.1197W}
{Woodall} J.,  {Ag{\'u}ndez} M.,  {Markwick-Kemper} A.~J.,   {Millar} T.~J.,
  2007, \mn@doi [\aap] {10.1051/0004-6361:20064981}, \href
  {http://adsabs.harvard.edu/abs/2007A%26A...466.1197W} {466, 1197}

\bibitem[\protect\citeauthoryear{{Yoshida}, {Omukai}, {Hernquist}  \&
  {Abel}}{{Yoshida} et~al.}{2006}]{Yoshida2006ApJ}
{Yoshida} N.,  {Omukai} K.,  {Hernquist} L.,   {Abel} T.,  2006, \mn@doi [\apj]
  {10.1086/507978}, \href {http://adsabs.harvard.edu/abs/2006ApJ...652....6Y}
  {652, 6}

\bibitem[\protect\citeauthoryear{{Zhao} et~al.,}{{Zhao}
  et~al.}{2004}]{Zhao2004}
{Zhao} L.~B.,  et~al., 2004, \mn@doi [\apj] {10.1086/424729}, \href
  {http://adsabs.harvard.edu/abs/2004ApJ...615.1063Z} {615, 1063}

\bibitem[\protect\citeauthoryear{{van Dishoeck}}{{van
  Dishoeck}}{1987}]{vanDishoeck1987}
{van Dishoeck} E.~F.,  1987, in {Vardya} M.~S.,  {Tarafdar} S.~P.,  eds,  IAU
  Symposium Vol. 120, Astrochemistry. pp 51--63

\bibitem[\protect\citeauthoryear{{van Dishoeck}, {Jonkheid}  \& {van
  Hemert}}{{van Dishoeck} et~al.}{2006}]{vanDishoeck2006}
{van Dishoeck} E.~F.,  {Jonkheid} B.,   {van Hemert} M.~C.,  2006, \mn@doi
  [Faraday Discussions] {10.1039/b517564j}, \href
  {http://adsabs.harvard.edu/abs/2006FaDi..133..231V} {133, 231}

\bibitem[\protect\citeauthoryear{{van de Hulst}}{{van de
  Hulst}}{1987}]{Hulst1987}
{van de Hulst} H.~C.,  1987, \aap, \href
  {http://adsabs.harvard.edu/abs/1987A%26A...173..115V} {173, 115}

\makeatother
\end{thebibliography}

\appendix

\section{Chemical network in Test~4 and Test~5}\label{sec:chemical-network-test-4}

    \begin{chemical_network_table}{\chemicalNetworkTableCaption
\label{\chemicalNetworkTableLabel}}
  \begin{chemical_network_tabular}

1 & \ch{H + e- -> H+ + e- + e-} & k$_{1}$ & $\begin{aligned}[t] = &\operatorname{exp}\Bigl(-32.71396 \\ 
 &+13.53655\,\ln(T_{\mathrm{e}}) \\ 
 &-5.73932\,{\ln(T_{\mathrm{e}})}^2 \\ 
 &+1.56315\,{\ln(T_{\mathrm{e}})}^3 \\ 
 &-0.28770\,{\ln(T_{\mathrm{e}})}^4 \\ 
 &+0.03482\,{\ln(T_{\mathrm{e}})}^5 \\ 
 &-0.00263\,{\ln(T_{\mathrm{e}})}^6 \\ 
 &+0.00011\,{\ln(T_{\mathrm{e}})}^7 \\ 
 &-2.03914\times10^{-6}\,{\ln(T_{\mathrm{e}})}^8\Bigr)\end{aligned}$ & $$ & 1 \\

2 & \ch{H+ + e- -> H + $\gamma$} & k$_{2}$ & $ = 3.92\times10^{-13}\,T_{\mathrm{e}}^{-0.6353}$ & $2.73 < T \leqslant 5500 \, \mathrm{K}$ & 2 \\

 &  &  & $\begin{aligned}[t] = &\operatorname{exp}\Bigl(-28.61303 \\ 
 &-0.72411\,{\ln(T_{\mathrm{e}}}) \\ 
 &-0.02026\,{\ln(T_{\mathrm{e}})}^2 \\ 
 &-0.00238\,{\ln(T_{\mathrm{e}})}^3 \\ 
 &-0.00032\,{\ln(T_{\mathrm{e}})}^4 \\ 
 &-1.42150\times10^{-5}\,{\ln(T_{\mathrm{e}})}^5 \\ 
 &+4.98910\times10^{-6}\,{\ln(T_{\mathrm{e}})}^6 \\ 
 &+5.75561\times10^{-7}\,{\ln(T_{\mathrm{e}})}^7 \\ 
 &-1.85676\times10^{-8}\,{\ln(T_{\mathrm{e}})}^8 \\ 
 &-3.07113\times10^{-9}\,{\ln(T_{\mathrm{e}})}^9\Bigr)\end{aligned}$ & $5500 < T < \num{e8} \, \mathrm{K}$ &  \\

3 & \ch{He + e- -> He+ + e- + e-} & k$_{3}$ & $\begin{aligned}[t] = &\operatorname{exp}\Bigl(-44.09864 \\ 
 &+23.91596\,{\ln(T_{\mathrm{e}}}) \\ 
 &-10.75323\,{\ln(T_{\mathrm{e}})}^2 \\ 
 &+3.05803\,{\ln(T_{\mathrm{e}})}^3 \\ 
 &-0.56851\,{\ln(T_{\mathrm{e}})}^4 \\ 
 &+0.06795\,{\ln(T_{\mathrm{e}})}^5 \\ 
 &-0.00500\,{\ln(T_{\mathrm{e}})}^6 \\ 
 &+0.00020\,{\ln(T_{\mathrm{e}})}^7 \\ 
 &-3.64916\times10^{-6}\,{\ln(T_{\mathrm{e}})}^8\Bigr)\end{aligned}$ & $$ & 1 \\

4 & \ch{He+ + e- -> He + $\gamma$} & k$_{4}$ & $ = 3.92\times10^{-13}\,T_{\mathrm{e}}^{-0.6353}$ & $2.73 < T \leqslant 9280 \, \mathrm{K}$ & 3 \\

 &  &  & $\begin{aligned}[t] = &1.54\times10^{-9}\Bigl(1+
 0.3\,\operatorname{exp}\Bigl(-\frac{8.09932}{T_{\mathrm{e}}}\Bigr)\Bigr)\\
&  \times \operatorname{exp}\Bigl(-\frac{40.49664}{T_{\mathrm{e}}}\Bigr)\,T_{\mathrm{e}}^{-1.5} \\ 
 &+3.92\times10^{-13}\,T_{\mathrm{e}}^{-0.6353}\end{aligned}$ & $9280 < T < \num{e8} \, \mathrm{K}$ &  \\

5 & \ch{He+ + e- -> He^{++} + e- + e-} & k$_{5}$ & $\begin{aligned}[t] = &\operatorname{exp}\Bigl(-68.71040 \\ 
 &+43.93347\,{\ln(T_{\mathrm{e}})} \\ 
 &-18.48066\,{\ln(T_{\mathrm{e}})}^2 \\ 
 &+4.70162\,{\ln(T_{\mathrm{e}})}^3 \\ 
 &-0.76924\,{\ln(T_{\mathrm{e}})}^4 \\ 
 &+0.08113\,{\ln(T_{\mathrm{e}})}^5 \\ 
 &-0.00532\,{\ln(T_{\mathrm{e}})}^6 \\ 
 &+0.00019\,{\ln(T_{\mathrm{e}})}^7 \\ 
 &-3.16558\times10^{-6}\,{\ln(T_{\mathrm{e}})}^8\Bigr)\end{aligned}$ & $$ & 4 \\

6 & \ch{He+ + H -> He + H+} & k$_{6}$ & $ = 3\times10^{-16}(T/300)$ & $$ & 5 \\

7 & \ch{He + H+ -> He+ + H} & k$_{7}$ & $ = 1.26\times10^{-9}\,{T}^{-0.75} \operatorname{exp}\left(-\frac{127500}{T}\right)$ & $T < \num{e4} \, \mathrm{K}$ & 6 \\

 &  &  & $ = 4\times10^{-37}{T}^{4.74}$ & $T > \num{e4} \, \mathrm{K}$ &  \\

8 & \ch{H2 + He -> H + H + He} & k$_{8}$ & $ = k_{\mathrm{h,He}}^{(1 - a_\mathrm{He})} 
k_{\mathrm{l,He}}^{a_\mathrm{He}}$ & $$ & 88 \\
  \end{chemical_network_tabular}
\end{chemical_network_table}

\begin{chemical_network_table}{\chemicalNetworkTableCaptionCont}
  \begin{chemical_network_tabular}

9 & \ch{H2 + He+ -> He + H2+} & k$_{9}$ & $ = 7.2\times10^{-15}$ & $$ & 8 \\

10 & \ch{H2 + He+ -> He + H + H+} & k$_{10}$ & $ = 3.7\times10^{-14}\operatorname{exp}\left(-\frac{35}{T}\right)$ & $$ & 8 \\

11 & \ch{H2 + He+ -> He+ + H + H} & k$_{11}$ & $ = 3\times10^{-11}\sqrt{(T/300)} \operatorname{exp}\left(-\frac{52000}{T}\right)$ & $$ & 9 \\

12 & \ch{He^{++} + e- -> He+ + $\gamma$} & k$_{12}$ & $\begin{aligned}[t] = &\frac{1.891\times10^{-10}}{\sqrt{\frac{T}{9.37}} {\Bigl(1+\sqrt{\frac{T}{9.37}}\Bigr)}^{0.2476} {\Bigl(1+\sqrt{\frac{T}{2774000}}\Bigr)}^{1.7524}}\end{aligned}$ & $$ & 10 \\

13 & \ch{H + e- -> H- + $\gamma$} & k$_{13}$ & $ = 1.4\times10^{-18}\,{T}^{0.928} \operatorname{exp}\left(-\frac{T}{16200}\right)$ & $$ & 10 \\

14 & \ch{H- + H -> H2 + e-} & k$_{14}$ & $\begin{aligned}[t] = &\frac{1.35\times10^{-9}\Bigl({T}^{0.09849}+
0.32852\,{T}^{0.5561}+2.771\times10^{-7}\,{T}^{2.1826}\Bigr)}{1+0.00619\,{T}^{1.0461}+8.9712\times10^{-11}\,{T}^{3.0424}+3.2576\times10^{-14}\, {T}^{3.7741}}\end{aligned}$ & $$ & 11 \\

15 & \ch{H + H+ -> H2+ + $\gamma$} & k$_{15}$ & $ = 2.1\times10^{-20}\,{\left(T/30\right)}^{-0.15}$ & $T < 30 \, \mathrm{K}$ & 12 \\

 &  &  & $ = {10}\, \hat{ }\left(-18.2-3.194\,\log(T)+1.786\,{\log(T)}^2-0.2072\,{\log(T)}^3\right)$ & $T \geqslant 30 \, \mathrm{K}$ &  \\

16 & \ch{H2+ + H -> H2 + H+} & k$_{16}$ & $ = 6\times10^{-10}$ & $$ & 13 \\

17 & \ch{H2 + H+ -> H2+ + H} & k$_{17}$ & $\begin{aligned}[t] = &{10}\, \hat{ }\, \Bigl(-8875.5774 \\ 
 &+10081.246\,\log(T) \\ 
 &-4860.6622\,{\log(T)}^2 \\ 
 &+1288.9659\,{\log(T)}^3 \\ 
 &-203.19575\,{\log(T)}^4 \\ 
 &+19.05749\,{\log(T)}^5 \\ 
 &-0.98530\,{\log(T)}^6 \\ 
 &+0.02167\,{\log(T)}^7\Bigr)\end{aligned}$ & $T \geqslant \num{e5} \, \mathrm{K}$ & 14 \\

18 & \ch{H2 + e- -> H + H + e-} & k$_{18}$ & $ = 4.38\times10^{-10}\operatorname{exp}\left(-\frac{102000}{T}\right) {T}^{0.35}$ & $$ & 15 \\

19 & \ch{H2 + H -> H + H + H} & k$_{19}$ & $= {10}\, \hat{ }\, \Bigl(\log(k_{\mathrm{h,H}})
 -\frac{\log(k_{\mathrm{h,H}}) - \log(k_{\mathrm{l,H}})}{1+\frac{n_{\mathrm{H}_\mathrm{tot}}}{n_{\mathrm{cr,H}}}}\Bigr)$ &  & 16 \\

20 & \ch{H- + e- -> H + e- + e-} & k$_{20}$ & $\begin{aligned}[t] = &\operatorname{exp}\Bigl(-18.01849 \\ 
 &+2.36085\,\ln(T_{\mathrm{e}}) \\ 
 &-0.28274\,{\ln(T_{\mathrm{e}})}^2 \\ 
 &+0.01623\,{\ln(T_{\mathrm{e}})}^3 \\ 
 &-0.03365\,{\ln(T_{\mathrm{e}})}^4 \\ 
 &+0.01178\,{\ln(T_{\mathrm{e}})}^5 \\ 
 &-0.00165\,{\ln(T_{\mathrm{e}})}^6 \\ 
 &+0.00010\,{\ln(T_{\mathrm{e}})}^7 \\ 
 &-2.63128\times10^{-6}\,{\ln(T_{\mathrm{e}})}^8\Bigr)\end{aligned}$ & $$ & 1 \\

21 & \ch{H- + H -> H + H + e-} & k$_{21}$ & $ = 2.56\times10^{-9}\,{T_{\mathrm{e}}}^{1.78186}$ & $T \leqslant 1160 \, \mathrm{K}$ & 17 \\

 &  &  & $\begin{aligned}[t] = &\operatorname{exp}\Bigl(-20.37260 \\ 
 &+1.13944\,\ln(T_{\mathrm{e}}) \\ 
 &-0.14210\,{\ln(T_{\mathrm{e}})}^2 \\ 
 &+0.00846\,{\ln(T_{\mathrm{e}})}^3 \\ 
 &-0.00143\,{\ln(T_{\mathrm{e}})}^4 \\ 
 &+0.00020\,{\ln(T_{\mathrm{e}})}^5 \\ 
 &+8.66396\times10^{-5}\,{\ln(T_{\mathrm{e}})}^6 \\ 
 &-2.58500\times10^{-5}\,{\ln(T_{\mathrm{e}})}^7 \\ 
 &+2.45550\times10^{-6}\,{\ln(T_{\mathrm{e}})}^8 \\ 
 &-8.06838\times10^{-8}\,{\ln(T_{\mathrm{e}})}^9\Bigr)\end{aligned}$ & $T > 1160 \, \mathrm{K}$ &  \\

22 & \ch{H- + H+ -> H + H} & k$_{22}$ & $ = \frac{2.96\times10^{-6}}{\sqrt{T}}-1.73\times10^{-9}+2.5\times10^{-10}\sqrt{T}$ & $$ & 18 \\

23 & \ch{H- + H+ -> H2+ + e-} & k$_{23}$ & $ = 10^{-8}\,{T}^{-0.4}$ & $$ & 19 \\

24 & \ch{H2+ + e- -> H + H} & k$_{24}$ & $ = 10^{-8}$ & $T \leqslant 617 \, \mathrm{K}$ & 20 \\

 &  &  & $ = 1.32\times10^{-6}\,{T}^{-0.76}$ & $T > 617 \, \mathrm{K}$ &  \\

25 & \ch{H2+ + H- -> H + H2} & k$_{25}$ & $ = 5\times10^{-7}\sqrt{\frac{100}{T}}$ & $$ & 21 \\
  \end{chemical_network_tabular}
\end{chemical_network_table}

\begin{chemical_network_table}{\chemicalNetworkTableCaptionCont}
  \begin{chemical_network_tabular}

26 & \ch{H2 + H2 -> H2 + H + H} & k$_{26}$ & $= {10}\, \hat{ }\, \Bigl(\log(k_{\mathrm{h,H}_2}) 
 -\frac{\log(k_{\mathrm{h,H}_2}) - \log(k_{\mathrm{l,H}_2})}{1+\frac{n_{\mathrm{H}_\mathrm{tot}}}{n_{\mathrm{crit,H}_2}}}\Bigr)$ & & 16 \\

27 & \ch{H + H + He -> H2 + He} & k$_{27}$ & $ = 6.9\times10^{-32}\,{T}^{-0.4}$ & $$ & 55 \\

28 & \ch{H + H + H -> H2 + H} & k$_{28}$ & $ = 6\times10^{-32}\,{T}^{-0.25}+2\times10^{-31}\,{T}^{-0.5}$ & $$ & 23 \\

29 & \ch{H2 + H + H -> H2 + H2} & k$_{29}$ & $ = 7.5\times10^{-33}\,{T}^{-0.25}+2.5\times10^{-32}\,{T}^{-0.5}$ & $$ & 24 \\

30 & \ch{C+ + e- -> C + $\gamma$} & k$_{30}$ & $ = 4.67\times10^{-12}\,{(T/300)}^{-0.6}$ & $T \leqslant 7950 \, \mathrm{K}$ & 25 \\

 &  &  & $ = 1.23\times10^{-17}\,{(T/300)}^{2.49} \operatorname{exp}\left(\frac{21845.6}{T}\right)$ & $7950 < T \leqslant \num{21140e0} \, \mathrm{K}$ &  \\

 &  &  & $ = 9.62\times10^{-8}\,{(T/300)}^{-1.37} \operatorname{exp}\left(-\frac{115786.2}{T}\right)$ & $T > \num{21140e0} \, \mathrm{K}$ &  \\

31 & \ch{O+ + e- -> O + $\gamma$} & k$_{31}$ & $ = 1.3\times10^{-10}\,{T}^{-0.64}$ & $T \leqslant 400 \, \mathrm{K}$ & 26 \\

 &  &  & $\,\begin{aligned}[t] = &1.41\times10^{-10}\,{T}^{-0.66} \\ 
 &+0.00074\,{T}^{-1.5} \operatorname{exp}\Bigl(-\frac{175000}{T}\Bigr) \\ 
 & \times\Bigl(1+0.062\,\operatorname{exp}\Bigl(-\frac{145000}{T}\Bigr)\Bigr)\end{aligned}$ & $T > 400 \, \mathrm{K}$ &  \\

32 & \ch{C + e- -> C+ + e- + e-} & k$_{32}$ & $\begin{aligned}[t] = &\frac{6.85\times10^{-8}{\Bigl(\frac{11.26}{T_{\mathrm{e}}}\Bigr)}^{0.25} \operatorname{exp}\Bigl(-\frac{11.26}{T_{\mathrm{e}}}\Bigr)}{0.193+11.26 / {T_{\mathrm{e}}}}\end{aligned}$ & $$ & 27 \\

33 & \ch{O + e- -> O+ + e- + e-} & k$_{33}$ & $\begin{aligned}[t] = &\frac{3.59\times10^{-8}{\Bigl(\frac{13.6}{T_{\mathrm{e}}}\Bigr)}^{0.34} \operatorname{exp}\Bigl(-\frac{13.6}{T_{\mathrm{e}}}\Bigr)}{0.073+13.6 / T_{\mathrm{e}}}\end{aligned}$ & $$ & 27 \\

34 & \ch{O+ + H -> O + H+} & k$_{34}$ & $ = 4.99\times10^{-11}\,{T}^{0.405}+7.54\times10^{-10}\,{T}^{-0.458}$ & $$ & 28 \\

35 & \ch{O + H+ -> O+ + H} & k$_{35}$ & $\begin{aligned}[t] = &\Bigl(1.08\times10^{-11}\,{T}^{0.517} \\ 
 &+4\times10^{-10}\,{T}^{0.00669}\Bigr) \operatorname{exp}\Bigl(-\frac{227}{T}\Bigr)\end{aligned}$ & $$ & 29 \\

36 & \ch{O + He+ -> O+ + He} & k$_{36}$ & $\begin{aligned}[t] = &4.991\times10^{-15}{\Bigl(\frac{T}{10000}\Bigr)}^{0.3794} \operatorname{exp}\Bigl(-8.9206\times10^{-7}\,T\Bigr) \\ 
 &+2.78\times10^{-15}{\Bigl(\frac{T}{10000}\Bigr)}^{-0.2163} \operatorname{exp}\Bigl(-1.2258\times10^{-6}\,T\Bigr)\end{aligned}$ & $$ & 30 \\

37 & \ch{C + H+ -> C+ + H} & k$_{37}$ & $ = 3.9\times10^{-16}\,{T}^{0.213}$ & $$ & 29 \\

38 & \ch{C+ + H -> C + H+} & k$_{38}$ & $ = 6.08\times10^{-14}{\left(\frac{T}{10000}\right)}^{1.96} \operatorname{exp}\left(-\frac{170000}{T}\right)$ & $$ & 29 \\

39 & \ch{C + He+ -> C+ + He} & k$_{39}$ & $ = 8.58\times10^{-17}\,{T}^{0.757}$ & $T \leqslant 200 \, \mathrm{K}$ & 31 \\

 &  &  & $ = 3.25\times10^{-17}\,{T}^{0.968}$ & $200 < T \leqslant 2000 \, \mathrm{K}$ &  \\

 &  &  & $ = 2.77\times10^{-19}\,{T}^{1.597}$ & $2000 < T < \num{e8} \, \mathrm{K}$ &  \\

40 & \ch{OH + H -> O + H + H} & k$_{40}$ & $ = 6\times10^{-9}\operatorname{exp}\left(-\frac{50900}{T}\right)$ & $$ & 32 \\

41 & \ch{HOC+ + H2 -> HCO+ + H2} & k$_{41}$ & $ = 3.8\times10^{-10}$ & $$ & 33 \\

42 & \ch{HOC+ + CO -> HCO+ + CO} & k$_{42}$ & $ = 1.604\times10^{-9}$ & $T < 10 \, \mathrm{K}$ & 34 \\

 &  &  & $ = 8.68\times10^{-10}\left(1+0.02427\sqrt{\frac{300}{T}}+\frac{7.1537}{T}\right)$ & $T > 10 \, \mathrm{K}$ &  \\

43 & \ch{C + H2 -> CH + H} & k$_{43}$ & $ = 6.64\times10^{-10}\operatorname{exp}\left(-\frac{11700}{T}\right)$ & $$ & 35 \\

44 & \ch{CH + H -> C + H2} & k$_{44}$ & $ = 1.31\times10^{-10}\operatorname{exp}\left(-\frac{80}{T}\right)$ & $$ & 36 \\

45 & \ch{CH + H2 -> CH2 + H} & k$_{45}$ & $ = 5.46\times10^{-10}\operatorname{exp}\left(-\frac{1943}{T}\right)$ & $$ & 32 \\

46 & \ch{CH + C -> C2 + H} & k$_{46}$ & $ = 2.4\times10^{-10}$ & $$ & 37 \\

47 & \ch{CH + O -> CO + H} & k$_{47}$ & $ = 1.02\times10^{-10}\operatorname{exp}\left(-\frac{914}{T}\right)$ & $$ & 38 \\

48 & \ch{CH + O -> HCO+ + e-} & k$_{48}$ & $ = 1.9\times10^{-11}{(T/300)}^{-2.2} \operatorname{exp}\left(-\frac{165.1}{T}\right)$ & $$ & 39 \\

49 & \ch{CH + O -> OH + C} & k$_{49}$ & $ = 2.52\times10^{-11}\operatorname{exp}\left(-\frac{2381}{T}\right)$ & $$ & 39 \\

50 & \ch{CH2 + H -> CH + H2} & k$_{50}$ & $ = 2.2\times10^{-10}$ & $$ & 40 \\

51 & \ch{CH2 + O -> CO + H + H} & k$_{51}$ & $ = 2.04\times10^{-10}\operatorname{exp}\left(-\frac{270}{T}\right)$ & $$ & 40 \\

52 & \ch{CH2 + O -> CO + H2} & k$_{52}$ & $ = 1.36\times10^{-10}\operatorname{exp}\left(-\frac{270}{T}\right)$ & $$ & 40 \\

53 & \ch{CH2 + O -> HCO + H} & k$_{53}$ & $ = 5.01\times10^{-11}$ & $$ & 40 \\

54 & \ch{CH2 + O -> CH + OH} & k$_{54}$ & $ = 4.98\times10^{-10}\operatorname{exp}\left(-\frac{6000}{T}\right)$ & $$ & 40 \\

55 & \ch{C2 + O -> CO + C} & k$_{55}$ & $ = 2\times10^{-12}\,{(T/300)}^{-0.12}$ & $T < 300 \, \mathrm{K}$ & 40 \\

 &  &  & $ = 2\times10^{-12}\,{(T/300)}^{0.757}$ & $T > 300 \, \mathrm{K}$ &  \\

56 & \ch{O + H2 -> OH + H} & k$_{56}$ & $ = 1.46\times10^{-12}\operatorname{exp}\left(-\frac{9650}{T}\right)$ & $$ & 40 \\

57 & \ch{OH + H -> O + H2} & k$_{57}$ & $ = 6.99\times10^{-14}\,{(T/300)}^{2.8} \operatorname{exp}\left(-\frac{1950}{T}\right)$ & $T < 280 \, \mathrm{K}$ & 56 \\

 &  &  & $ = 5.45\times10^{-17}$ & $T > 280 \, \mathrm{K}$ &  \\

58 & \ch{H2 + OH -> H2O + H} & k$_{58}$ & $ = 3.6\times10^{-16}\,{T}^{1.52} \operatorname{exp}\left(-\frac{1740}{T}\right)$ & $$ & 41 \\
  \end{chemical_network_tabular}
\end{chemical_network_table}

\begin{chemical_network_table}{\chemicalNetworkTableCaptionCont}
  \begin{chemical_network_tabular}

59 & \ch{C + OH -> H + CO} & k$_{59}$ & $ = 7.051\times10^{-11}$ & $T < 10 \, \mathrm{K}$ & 40 \\

 &  &  & $ = 2.25\times10^{-11}\,{(T/300)}^{-0.339} \operatorname{exp}\left(-\frac{0.108}{T}\right)$ & $T > 10 \, \mathrm{K}$ &  \\

60 & \ch{O + OH -> H + O2} & k$_{60}$ & $ = 2.4\times10^{-11}\operatorname{exp}\left(\frac{110}{T}\right)$ & $T > 150 \, \mathrm{K}$ & 40 \\

 &  &  & $ = 4.997\times10^{-11}$ & $T < 150 \, \mathrm{K}$ &  \\

61 & \ch{OH + OH -> H2O + O} & k$_{61}$ & $ = 1.65\times10^{-12}\,{(T/300)}^{1.14} \operatorname{exp}\left(-\frac{50}{T}\right)$ & $$ & 57 \\

62 & \ch{H2O + H -> H2 + OH} & k$_{62}$ & $ = 1.59\times10^{-11}\,(T/300)\times1.2 \operatorname{exp}\left(-\frac{9610}{T}\right)$ & $$ & 58 \\

63 & \ch{O2 + H -> OH + O} & k$_{63}$ & $ = 3.13\times10^{-10} \operatorname{exp}\left(-\frac{8156}{T}\right)$ & $$ & 22 \\

64 & \ch{O2 + H2 -> OH + OH} & k$_{64}$ & $ = 3.16\times10^{-10}\operatorname{exp}\left(-\frac{21890}{T}\right)$ & $$ & 59 \\

65 & \ch{O2 + C -> CO + O} & k$_{65}$ & $ = 4.7\times10^{-11}\,{(T/300)}^{-0.34}$ & $T < 1052 \, \mathrm{K}$ & 57 \\

 &  &  & $ = 2.48\times10^{-12}\,{(T/300)}^{1.54} \operatorname{exp}\left(\frac{613}{T}\right)$ & $T > 1052 \, \mathrm{K}$ & 22 \\

66 & \ch{CO + H -> C + OH} & k$_{66}$ & $ = 1.1\times10^{-10}\,{(T/300)}^{0.5} \operatorname{exp}\left(-\frac{77700}{T}\right)$ & $$ & 32 \\

67 & \ch{H2+ + H2 -> H3+ + H} & k$_{67}$ & $ = 2.24\times10^{-9}\,{(T/300)}^{0.042} \operatorname{exp}\left(-\frac{T}{46600}\right)$ & $$ & 60 \\

68 & \ch{H3+ + H -> H2+ + H2} & k$_{68}$ & $ = 7.7\times10^{-9}\operatorname{exp}\left(-\frac{17560}{T}\right)$ & $$ & 61 \\

69 & \ch{C + H2+ -> CH+ + H} & k$_{69}$ & $ = 2.4\times10^{-9}$ & $$ & 32 \\

70 & \ch{C + H3+ -> CH+ + H2} & k$_{70}$ & $\begin{aligned}[t] = &\frac{1.0218\times10^{-9}+7.2733\times10^{-11}\sqrt{T}+5.9203\times10^{-14}\,T}{{T}^{0.1667}+0.04491\sqrt{T}-5.9203\times10^{-14}\,T+2.6397\times10^{-6}\,{T}^{1.5}}\end{aligned}$ & $$ & 42 \\

71 & \ch{C + H3+ -> CH2+ + H} & k$_{71}$ & $\begin{aligned}[t] = &\frac{8.5145\times10^{-10}}{{T}^{0.1667}+0.00095\sqrt{T}-4.404\times10^{-5}\,T+2.3496\times10^{-6}\,{T}^{1.5}}\end{aligned}$ & $$ & 42 \\

72 & \ch{C+ + H2 -> CH+ + H} & k$_{72}$ & $ = 10^{-10}\operatorname{exp}\left(-\frac{4640}{T}\right)$ & $$ & 62 \\

73 & \ch{CH+ + H -> C+ + H2} & k$_{73}$ & $ = 7.5\times10^{-10}$ & $$ & 63 \\

74 & \ch{CH+ + H2 -> CH2+ + H} & k$_{74}$ & $ = 1.2\times10^{-9}$ & $$ & 63 \\

75 & \ch{CH+ + O -> CO+ + H} & k$_{75}$ & $ = 3.5\times10^{-10}$ & $$ & 64 \\

76 & \ch{CH2+ + H -> CH+ + H2} & k$_{76}$ & $ = 10^{-9}\operatorname{exp}\left(-\frac{7080}{T}\right)$ & $$ & 64 \\

77 & \ch{CH2+ + H2 -> CH3+ + H} & k$_{77}$ & $ = 1.6\times10^{-9}$ & $$ & 65 \\

78 & \ch{CH2+ + O -> HCO+ + H} & k$_{78}$ & $ = 7.5\times10^{-10}$ & $$ & 64 \\

79 & \ch{CH3+ + H -> CH2+ + H2} & k$_{79}$ & $ = 7\times10^{-10}\operatorname{exp}\left(-\frac{10560}{T}\right)$ & $$ & 64 \\

80 & \ch{CH3+ + O -> HOC+ + H2} & k$_{80}$ & $ = 2.5\times10^{-10}$ & $$ & 40 \\

81 & \ch{CH3+ + O -> HCO+ + H2} & k$_{81}$ & $ = 2.5\times10^{-10}$ & $$ & 40 \\

82 & \ch{C2 + O+ -> CO+ + C} & k$_{82}$ & $ = 4.8\times10^{-10}$ & $$ & 40 \\

83 & \ch{O+ + H2 -> H + OH+} & k$_{83}$ & $ = 1.69\times10^{-9}$ & $$ & 40 \\

84 & \ch{O + H2+ -> H + OH+} & k$_{84}$ & $ = 1.5\times10^{-9}$ & $$ & 64 \\

85 & \ch{O + H3+ -> H2 + OH+} & k$_{85}$ & $ = 7.98\times10^{-10}\,{(T/300)}^{-0.156} \operatorname{exp}\left(-\frac{1.41}{T}\right)$ & $$ & 40 \\

86 & \ch{O + H3+ -> H + H2O+} & k$_{86}$ & $ = 3.42\times10^{-10}\,{(T/300)}^{-0.156} \operatorname{exp}\left(-\frac{1.41}{T}\right)$ & $$ & 40 \\

87 & \ch{OH + H3+ -> H2 + H2O+} & k$_{87}$ & $ = 2.277\times10^{-8}$ & $T < 10 \, \mathrm{K}$ & 40 \\

 &  &  & $ = 1.52\times10^{-9}\left(0.62+2.62185\,{\left(T/300\right)}^{-0.5}\right)$ & $T > 10 \, \mathrm{K}$ &  \\

88 & \ch{OH + C+ -> H + CO+} & k$_{88}$ & $ = 1.371\times10^{-8}$ & $T < 10 \, \mathrm{K}$ & 40 \\

 &  &  & $ = 9.15\times10^{-10}\left(0.62+2.62185\,{\left(T/300\right)}^{-0.5}\right)$ & $T > 10 \, \mathrm{K}$ &  \\

89 & \ch{OH+ + H2 -> H2O+ + H} & k$_{89}$ & $ = 1.01\times10^{-9}$ & $$ & 66 \\

90 & \ch{H2O+ + H2 -> H3O+ + H} & k$_{90}$ & $ = 6.4\times10^{-10}$ & $$ & 67 \\

91 & \ch{H2O + H3+ -> H2 + H3O+} & k$_{91}$ & $ = 2.55\times10^{-8}$ & $T < 10 \, \mathrm{K}$ & 40 \\

 &  &  & $ = 1.73\times10^{-9}\left(0.62+2.57894\,{\left(T/300\right)}^{-0.5}\right)$ & $T > 10 \, \mathrm{K}$ &  \\

92 & \ch{H2O + C+ -> HOC+ + H} & k$_{92}$ & $ = 1.8\times10^{-9}$ & $$ & 40 \\

93 & \ch{H2O + C+ -> HCO+ + H} & k$_{93}$ & $ = 5.027\times10^{-9}$ & $T < 10 \, \mathrm{K}$ & 40 \\

 &  &  & $ = 3.4093\times10^{-10}\left(0.62+2.57894\,{\left(T/300\right)}^{-0.5}\right)$ & $T > 10 \, \mathrm{K}$ &  \\

94 & \ch{H2O + C+ -> H2O+ + C} & k$_{94}$ & $ = 2.4\times10^{-10}$ & $$ & 40 \\

95 & \ch{H3O+ + C -> HCO+ + H2} & k$_{95}$ & $ = 10^{-11}$ & $$ & 64 \\

96 & \ch{O2 + C+ -> CO+ + O} & k$_{96}$ & $ = 3.42\times10^{-10}$ & $$ & 40 \\

97 & \ch{O2 + C+ -> CO + O+} & k$_{97}$ & $ = 4.53\times10^{-10}$ & $$ & 40 \\

98 & \ch{O2 + CH2+ -> HCO+ + OH} & k$_{98}$ & $ = 9.1\times10^{-10}$ & $$ & 65 \\

99 & \ch{C + O2+ -> O + CO+} & k$_{99}$ & $ = 5.2\times10^{-11}$ & $$ & 32 \\

100 & \ch{C + O2+ -> O2 + C+} & k$_{100}$ & $ = 5.2\times10^{-11}$ & $$ & 40 \\
  \end{chemical_network_tabular}
\end{chemical_network_table}

\begin{chemical_network_table}{\chemicalNetworkTableCaptionCont}
  \begin{chemical_network_tabular}

101 & \ch{CO + H3+ -> H2 + HCO+} & k$_{101}$ & $ = 2.468\times10^{-9}$ & $T < 10 \, \mathrm{K}$ & 40 \\

 &  &  & $ = 1.88055\times10^{-9}\left(1+0.02427\,{\left(T/300\right)}^{-0.5}+\frac{1.79558}{T}\right)$ & $T > 10 \, \mathrm{K}$ &  \\

102 & \ch{CO + H3+ -> H2 + HOC+} & k$_{102}$ & $ = 1.421\times10^{-10}$ & $T < 10 \, \mathrm{K}$ & 40 \\

 &  &  & $ = 1.08256\times10^{-10}\left(1+0.02427\,{\left(T/300\right)}^{-0.5}+\frac{1.79558}{T}\right)$ & $T > 10 \, \mathrm{K}$ &  \\

103 & \ch{HCO+ + C -> CO + CH+} & k$_{103}$ & $ = 1.1\times10^{-9}$ & $$ & 40 \\

104 & \ch{HCO+ + H2O -> CO + H3O+} & k$_{104}$ & $ = 7.279\times10^{-8}$ & $T < 10 \, \mathrm{K}$ & 40 \\

 &  &  & $ = 8.34\times10^{-10}\left(1+0.5232\,{\left(T/300\right)}^{-0.5}+\frac{834.16588}{T}\right)$ & $T > 10 \, \mathrm{K}$ &  \\

105 & \ch{CH + H+ -> CH+ + H} & k$_{105}$ & $ = 3.297\times10^{-8}$ & $T < 10 \, \mathrm{K}$ & 40 \\

 &  &  & $ = 3.54\times10^{-9}\left(0.62+1.58741\,{\left(T/300\right)}^{-0.5}\right)$ & $T > 10 \, \mathrm{K}$ &  \\

106 & \ch{CH2 + H+ -> H2 + CH+} & k$_{106}$ & $ = 1.765\times10^{-9}\left(0.62+0.67214\,{\left(T/300\right)}^{-0.5}\right)$ & $T < 150 \, \mathrm{K}$ & 40 \\

 &  &  & $ = 1.765\times10^{-9}\left(1+0.13634\,{\left(T/300\right)}^{-0.5}+\frac{56.66255}{T}\right)$ & $T > 150 \, \mathrm{K}$ &  \\

107 & \ch{CH2 + H+ -> H + CH2+} & k$_{107}$ & $ = 1.765\times10^{-9}\left(0.62+0.67214\,{\left(T/300\right)}^{-0.5}\right)$ & $T < 150 \, \mathrm{K}$ & 40 \\

 &  &  & $ = 1.765\times10^{-9}\left(1+0.13634\,{\left(T/300\right)}^{-0.5}+\frac{56.66255}{T}\right)$ & $T > 150 \, \mathrm{K}$ &  \\

108 & \ch{CH2 + He+ -> He + H2 + C+} & k$_{108}$ & $ = 9.65\times10^{-10}\left(0.62+0.67214\,{\left(T/300\right)}^{-0.5}\right)$ & $T < 150 \, \mathrm{K}$ & 40 \\

 &  &  & $ = 9.65\times10^{-10}\left(1+0.13634\,{\left(T/300\right)}^{-0.5}+\frac{56.66254}{T}\right)$ & $T > 150 \, \mathrm{K}$ &  \\

109 & \ch{CH2 + He+ -> He + H + CH+} & k$_{109}$ & $ = 9.65\times10^{-10}\left(0.62+0.67214\,{\left(T/300\right)}^{-0.5}\right)$ & $T < 150 \, \mathrm{K}$ & 40 \\

 &  &  & $ = 9.65\times10^{-10}\left(1+0.13634\,{\left(T/300\right)}^{-0.5}+\frac{56.66254}{T}\right)$ & $T > 150 \, \mathrm{K}$ &  \\

110 & \ch{C2 + He+ -> C+ + C + He} & k$_{110}$ & $ = 1.6\times10^{-9}$ & $$ & 40 \\

111 & \ch{OH + H+ -> OH+ + H} & k$_{111}$ & $ = 3.745\times10^{-8}$ & $T < 10 \, \mathrm{K}$ & 40 \\

 &  &  & $ = 2.5\times10^{-9}\left(0.62+2.62185\,{\left(T/300\right)}^{-0.5}\right)$ & $T > 10 \, \mathrm{K}$ &  \\

112 & \ch{OH + He+ -> O+ + He + H} & k$_{112}$ & $ = 2.022\times10^{-8}$ & $T < 10 \, \mathrm{K}$ & 40 \\

 &  &  & $ = 1.35\times10^{-9}\left(0.62+2.62185\,{\left(T/300\right)}^{-0.5}\right)$ & $T > 10 \, \mathrm{K}$ &  \\

113 & \ch{H2O + H+ -> H + H2O+} & k$_{113}$ & $ = 4.202\times10^{-8}$ & $T < 10 \, \mathrm{K}$ & 40 \\

 &  &  & $ = 2.85\times10^{-9}\left(0.62+2.57894\,{\left(T/300\right)}^{-0.5}\right)$ & $T > 10 \, \mathrm{K}$ &  \\

114 & \ch{H2O + He+ -> He + OH + H+} & k$_{114}$ & $ = 7.562\times10^{-9}$ & $T < 10 \, \mathrm{K}$ & 40 \\

 &  &  & $ = 5.1282\times10^{-10}\left(0.62+2.57894\,{\left(T/300\right)}^{-0.5}\right)$ & $T > 10 \, \mathrm{K}$ &  \\

115 & \ch{H2O + He+ -> He + OH+ + H} & k$_{115}$ & $ = 7.562\times10^{-9}$ & $T < 10 \, \mathrm{K}$ & 40 \\

 &  &  & $ = 5.1282\times10^{-10}\left(0.62+2.57894\,{\left(T/300\right)}^{-0.5}\right)$ & $T > 10 \, \mathrm{K}$ &  \\

116 & \ch{H2O + He+ -> He + H2O+} & k$_{116}$ & $ = 7.56\times10^{-9}$ & $T < 10 \, \mathrm{K}$ & 40 \\

 &  &  & $ = 5.1282\times10^{-10}\left(0.62+2.57894\,{\left(T/300\right)}^{-0.5}\right)$ & $T > 10 \, \mathrm{K}$ &  \\

117 & \ch{O2 + H+ -> O2+ + H} & k$_{117}$ & $ = 2\times10^{-9}$ & $$ & 68 \\

118 & \ch{O2 + He+ -> O2+ + He} & k$_{118}$ & $ = 3.3\times10^{-11}$ & $$ & 65 \\

119 & \ch{O2 + He+ -> O+ + He + O} & k$_{119}$ & $ = 1.1\times10^{-9}$ & $$ & 65 \\

120 & \ch{CO + He+ -> C+ + He + O} & k$_{120}$ & $ = 1.4\times10^{-9}{(T/300)}^{-0.5}$ & $$ & 51 \\

121 & \ch{CO + He+ -> C + He + O+} & k$_{121}$ & $ = 1.4\times10^{-16}{(T/300)}^{-0.5}$ & $$ & 51 \\

122 & \ch{CO+ + H -> CO + H+} & k$_{122}$ & $ = 7.5\times10^{-10}$ & $$ & 69 \\

123 & \ch{C- + H+ -> C + H} & k$_{123}$ & $ = 2.3\times10^{-7}{(T/300)}^{-0.5}$ & $$ & 32 \\

124 & \ch{O- + H+ -> O + H} & k$_{124}$ & $ = 2.3\times10^{-7}{(T/300)}^{-0.5}$ & $$ & 32 \\

125 & \ch{He+ + H- -> H + He} & k$_{125}$ & $ = 2.3\times10^{-7}\,{(T/300)}^{-0.5}$ & $$ & 70 \\

126 & \ch{H3+ + e- -> H2 + H} & k$_{126}$ & $ = 2.34\times10^{-8}\,{(T/300)}^{-0.52}$ & $$ & 71 \\

127 & \ch{H3+ + e- -> H + H + H} & k$_{127}$ & $ = 4.36\times10^{-8}\,{(T/300)}^{-0.52}$ & $$ & 71 \\

128 & \ch{CH+ + e- -> C + H} & k$_{128}$ & $ = 7\times10^{-8}\,{(T/300)}^{-0.5}$ & $$ & 72 \\

129 & \ch{CH2+ + e- -> CH + H} & k$_{129}$ & $ = 1.6\times10^{-7}\,{(T/300)}^{-0.6}$ & $$ & 73 \\

130 & \ch{CH2+ + e- -> C + H2} & k$_{130}$ & $ = 7.68\times10^{-8}\,{(T/300)}^{-0.6}$ & $$ & 73 \\

131 & \ch{CH2+ + e- -> C + H + H} & k$_{131}$ & $ = 4.03\times10^{-7}\,{(T/300)}^{-0.6}$ & $$ & 73 \\

132 & \ch{CH3+ + e- -> CH2 + H} & k$_{132}$ & $ = 7.75\times10^{-8}\,{(T/300)}^{-0.5}$ & $$ & 74 \\

133 & \ch{CH3+ + e- -> CH + H2} & k$_{133}$ & $ = 1.95\times10^{-7}\,{(T/300)}^{-0.5}$ & $$ & 74 \\

134 & \ch{CH3+ + e- -> CH + H + H} & k$_{134}$ & $ = 2\times10^{-7}\,{(T/300)}^{-0.5}$ & $$ & 32 \\

135 & \ch{OH+ + e- -> O + H} & k$_{135}$ & $ = 6.3\times10^{-9}\,{(T/300)}^{-0.48}$ & $$ & 75 \\

136 & \ch{H2O+ + e- -> O + H2} & k$_{136}$ & $ = 3.9\times10^{-8}\,{(T/300)}^{-0.5}$ & $$ & 76 \\

137 & \ch{H2O+ + e- -> OH + H} & k$_{137}$ & $ = 8.6\times10^{-8}\,{(T/300)}^{-0.5}$ & $$ & 76 \\

138 & \ch{H2O+ + e- -> O + H + H} & k$_{138}$ & $ = 3.05\times10^{-7}\,{(T/300)}^{-0.5}$ & $$ & 76 \\
  \end{chemical_network_tabular}
\end{chemical_network_table}

\begin{chemical_network_table}{\chemicalNetworkTableCaptionCont}
  \begin{chemical_network_tabular}

139 & \ch{H3O+ + e- -> OH + H + H} & k$_{139}$ & $ = 2.58\times10^{-7}\,{(T/300)}^{-0.5}$ & $$ & 77 \\

140 & \ch{H3O+ + e- -> O + H + H2} & k$_{140}$ & $ = 5.6\times10^{-9}\,{(T/300)}^{-0.5}$ & $$ & 77 \\

141 & \ch{H3O+ + e- -> H + H2O} & k$_{141}$ & $ = 1.08\times10^{-7}\,{(T/300)}^{-0.5}$ & $$ & 77 \\

142 & \ch{H3O+ + e- -> OH + H2} & k$_{142}$ & $ = 6.02\times10^{-8}\,{(T/300)}^{-0.5}$ & $$ & 77 \\

143 & \ch{O2+ + e- -> O + O} & k$_{143}$ & $ = 1.95\times10^{-7}\,{(T/300)}^{-0.7}$ & $$ & 78 \\

144 & \ch{CO+ + e- -> C + O} & k$_{144}$ & $ = 2.75\times10^{-7}\,{(T/300)}^{-0.55}$ & $$ & 79 \\

145 & \ch{HCO+ + e- -> CO + H} & k$_{145}$ & $ = 2.76\times10^{-7}\,{(T/300)}^{-0.64}$ & $$ & 80 \\

146 & \ch{HCO+ + e- -> OH + C} & k$_{146}$ & $ = 2.4\times10^{-8}\,{(T/300)}^{-0.64}$ & $$ & 80 \\

147 & \ch{HOC+ + e- -> CO + H} & k$_{147}$ & $ = 1.1\times10^{-7}\,\left(T/300\right)^{-1}$ & $$ & 32 \\

148 & \ch{H- + C -> CH + e-} & k$_{148}$ & $ = 10^{-9}$ & $$ & 32 \\

149 & \ch{H- + O -> OH + e-} & k$_{149}$ & $ = 10^{-10}$ & $$ & 32 \\

150 & \ch{H- + OH -> H2O + e-} & k$_{150}$ & $ = 5\times10^{-10}$ & $$ & 32 \\

151 & \ch{C- + H -> CH + e-} & k$_{151}$ & $ = 10^{-13}$ & $$ & 32 \\

152 & \ch{C- + H2 -> CH2 + e-} & k$_{152}$ & $ = 5\times10^{-10}$ & $$ & 32 \\

153 & \ch{C- + O -> CO + e-} & k$_{153}$ & $ = 5\times10^{-10}$ & $$ & 32 \\

154 & \ch{O- + H -> OH + e-} & k$_{154}$ & $ = 7\times10^{-10}$ & $$ & 32 \\

155 & \ch{O- + H2 -> H2O + e-} & k$_{155}$ & $ = 7\times10^{-10}$ & $$ & 32 \\

156 & \ch{O- + C -> CO + e-} & k$_{156}$ & $ = 5\times10^{-10}$ & $$ & 32 \\

157 & \ch{H2 + H+ -> H + H + H+} & k$_{157}$ & $ = 3\times10^{-11}\,{(T/300)}^{0.5} \operatorname{exp}\left(-\frac{52000}{T}\right)$ & $$ & 22 \\

158 & \ch{H2 + H+ -> H3+ + $\gamma$} & k$_{158}$ & $ = 10^{-16}$ & $$ & 81 \\

159 & \ch{C + e- -> C- + $\gamma$} & k$_{159}$ & $ = 2.25\times10^{-15}$ & $$ & 82 \\

160 & \ch{C + H -> CH + $\gamma$} & k$_{160}$ & $ = 10^{-17}$ & $$ & 83 \\

161 & \ch{C + H2 -> CH2 + $\gamma$} & k$_{161}$ & $ = 10^{-17}$ & $$ & 83 \\

162 & \ch{C + C -> C2 + $\gamma$} & k$_{162}$ & $ = 4.36\times10^{-18}\,{(T/300)}^{0.35} \operatorname{exp}\left(-\frac{161.3}{T}\right)$ & $$ & 84 \\

163 & \ch{C + O -> CO + $\gamma$} & k$_{163}$ & $ = 3.09\times10^{-17}\,{(T/300)}^{0.33} \operatorname{exp}\left(-\frac{1629}{T}\right)$ & $$ & 85 \\

164 & \ch{C+ + H -> CH+ + $\gamma$} & k$_{164}$ & $ = 4.46\times10^{-16}\,{T}^{-0.5} \operatorname{exp}\left(-4.93\,{T}^{-0.6667}\right)$ & $$ & 86 \\

165 & \ch{C+ + H2 -> CH2+ + $\gamma$} & k$_{165}$ & $ = 2\times10^{-16}\,{(T/300)}^{-1.3} \operatorname{exp}\left(-\frac{23}{T}\right)$ & $$ & 40 \\

166 & \ch{C+ + O -> CO+ + $\gamma$} & k$_{166}$ & $ = 2.5\times10^{-18}$ & $T < 300 \, \mathrm{K}$ & 40 \\

 &  &  & $ = 3.14\times10^{-18}\,{(T/300)}^{-0.15} \operatorname{exp}\left(-\frac{68}{T}\right)$ & $T > 300 \, \mathrm{K}$ &  \\

167 & \ch{O + e- -> O- + $\gamma$} & k$_{167}$ & $ = 1.5\times10^{-15}$ & $$ & 40 \\

168 & \ch{O + H -> OH + $\gamma$} & k$_{168}$ & $ = 9.9\times10^{-19}\,{(T/300)}^{-0.38}$ & $$ & 40 \\

169 & \ch{O + O -> O2 + $\gamma$} & k$_{169}$ & $ = 4.9\times10^{-20}\,{(T/300)}^{1.58}$ & $$ & 40 \\

170 & \ch{OH + H -> H2O + $\gamma$} & k$_{170}$ & $ = 5.26\times10^{-18}\,{(T/300)}^{-5.22} \operatorname{exp}\left(-\frac{90}{T}\right)$ & $$ & 40 \\

171 & \ch{H + $\gamma$ -> H+ + e-} & k$_{171}$ & $\begin{aligned}[t] = &\operatorname{\int \mathrm{d}E\, J(E)\, \sigma_{\mathrm{v96}}}\Bigl(E, 0.4298, 54750, 32.88, 2.963, 0, 0, 0\Bigr)\end{aligned}$ & $$ & 10 \\

172 & \ch{He + $\gamma$ -> He+ + e-} & k$_{172}$ & $\begin{aligned}[t] = &\operatorname{\int \mathrm{d}E\, J(E)\, \sigma_{\mathrm{v96}}}\Bigl(E, 13.61, 949.2, 1.469, 3.188, 2.039, 0.4434, 2.136\Bigr)\end{aligned}$ & $$ & 10 \\

173 & \ch{He+ + $\gamma$ -> He^{++} + e-} & k$_{173}$ & $\begin{aligned}[t] = &\operatorname{\int \mathrm{d}E\, J(E)\, \sigma_{\mathrm{v96}}}\Bigl(E, 1.72, 13690, 32.88, 2.963, 0, 0, 0\Bigr)\end{aligned}$ & $$ & 10 \\

174 & \ch{O + $\gamma$ -> O+ + e-} & k$_{174}$ & $\begin{aligned}[t] = &\operatorname{\int \mathrm{d}E\, J(E)\, \sigma_{\mathrm{v96}}}\Bigl(E, 1.24, 1745, 3.784, 17.64, 0.07589, 8.698, 0.1271\Bigr)\end{aligned}$ & $$ & 10 \\
175 & \ch{C + $\gamma$ -> C+ + e-} & k$_{175}$ & $= \textrm{table}$ & $$ & 43 \\
176 & \ch{H2 + $\gamma$ -> H2+ + e-} & k$_{176}$ & $= \textrm{table}$ & $$ & 43 \\
177 & \ch{H- + $\gamma$ -> H + e-} & k$_{177}$ & $= \textrm{table}$ & $$ & 43 \\
178 & \ch{CH + $\gamma$ -> C + H} & k$_{178}$ & $= \textrm{table}$ & $$ & 43 \\
179 & \ch{CH + $\gamma$ -> CH+ + e-} & k$_{179}$ & $= \textrm{table}$ & $$ & 43 \\
180 & \ch{C2 + $\gamma$ -> C + C} & k$_{180}$ & $= \textrm{table}$ & $$ & 43 \\
181 & \ch{OH + $\gamma$ -> O + H} & k$_{181}$ & $= \textrm{table}$ & $$ & 43 \\
182 & \ch{OH + $\gamma$ -> OH+ + e-} & k$_{182}$ & $= \textrm{table}$ & $$ & 43 \\
183 & \ch{H2O + $\gamma$ -> OH + H} & k$_{183}$ & $= \textrm{table}$ & $$ & 43 \\
184 & \ch{H2O + $\gamma$ -> H2O+ + e-} & k$_{184}$ & $= \textrm{table}$ & $$ & 43 \\
185 & \ch{O2 + $\gamma$ -> O2+ + e-} & k$_{185}$ & $= \textrm{table}$ & $$ & 43 \\
186 & \ch{O2 + $\gamma$ -> O + O} & k$_{186}$ & $= \textrm{table}$ & $$ & 43 \\
187 & \ch{H2 + $\gamma$ -> H+ + H + e-} & k$_{187}$ & $= \textrm{table}$ & $$ & 43 \\

188 & \ch{CO -> C + O} & k$_{188}$ & $ = \mathrm{(see\, sec.\, \ref{sec:external-isrf-as-diffuse-emission})}$ & $$ & 44 \\

189 & \ch{H2 -> H + H} & k$_{189}$ & $ = \mathrm{(see\, sec.\, \ref{sec:external-isrf-as-diffuse-emission})}$ & $$ & 44 \\
  \end{chemical_network_tabular}
\end{chemical_network_table}

\begin{chemical_network_table}{\chemicalNetworkTableCaptionCont}
  \begin{chemical_network_tabular}

190 & \ch{H2+ -> H + H+} & k$_{190}$ & $ = 1.1\times10^{-9}\,G_0 \operatorname{exp}\left(-1.9A_{\mathrm{v}}\right)$ & $$ & 45 \\

191 & \ch{H3+ -> H2 + H+} & k$_{191}$ & $ = 4.9\times10^{-13}\,G_0 \operatorname{exp}\left(-1.8A_{\mathrm{v}}\right)$ & $$ & 46 \\

192 & \ch{H3+ -> H2+ + H} & k$_{192}$ & $ = 4.9\times10^{-13}\,G_0 \operatorname{exp}\left(-2.3A_{\mathrm{v}}\right)$ & $$ & 46 \\

193 & \ch{C- -> C + e-} & k$_{193}$ & $ = 2.4\times10^{-7}\,G_0 \operatorname{exp}\left(-0.9A_{\mathrm{v}}\right)$ & $$ & 32 \\

194 & \ch{CH+ -> C + H+} & k$_{194}$ & $ = 2.6\times10^{-10}\,G_0 \operatorname{exp}\left(-2.5A_{\mathrm{v}}\right)$ & $$ & 47 \\

195 & \ch{CH2 -> CH + H} & k$_{195}$ & $ = 7.1\times10^{-10}\,G_0 \operatorname{exp}\left(-1.7A_{\mathrm{v}}\right)$ & $$ & 47 \\

196 & \ch{CH2 -> CH2+ + e-} & k$_{196}$ & $ = 5.9\times10^{-10}\,G_0 \operatorname{exp}\left(-2.3A_{\mathrm{v}}\right)$ & $$ & 32 \\

197 & \ch{CH2+ -> CH+ + H} & k$_{197}$ & $ = 4.6\times10^{-10}\,G_0 \operatorname{exp}\left(-1.7A_{\mathrm{v}}\right)$ & $$ & 48 \\

198 & \ch{CH3+ -> CH2+ + H} & k$_{198}$ & $ = 10^{-9}\,G_0 \operatorname{exp}\left(-1.7A_{\mathrm{v}}\right)$ & $$ & 32 \\

199 & \ch{CH3+ -> CH+ + H2} & k$_{199}$ & $ = 10^{-9}\,G_0 \operatorname{exp}\left(-1.7A_{\mathrm{v}}\right)$ & $$ & 32 \\

200 & \ch{O- -> O + e-} & k$_{200}$ & $ = 2.4\times10^{-7}\,G_0 \operatorname{exp}\left(-0.5A_{\mathrm{v}}\right)$ & $$ & 32 \\

201 & \ch{OH+ -> O + H+} & k$_{201}$ & $ = 10^{-12}\,G_0 \operatorname{exp}\left(-1.8A_{\mathrm{v}}\right)$ & $$ & 46 \\

202 & \ch{H2O+ -> H2+ + O} & k$_{202}$ & $ = 5\times10^{-11}\,G_0 \operatorname{f_{\mathrm{H}_{\mathrm{n}}\mathrm{O}^+}}\left(A_{\mathrm{v}}\right)$ & $$ & 49 \\

203 & \ch{H2O+ -> H+ + OH} & k$_{203}$ & $ = 5\times10^{-11}\,G_0 \operatorname{f_{\mathrm{H}_{\mathrm{n}}\mathrm{O}^+}}\left(A_{\mathrm{v}}\right)$ & $$ & 49 \\

204 & \ch{H2O+ -> O+ + H2} & k$_{204}$ & $ = 5\times10^{-11}\,G_0 \operatorname{f_{\mathrm{H}_{\mathrm{n}}\mathrm{O}^+}}\left(A_{\mathrm{v}}\right)$ & $$ & 49 \\

205 & \ch{H2O+ -> OH+ + H} & k$_{205}$ & $\begin{aligned}[t] = &1.5\times10^{-10}\,G_0 \operatorname{f_{\mathrm{H}_{\mathrm{n}}\mathrm{O}^+}}\Bigl(A_{\mathrm{v}}\Bigr)\end{aligned}$ & $$ & 49 \\

206 & \ch{H3O+ -> H+ + H2O} & k$_{206}$ & $\begin{aligned}[t] = &2.5\times10^{-11}\,G_0 \operatorname{f_{\mathrm{H}_{\mathrm{n}}\mathrm{O}^+}}\Bigl(A_{\mathrm{v}}\Bigr)\end{aligned}$ & $$ & 49 \\

207 & \ch{H3O+ -> H2+ + OH} & k$_{207}$ & $\begin{aligned}[t] = &2.5\times10^{-11}\,G_0 \operatorname{f_{\mathrm{H}_{\mathrm{n}}\mathrm{O}^+}}\Bigl(A_{\mathrm{v}}\Bigr)\end{aligned}$ & $$ & 49 \\

208 & \ch{H3O+ -> H2O+ + H} & k$_{208}$ & $\begin{aligned}[t] = &7.5\times10^{-12}\,G_0 \operatorname{f_{\mathrm{H}_{\mathrm{n}}\mathrm{O}^+}}\Bigl(A_{\mathrm{v}}\Bigr)\end{aligned}$ & $$ & 49 \\

209 & \ch{H3O+ -> OH+ + H2} & k$_{209}$ & $\begin{aligned}[t] = &2.5\times10^{-11}\,G_0 \operatorname{f_{\mathrm{H}_{\mathrm{n}}\mathrm{O}^+}}\Bigl(A_{\mathrm{v}}\Bigr)\end{aligned}$ & $$ & 49 \\

210 & \ch{H ->[CR] H+ + e-} & k$_{210}$ & $ = 0.46\,\zeta_{\mathrm{H}_{\mathrm{I}}}$ & $$ & 40 \\

211 & \ch{He ->[CR] He+ + e-} & k$_{211}$ & $ = 0.5\,\zeta_{\mathrm{H}_{\mathrm{I}}}$ & $$ & 40 \\

212 & \ch{O ->[CR] O+ + e-} & k$_{212}$ & $ = 2.8\,\zeta_{\mathrm{H}_{\mathrm{I}}}$ & $$ & 40 \\

213 & \ch{CO ->[CR] C + O} & k$_{213}$ & $ = 5\,\zeta_{\mathrm{H}_{\mathrm{I}}}$ & $$ & 40 \\

214 & \ch{CO ->[CR] CO+ + e-} & k$_{214}$ & $ = 3\,\zeta_{\mathrm{H}_{\mathrm{I}}}$ & $$ & 40 \\

215 & \ch{C2 ->[CR] C + C} & k$_{215}$ & $ = 237\,\zeta_{\mathrm{H}_{\mathrm{I}}}$ & $$ & 40 \\

216 & \ch{H2 ->[CR] H + H} & k$_{216}$ & $ = 0.1\,\zeta_{\mathrm{H}_{\mathrm{I}}}$ & $$ & 40 \\

217 & \ch{H2 ->[CR] H+ + H-} & k$_{217}$ & $ = 0.0003\,\zeta_{\mathrm{H}_{\mathrm{I}}}$ & $$ & 40 \\

218 & \ch{H2 ->[CR] H2+ + e-} & k$_{218}$ & $ = 0.93\,\zeta_{\mathrm{H}_{\mathrm{I}}}$ & $$ & 40 \\

219 & \ch{C ->[CR] C+ + e-} & k$_{219}$ & $ = 1020\,\zeta_{\mathrm{H}_{\mathrm{I}}}$ & $$ & 40 \\

220 & \ch{CH ->[CR] C + H} & k$_{220}$ & $ = 730\,\zeta_{\mathrm{H}_{\mathrm{I}}}$ & $$ & 40 \\

221 & \ch{O2 ->[CR] O + O} & k$_{221}$ & $ = 750\,\zeta_{\mathrm{H}_{\mathrm{I}}}$ & $$ & 40 \\

222 & \ch{O2 ->[CR] O2+ + e-} & k$_{222}$ & $ = 117\,\zeta_{\mathrm{H}_{\mathrm{I}}}$ & $$ & 40 \\

223 & \ch{OH ->[CR] O + H} & k$_{223}$ & $ = 510\,\zeta_{\mathrm{H}_{\mathrm{I}}}$ & $$ & 40 \\

224 & \ch{CH2 ->[CR] CH2+ + e-} & k$_{224}$ & $ = 500\,\zeta_{\mathrm{H}_{\mathrm{I}}}$ & $$ & 40 \\

225 & \ch{H2O ->[CR] OH + H} & k$_{225}$ & $ = 970\,\zeta_{\mathrm{H}_{\mathrm{I}}}$ & $$ & 40 \\

226 & \ch{HCO ->[CR] CO + H} & k$_{226}$ & $ = 421\,\zeta_{\mathrm{H}_{\mathrm{I}}}$ & $$ & 40 \\

227 & \ch{HCO ->[CR] HCO+ + e-} & k$_{227}$ & $ = 1170\,\zeta_{\mathrm{H}_{\mathrm{I}}}$ & $$ & 40 \\

228 & \ch{H2 ->[CR] H + H+ + e-} & k$_{228}$ & $ = 0.93\,\zeta_{\mathrm{H}_{\mathrm{I}}}$ & $$ & 40 \\

229 & \ch{C + C + M -> C2 + M} & k$_{229}$ & $ = 5.99\times10^{-33}\,{\left(T/5000\right)}^{-1.6} n_{\mathrm{H}_\mathrm{tot}}$ & $T < 5000 \, \mathrm{K}$ & 87 \\

 &  &  & $\begin{aligned}[t] = &5.99\times10^{-33}\,{\left(T/5000\right)}^{-0.64} \operatorname{exp}\Bigl(\frac{5255}{T}\Bigr)\, n_{\mathrm{H}_\mathrm{tot}}\end{aligned}$ & $T > 5000 \, \mathrm{K}$ &  \\

230 & \ch{C + O + M -> CO + M} & k$_{230}$ & $ = 6.16\times10^{-29}\,{\left(T/300\right)}^{-3.08} n_{\mathrm{H}_\mathrm{tot}}$ & $T < 2000 \, \mathrm{K}$ & 50 \\

 &  &  & $\begin{aligned}[t] = &2.14\times10^{-29}\,{\left(T/300\right)}^{-3.08} \operatorname{exp}\Bigl(\frac{2114}{T}\Bigr)\, n_{\mathrm{H}_\mathrm{tot}}\end{aligned}$ & $T > 2000 \, \mathrm{K}$ &  \\

231 & \ch{C+ + O + M -> CO+ + M} & k$_{231}$ & $ = 6.16\times10^{-27}\,{\left(T/300\right)}^{-3.08} n_{\mathrm{H}_\mathrm{tot}}$ & $T < 2000 \, \mathrm{K}$ & 51 \\

 &  &  & $\begin{aligned}[t] = &2.14\times10^{-27}\,{\left(T/300\right)}^{-3.08} \operatorname{exp}\Bigl(\frac{2114}{T}\Bigr)\, n_{\mathrm{H}_\mathrm{tot}}\end{aligned}$ & $T > 2000 \, \mathrm{K}$ &  \\

232 & \ch{C + O+ + M -> CO+ + M} & k$_{232}$ & $ = 6.16\times10^{-27}\,{\left(T/300\right)}^{-3.08} n_{\mathrm{H}_\mathrm{tot}}$ & $T < 2000 \, \mathrm{K}$ & 51 \\

 &  &  & $\begin{aligned}[t] = &2.14\times10^{-27}\,{\left(T/300\right)}^{-3.08} \operatorname{exp}\Bigl(\frac{2114}{T}\Bigr)\, n_{\mathrm{H}_\mathrm{tot}}\end{aligned}$ & $T > 2000 \, \mathrm{K}$ &  \\

233 & \ch{H + O + M -> OH + M} & k$_{233}$ & $ = 4.33\times10^{-32}\,{(T/300)}^{-1}\, n_{\mathrm{H}_\mathrm{tot}}$ & $$ & 52 \\

234 & \ch{OH + H + M -> H2O + M} & k$_{234}$ & $ = 2.56\times10^{-31}\,{(T/300)}^{-2}\, n_{\mathrm{H}_\mathrm{tot}}$ & $$ & 50 \\

235 & \ch{O + O + M -> O2 + M} & k$_{235}$ & $ = 9.2\times10^{-34}\,{(T/300)}^{-1}\, n_{\mathrm{H}_\mathrm{tot}}$ & $$ & 53 \\
  \end{chemical_network_tabular}
\end{chemical_network_table}

\begin{chemical_network_table}{\chemicalNetworkTableCaptionCont}
  \begin{chemical_network_tabular}

236 & \ch{H + H ->[dust] H2} & k$_{236}$ & $ = \mathrm{(see\, sec.\, \ref{sec:thermal-balance-dust-grains})}$ & $$ & 54 \\

237 & \ch{CO}$_\textrm{(gas)}$ \ch{-> CO}$_\textrm{(ice)}$ & k$_{237}$ & $ = \mathrm{(see\, sec.\, \ref{sect:freeze_out})}$ & $$ &  \\

238 & \ch{CO}$_\textrm{(ice)}$ \ch{-> CO}$_\textrm{(gas)}$ & k$_{238}$ & $ = \mathrm{(see\, sec.\, \ref{sect:freeze_out})}$ & $$ &  \\

239 & \ch{H2O}$_\textrm{(gas)}$ \ch{-> H2O}$_\textrm{(ice)}$ & k$_{239}$ & $ = \mathrm{(see\, sec.\, \ref{sect:freeze_out})}$ & $$ &  \\

240 & \ch{H2O}$_\textrm{(ice)}$ \ch{-> H2O}$_\textrm{(gas)}$ & k$_{240}$ & $ = \mathrm{(see\, sec.\, \ref{sect:freeze_out})}$ & $$ &  \\

  \end{chemical_network_tabular}

  \begin{minipage}{13cm}
  \begin{tabular}{@{}ll}
    \ontop{\textbf{Parameters:}} & 
    \ontop{
      \begin{tabular}{@{}r@{\:}l@{\:}@{}}
        $T_{\mathrm{e}}$ & $ = 8.617343\times10^{-5}T\,,\,\,T_{\mathrm{e}}$ is the temperature measured in eV \\ 
$k_{\mathrm{l,H}}$ & $ = 6.67\times10^{-12}\sqrt{T} \operatorname{exp}\left(-\left(1+63590\frac{1}{T}\right)\right)$ \\ 
$k_{\mathrm{h,H}}$ & $ = 3.52\times10^{-9}\operatorname{exp}\left(-43900\frac{1}{T}\right)$ \\ 
$n_{\mathrm{crit,H}}$ & $ = {10}\, \hat{ }\,\Bigl(3-0.416\ln(\frac{T}{10000})-0.327{\ln(\frac{T}{10000})}^2\Bigr)$ \\ 
$k_{\mathrm{l,H}_2}$ & $ = \frac{5.996\times10^{-30}{T}^{4.1881}}{{\left(1+6.761\times10^{-6}T\right)}^{5.6881}} \operatorname{exp}\left(-54657.4\frac{1}{T}\right)$ \\ 
$k_{\mathrm{h,H}_2}$ & $ = 1.3\times10^{-9}\operatorname{exp}\left(-53300\frac{1}{T}\right)$ \\ 
$n_{\mathrm{crit,H}_2}$ & $ = {10}\, \hat{ }\,\Bigl(4.845-1.3\ln(\frac{T}{10000})+1.62{\ln(\frac{T}{10000})}^2\Bigr)$ \\ 
$k_{\mathrm{l,He}}$ & $ = {10}\, \hat{ }\,\Bigl((-27.029+3.801\log(T)-29487 / T )\, a_\mathrm{He}\Bigr)$ \\
$k_{\mathrm{h,He}}$ & $ = {10}\, \hat{ }\,\Bigl((-2.729-1.75\,\log(T)-23474 / T)(1-a_\mathrm{He})\Bigr)$ \\
$n_{\mathrm{crit,He}}$ & $ = {10}\, \hat{ }\,\Bigl(5.0792 \left(1 - 1.23\times 10^{-5}\, (T - 2000)\right)\Bigr)$\\
$a_\mathrm{He}$ & $ = \left(1+n_{\mathrm{H}_\mathrm{tot}} / n_{\mathrm{crit,He}} \right)^{-1}$ \\ 

        $\sigma_{\mathrm{v96}}\left( E, E_0, \sigma_0, y_{\mathrm{a}}, P, y_{\mathrm{w}}, y_0, y_1 \right)$ & $ = 10^{-18} \sigma_0 \left( \left( x - 1\right)^2 + y_{\mathrm{w}} \right) y^{0.5P-5.5} \left( 1 + \sqrt{y / y_{\mathrm{a}}} \right)^{-P}$ \\
$x$ & $ = \frac{E}{E_0} - y_0$ \\
$y$ & $ = \sqrt{x^2 + y_1^2}$

      \end{tabular} } \\
  \end{tabular}

  \textbf{References:} (1) \cite{Janev1987Springer}, (2) \cite{Abel1997NewAstronomy,Ferland1992ApJ}, (3) \cite{Cen1992ApJS,Aldrovandi1973A&A}, (4) \ALADDIN; \cite{Abel1997NewAstronomy}, (5) \cite{Yoshida2006ApJ}, (6) \cite{Glover2007}, (7) \cite{Omukai2000}, (8) \cite{Millar1997}, (9) \cite{Harada2010ApJ}, (10) \cite{Verner1996ApJ}, (11) \cite{Kreckel2010Sci}, (12) \cite{Coppola2011ApJS,Ramaker1976PhRvA}, (13) \cite{Karpas1979JChPh}, (14) \cite{Grassi2011A&A,Savin2004ApJ}, (15) \cite{Mitchell1983ApJ,Corrigan1965JChPh}, (16) \cite{Lepp1983ApJ,Glover2010MNRAS}, (17) \cite{Abel1997NewAstronomy,Janev1987Springer}, (18) \cite{Stenrup2009PhRvA}, (19) \cite{Poulaert1978}, (20) \cite{Abel1997NewAstronomy,Schneider1994ApJ}, (21) \cite{Dalgarno1987IAUS}, (22) \cite{Glover2010MNRAS}, (23) \cite{Forrey2013ApJ}, (24) \cite{Glover2008MNRAS}, (25) \cite{Nahar1997}, (26) \cite{Nahar1999}, (27) \cite{Voronov1997}, (28) \cite{Stancil1999}, (29) \cite{Stancil1998}, (30) \cite{Zhao2004}, (31) \cite{Kimura1993}, (32) \cite{LeTeuff2000}, (33) \cite{Smith2002ApJ}, (34) \cite{Anicich2003}, (35) \cite{Dean1991}, (36) \cite{Harding1993}, (37) \cite{Loison2014MNRAS}, (38) \cite{Murrell1986}, (39) \UMIST, (40) \KIDA, (41) \cite{Baulch2005JPCRD}, (42) \cite{OConnor2015ApJS}, (43) ~PHIDRATES database\footnote{\url{http://phidrates.space.swri.edu}} \citep{Huebner1992ApSS}, (44) \cite{Richings2014II}, (45) \cite{Dunn1968}, (46) \cite{vanDishoeck1987}, (47) \cite{Roberge1991}, (48) \cite{vanDishoeck2006}, (49) \cite{Sternberg1995}, (50) \cite{Baulch1992JPCRD}, (51) \cite{Petuchowski1989}, (52) \cite{Tsang1986}, (53) \cite{Warnatz1984}, (54) \cite{Cazaux2009A&A,Grassi2017MNRAS}, (55) \cite{WALKAUSKAS1975691}, (56) \cite{1986JPCRD..15.1087T}, (57) \cite{2004MNRAS.350..323S}, (58) \cite{cohen1979evaluation}, (59) \cite{osti_4089545}, (60) \cite{linder1995reactive}, (61) \cite{1992A&A...255..453S}, (62) \cite{1984MNRAS.211..857A}, (63) \cite{1999ApJ...513..287M}, (64) \cite{1980ApJ...236..492V}, (65) \cite{1977IJMSI..23..123S,1977CPL....47..383A}, (66) \cite{1981CPL....77..484J}, (67) \cite{F29807601084}, (68) \cite{1992IJMSI.117..457S}, (69) \cite{1984PhRvL..52.2084F}, (70) \cite{1994JPhB...27.2551P}, (71) \cite{2007A&A...466.1197W,2004PhRvA..70e2716M}, (72) \cite{0953-4075-24-3-026}, (73) \cite{1998ApJ...505..459L}, (74) \cite{1990PhR...186..215M}, (75) \cite{1995JChPh.102.1699G}, (76) \cite{2000FaDi..115..295R}, (77) \cite{2000ApJ...543..764J}, (78) \cite{1983JPhB...16.1433A}, (79) \cite{1998PhRvA..57.4462R}, (80) \cite{2005JPhCS...4...26G}, (81) \cite{gerlich1992experimental}, (82) \cite{1998FaDi..109...61S}, (83) \cite{Prasad1980ApJS}, (84) \cite{1997MNRAS.287..287A}, (85) \cite{1999MNRAS.303..235S}, (86) \cite{2006ApJ...636..923B}, (87) \cite{LeTeuff2000,1969RSPSA.312..207F,1976JChPh..64..228S}, (88) \cite{2009MNRAS.393..911G}

  \end{minipage}
  
\end{chemical_network_table}

\bsp
\label{lastpage}
\end{document}